\documentstyle[twoside,12pt,epsf,aps]{revtex}
\relax

\newcommand{\MET}{\mbox{$\protect \raisebox{.3ex}{$\not$}\et$}}

\def\W{{\em W\/ }}
\def\Z0{${\em Z^0\/}$}

\def\r#1 {$^{#1}$}

\newcommand{\et}{{\rm E}_{\scriptscriptstyle\rm T}}

\newcommand{\met}{\mbox{$\protect \raisebox{.3ex}{$\not$}\et \ $}}

\newcommand{\ttbar}{t\bar{t}}
\newcommand{\bbbar}{b\bar{b}}
\newcommand{\ccbar}{c\bar{c}}

\newcommand{\gev}  { {\rm GeV}}

\newcommand{\gevc} { {\rm GeV/c  }}
\newcommand{\gevcc}{ {\rm GeV/c^2}}

\def\gepsfcentered#1{
  \def\testit{#1}
  \def\lbracket{[}
  \ifx\testit\lbracket
    \let\dofilecmd=\gepsfwithopt
  \else
    \let\dofilecmd=\gepsfnoopt
  \fi
  \dofilecmd}

\def\gepsfnoopt#1{
  \begin{center}
  \leavevmode
  \epsffile{#1}
  \end{center}}

\def\gepsfwithopt#1 #2 #3 #4]#5{
  \begin{center}
  \leavevmode
  \gepsfmaxx=0.94\textwidth
  \epsffile[#1 #2 #3 #4]{#5}
  \end{center}}

\newdimen\gepsfmaxx
\def\epsfsize#1#2{
  \ifnum \epsfxsize=0
    \ifnum \epsfysize=0
      \ifnum #1 > \gepsfmaxx
        \gepsfmaxx
      \else
        #1
      \fi
    \else
      \epsfxsize
    \fi
  \else
    \epsfxsize
  \fi
}

\begin{document}
\bibliographystyle{unsrt}
\title{Measurement of the $t\overline{t}$ production cross section 
 in $p\bar{p}$ collisions at $\sqrt{s}$ = 1.8 TeV}
\maketitle
\font\eightit=cmti8
\def\r#1{\ignorespaces $^{#1}$}
\hfilneg
\begin{sloppypar}
\noindent
T.~Affolder,\r {23} H.~Akimoto,\r {45}
A.~Akopian,\r {38} M.~G.~Albrow,\r {11} P.~Amaral,\r 8 S.~R.~Amendolia,\r {34} 
D.~Amidei,\r {26} K.~Anikeev,\r {24} J.~Antos,\r 1 
G.~Apollinari,\r {11}
% ...
\end{sloppypar}
\vskip .026in
\begin{center}
(CDF Collaboration)
\end{center}

\vskip .026in
\begin{center}
\r 1  {\eightit Institute of Physics, Academia Sinica, Taipei, Taiwan 11529, 
Republic of China} \\
\r 2  {\eightit Argonne National Laboratory, Argonne, Illinois 60439} \\
\r 3  {\eightit Istituto Nazionale di Fisica Nucleare, University of Bologna,
I-40127 Bologna, Italy} \\
\r 4  {\eightit Brandeis University, Waltham, Massachusetts 02254} \\
\r 5  {\eightit University of California at Davis, Davis, California  95616} \\
\r 6  {\eightit University of California at Los Angeles, Los 
Angeles, California  90024} \\  
\r 7  {\eightit Instituto de Fisica de Cantabria, CSIC-University of Cantabria, 
39005 Santander, Spain} \\
\r 8  {\eightit Enrico Fermi Institute, University of Chicago, Chicago, 
Illinois 60637} \\
\r 9  {\eightit Joint Institute for Nuclear Research, RU-141980 Dubna, Russia}
\\
\r {10} {\eightit Duke University, Durham, North Carolina  27708} \\
\r {11} {\eightit Fermi National Accelerator Laboratory, Batavia, Illinois 
60510} \\
\r {12} {\eightit University of Florida, Gainesville, Florida  32611} \\
\r {13} {\eightit Laboratori Nazionali di Frascati, Istituto Nazionale di Fisica
               Nucleare, I-00044 Frascati, Italy} \\
\r {14} {\eightit University of Geneva, CH-1211 Geneva 4, Switzerland} \\
\r {15} {\eightit Glasgow University, Glasgow G12 8QQ, United Kingdom}\\
\r {16} {\eightit Harvard University, Cambridge, Massachusetts 02138} \\
\r {17} {\eightit Hiroshima University, Higashi-Hiroshima 724, Japan} \\
\r {18} {\eightit University of Illinois, Urbana, Illinois 61801} \\
\r {19} {\eightit The Johns Hopkins University, Baltimore, Maryland 21218} \\
\r {20} {\eightit Institut f\"{u}r Experimentelle Kernphysik, 
Universit\"{a}t Karlsruhe, 76128 Karlsruhe, Germany} \\
\r {21} {\eightit Center for High Energy Physics: Kyungpook National
University, Taegu 702-701; Seoul National University, Seoul 151-742; and
SungKyunKwan University, Suwon 440-746; Korea} \\
\r {22} {\eightit High Energy Accelerator Research Organization (KEK), Tsukuba, 
Ibaraki 305, Japan} \\
\r {23} {\eightit Ernest Orlando Lawrence Berkeley National Laboratory, 
Berkeley, California 94720} \\
\r {24} {\eightit Massachusetts Institute of Technology, Cambridge,
Massachusetts  02139} \\   
\r {25} {\eightit Institute of Particle Physics: McGill University, Montreal 
H3A 2T8; and University of Toronto, Toronto M5S 1A7; Canada} \\
\r {26} {\eightit University of Michigan, Ann Arbor, Michigan 48109} \\
\r {27} {\eightit Michigan State University, East Lansing, Michigan  48824} \\
\r {28} {\eightit University of New Mexico, Albuquerque, New Mexico 87131} \\
\r {29} {\eightit The Ohio State University, Columbus, Ohio  43210} \\
\r {30} {\eightit Osaka City University, Osaka 588, Japan} \\
\r {31} {\eightit University of Oxford, Oxford OX1 3RH, United Kingdom} \\
\r {32} {\eightit Universita di Padova, Istituto Nazionale di Fisica 
          Nucleare, Sezione di Padova, I-35131 Padova, Italy} \\
\r {33} {\eightit University of Pennsylvania, Philadelphia, 
        Pennsylvania 19104} \\   
\r {34} {\eightit Istituto Nazionale di Fisica Nucleare, University and Scuola
               Normale Superiore of Pisa, I-56100 Pisa, Italy} \\
\r {35} {\eightit University of Pittsburgh, Pittsburgh, Pennsylvania 15260} \\
\r {36} {\eightit Purdue University, West Lafayette, Indiana 47907} \\
\r {37} {\eightit University of Rochester, Rochester, New York 14627} \\
\r {38} {\eightit Rockefeller University, New York, New York 10021} \\
\r {39} {\eightit Rutgers University, Piscataway, New Jersey 08855} \\
\r {40} {\eightit Texas A\&M University, College Station, Texas 77843} \\
\r {41} {\eightit Texas Tech University, Lubbock, Texas 79409} \\
\r {42} {\eightit Istituto Nazionale di Fisica Nucleare, University of Trieste/
Udine, Italy} \\
\r {43} {\eightit University of Tsukuba, Tsukuba, Ibaraki 305, Japan} \\
\r {44} {\eightit Tufts University, Medford, Massachusetts 02155} \\
\r {45} {\eightit Waseda University, Tokyo 169, Japan} \\
\r {46} {\eightit University of Wisconsin, Madison, Wisconsin 53706} \\
\r {47} {\eightit Yale University, New Haven, Connecticut 06520} \\
\r {(\ast)} {\eightit Now at Carnegie Mellon University, Pittsburgh,
Pennsylvania  15213}
\end{center}

\vspace{0.2em}
\begin{abstract}
      We update the measurement of the top production cross section  
       using the CDF detector at the Fermilab Tevatron.
      This measurement uses $t\bar{t}$ decays to the final states $e+\nu$+jets
      and $\mu+\nu$+jets. We search for $b$ quarks from $t$ decays
      via secondary-vertex identification or the identification
      of  semileptonic decays of the 
      $b$ and cascade $c$ quarks. The background to the $t\bar{t}$ production
      is determined primarily through a Monte Carlo simulation. However, we 
      calibrate the simulation and evaluate its uncertainty using several 
      independent data samples. 
      For a top mass of 175 $\gevcc$, we measure $\sigma_{t\bar{t}}=5.1 \pm 1.5$ pb and 
      $\sigma_{t\bar{t}}=9.2 \pm 4.3$ pb using the secondary vertex and 
      the lepton tagging algorithms, respectively. Finally, we combine these
      results with those from other $t\bar{t}$ decay channels and obtain
     $\sigma_{t\bar{t}} = 6.5^{+1.7}_{-1.4}$ pb.  \\

 PACS number(s): 14.65.Ha, 13.85.Ni, 13.85.Qk
\end{abstract}
\section{Introduction}~\label{intro}
      In the Standard Model (SM) the top quark completes the third fermion
      generation. The measurement of the top-quark pair production cross
      section $\sigma_{t\bar{t}}$ is of interest as a test of quantum chromodynamics (QCD) 
      predictions. Aside from the obvious observation that a 
      deviation from these
      predictions could be indicative of new physics, recent QCD calculations
      predict $\sigma_{t\bar{t}}$ with an uncertainty smaller than 
      15\%~\cite{catani} which motivates measurements of
      comparable precision.

      In $p\bar{p}$ collisions at $\sqrt{s}$ = 1.8 TeV top quarks  are
      pair produced through $q\bar{q}$ annihilation ($\simeq$ 90\%)
      or gluon fusion ($\simeq$ 10\%). In the SM framework the top quark
      decays into a $\W$ boson and a $b$ quark. When one of the $\W$ bosons
      decays to an electron or a muon, the final state includes a charged 
      lepton with high transverse momentum ($p_T$), a large transverse energy 
      imbalance from the undetected neutrino, referred to as $\MET$, and 
      four jets from the hadronized quarks. However, due to gluon radiation 
      or jet merging, the number of detected jets may vary. We measure 
      $\sigma_{t\bar{t}}$ using this final state, referred to in
      this paper as $\W+$ jets
      and selected from the data (105.1 pb$^{-1}$) 
      collected by CDF in the $1992-1995$  collider run.

      The same data set has been used in the previous CDF measurement of 
      $\sigma_{t\bar{t}}$~\cite{cdf_xsec}. 
      This paper revises that measurement and expands on many of
      the analysis details.
      The selection of the $\W+$ jet sample follows the 
      guide-lines used in all previous CDF measurements of the
      top mass and production cross section~\cite{topxs,topmass}.

      As done in previous analyses, we employ two techniques to enhance the 
      relative fraction of events coming from top quark decays with respect 
      to the background. The first method searches a jet for the presence 
      of a secondary vertex reconstructed using the silicon vertex detector
      (SVX) and displaced from the primary event vertex due to the long 
      $b$-quark lifetime (SECVTX tag). The second method searches a jet for the 
      presence of a lepton, indicative of a semileptonic $b$-decay. Since 
      these leptons typically have low momentum compared to the lepton from
      the $W$ decay, they are referred to as soft lepton tags (SLT).

      In this analysis we use the same SECVTX and SLT algorithms as in 
      Ref.~\cite{cdf_xsec}. Differently from 
      Refs.~\cite{cdf_xsec,topxs,top_prd}, we search jets and not events for 
      soft lepton tags; this approach provides the rates of backgrounds for the top 
      mass measurement which uses the same technique~\cite{topmass}.

      As a cross-check,
      we take advantage of a third algorithm, jet-probability, which 
      uses the impact parameter significance of all tracks in a jet to derive
      a probability that the jet originates from the primary event 
      vertex~\cite{aleph}. Jets with small probability of having 
      zero lifetime are considered jet-probability tags (JPB).
      The value of the jet-probability threshold  is tuned to 
      have higher tagging efficiency than SECVTX in jets originating from
      $c$ quarks and a higher rate of fake tags in jets without heavy flavor.
      Although similar to SECVTX in that it is based on
      track information from displaced vertices, it is used here to provide
      important cross-checks of the background determination and measured
      cross section.

      The method used to measure $\sigma_{t\bar{t}}$ is
      outlined in Ref.~\cite{top_prd} and has since been improved.
      As  summarized in Ref.~\cite{cdf_xsec}, the method relies
      on the  calculation  of all the background contributions to 
      the tagged $W$+jet sample. The excess
      over background of the $W+3,4$ jet events with 
      at least one tag  is attributed to $t\bar{t}$ production and 
      used to derive $\sigma_{t\bar{t}}$.
  
      The major sources of background are the processes 
      $p\bar{p} \rightarrow Wg$ with $g \rightarrow b\bar{b}$, $c\bar{c}$ 
      (referred to as gluon splitting) and $p\bar{p}\rightarrow Wc$.
      The second largest source of background is mistags (tags in jets which
       do not contain heavy flavor). Smaller contributions come from 
      other processes like non-$W$ production, single top production, 
      $WW$, $WZ$, $ZZ$ and $Z\rightarrow \tau \tau$. 

      The method used to measure $\sigma_{t\bar{t}}$ relies on the correct 
      calibration of the Monte 
      Carlo generators and  the detector simulation. 
      Simulated events are produced with the Herwig~\cite{herwig}
      or Pythia~\cite{pythia} Monte Carlo generators. Hadrons with heavy 
      flavor ($b$ and $c$)
      are decayed using the {\sc CLEO} Monte Carlo (QQ)~\cite{cleomc}.
      All other particles are decayed, when appropriate, by the
      CDF detector 
      simulation (QFL) which uses its own lifetime table
      for $b$ and $c$-hadrons. QFL simulates the interaction
      of all particles in the final state with the CDF detector; the 
      detector response is based on parametrizations
      that are  functions of the particle kinematics and have been derived 
      using the data.
 
      This paper describes the work done to understand
      and improve the calibrations  used in 
      the method  to calculate the background to $t\bar{t}$
      events using independent data samples and the corresponding 
      simulations. This work was primarily focussed on the components 
      with the largest influence on the determination of $\sigma_{t\bar{t}}$; 
      mistags, the efficiencies of the tagging algorithms, and the 
      fraction of $W+$ jet direct production 
      which contains heavy flavor.
      We summarize here the relevant conclusions:
\begin{itemize}
\item  we find that the SECVTX tagging efficiency for $b$-quark
       jets ($b$-jets) is ($25\pm13$)\% higher in data control samples
       than it is in the Monte Carlo simulation of the same processes.
       Therefore we conclude that the $b$-jet tagging rate in
       Refs.~\cite{cdf_xsec,topxs} is underestimated by this factor.
       This data-to-simulation discrepancy is largely due to errors in the
       simulation that were found {\it a~posteriori}.  Instead of remaking
       the large
       Monte Carlo samples used in this paper we chose to rescale the simulated
       $b$-quark tagging rate by the factor of 1.25$\pm$0.13.
\item  we find that the rate of SECVTX mistags in jets without heavy flavor
       is (50$\pm$5)\% smaller than what we estimated 
       in Refs.~\cite{cdf_xsec,topxs}.
\item  we find the fraction of $g \rightarrow b\bar{b}$ and 
       $g \rightarrow c\bar{c}$ in the $\W+$ jet direct production evaluated 
       with Herwig needs to be increased by  ($39 \pm 19$)\% and 
       ($35 \pm 36$)\%, respectively.
\end{itemize}
      These last two effects tend to cancel, leaving the net background to top
      approximately unchanged from our previous results in Refs.~\cite{cdf_xsec,topxs}.
      In the $W+3,4$ jets sample we observe 29 events with
      one or more  SECVTX tags and 25 events with one or more SLT 
      tags. The expected backgrounds 
      are $8.0\pm 1.0$ and $13.2\pm1.2$ events, respectively.
      The excess of SECVTX tags yields the cross section
      $\sigma_{t\bar{t}}$ = $5.08\pm1.54$ pb  and the excess of SLT tags
      yields $\sigma_{t\bar{t}}$ = $9.18\pm4.26$ pb for a top quark mass 
      of $175$ GeV/$c^2$.

      Following a brief description of the CDF detector in 
      Section~\ref{sec:s-det}, Section~\ref{sec:data} 
      describes the triggers and the reconstruction of leptons, 
      jets and the missing transverse energy.  The selection of 
      $\W+$ jet events is detailed in Section~\ref{sec:W-sel}, along
      with the selection of the $Z+$ jet sample, which will be used to
      check the background calculation. 
      The selection of other data  samples  used to calibrate the event generators 
      and the detector simulation is described in Section~\ref{sec:other_samples}. 
      Sections~\ref{sec:degrade} and~\ref{sec:Simul} are dedicated to the 
      Monte Carlo generators and the CDF detector simulation (QFL), respectively.
      Section~\ref{sec:taggers} describes the algorithms used for
      the identification of the heavy 
      flavor. The efficiency of those algorithms is calculated
      in Section~\ref{sec:simu_mod}, which also includes numerous checks of the
      result and an evaluation of its systematic uncertainty.
      The new method for evaluating 
      mistags and the determination of its accuracy is described in
Section~\ref{sec:mistags}.
      Section~\ref{sec:herwig_norm} details the calibration 
      of the $g \rightarrow b\bar{b}$ and $g \rightarrow c\bar{c}$ cross 
      sections predicted by the Herwig generator. Section~\ref{sec:phys_bkgd} 
      describes the calculation of the backgrounds to the $t\bar{t}$ production.
      In Section~\ref{sec:zjets_Tags}, we check the background calculation 
      using the $Z+$ jet sample. Additional checks of the background calculation 
      are described in Section~\ref{sec:w_neg_Tags}.
      Finally, $\sigma_{t\bar{t}}$ is derived in Section~\ref{sec:xsec}.
      In Section~\ref{sec:comb}, we combine the present results with
      previous CDF measurements of $\sigma_{t\bar{t}}$ that are 
	derived using different data sets.  We conclude in Section~\ref{sec:concl}.
\section{ The CDF detector}~\label{sec:s-det}
      CDF is a general purpose detector with  azimuthal and 
      forward-backward symmetry designed to study $p\bar{p}$ interactions.
      The CDF coordinate system has the $z$-axis pointing along the proton 
      momentum and the $x$-axis located in the horizontal plane of 
      the Tevatron storage ring pointing radially outward so that the $y$-axis
      points up. The coordinates $r$-$\phi$ are the standard cylindrical 
      coordinates. A complete description of CDF can be found in 
      Refs.~\cite{top_prd,cdf_nim}. The detector components most relevant
      to this analysis are summarized below. 

      A superconducting solenoid of length 4.8 m and radius 1.5 m 
      generates a 1.4-T magnetic field. The solenoid contains 
      three types of tracking chambers for detecting charged 
      particles and measuring their momenta. A four layer
      silicon microstrip vertex detector (SVX) surrounds the
      beryllium beam pipe of radius 1.9 cm. The SVX has an active 
      length of 51 cm; the four layers of the SVX are at distances 
      of approximately 2.9, 4.2, 5.5 and 7.9 cm from the beamline. 
      Axial micro-strips
      with 60-$\mu$m pitch provide accurate track reconstruction in 
      the $r$-$\phi$ plane transverse to the beam~\cite{SVX_nim}. Outside the 
      SVX there is a vertex drift chamber (VTX) which provides track 
      information up to a radius of 22 cm and for pseudo-rapidities 
      $|\eta |\leq$ 3.5. The VTX measures the $z$-position 
      of the primary vertex. Both the SVX 
      and the VTX are mounted inside the CTC, a 3.2 m long drift chamber
      with an outer radius of 132 cm containing 84 concentric,
      cylindrical layers of sense wires, which are grouped into 
      8 alternating axial and stereo superlayers. The solenoid is 
      surrounded by sampling calorimeters used to measure the 
      electromagnetic and hadronic energy of jets and electrons.
      The calorimeters cover the  pseudo-rapidity range 
      $|\eta|\leq$ 4.2. The calorimeters are segmented in 
      $\eta$-$\phi$ towers pointing to the nominal interaction 
      point. There are three separate $\eta$-regions of calorimeters. 
      Each region has an electromagnetic calorimeter [central (CEM), 
      plug (PEM) and forward (FEM)] and behind it a hadron calorimeter
      [CHA, PHA and FHA, respectively]. Located six radiation lengths 
      inside the CEM calorimeter, proportional wire chambers (CES) 
      provide shower-position measurements in the $z$ and $r-\phi$ view.
      Proportional chambers (CPR) located between the solenoid 
      and the CEM detect early development of electromagnetic showers 
      in the solenoid coil. These chambers provide $r$-$\phi$ 
      information only.

      The calorimeter acts as a hadron absorber for the central muon
      detection system (CMU).
      The CMU consists of four layers of drift chambers located outside 
      the CHA calorimeter. The CMU system covers the pseudo-rapidity 
      $|\eta| \leq$0.6 and can be reached by muons with 
      $p_T \geq 1.4$ GeV/c. The CMU system is followed by 0.6 m of steel 
      and four additional layers of drift chambers (CMP).
      The  system of drift chambers CMX 
      extends the  muon detection to $|\eta| \leq $ 1.0. 
\clearpage
\section{Data collection and identification of high $p_T$ leptons and jets}
        ~\label{sec:data}
      The  last collider run, called Run I, lasted from August of 1992 till 
      July of 1993 (Run  1A) and from January of 1994 till July of 1995, 
      (Run 1B).  The data collected during this run correspond to an 
      integrated luminosity of $105.1 \pm 4.0$~ pb$^{-1}$ when using 
      the total $p\bar{p}$ cross section value $80.03 \pm 2.24$ mb 
      ~\cite{paolo_pp}. We begin this section with a description of the
      triggers used in this analysis. This is followed by subsections
      on the reconstruction and identification of electrons, muons, jets
      and neutrinos.
 \subsection{Triggers}
      A three-level trigger system is used  to  select
      events originating from $p \bar{p}$ interactions and containing
      electrons, muons, jets or missing transverse energy ($\met$). 

      The first-level trigger (L1) accepts events based on the identification 
      of energy clusters in the calorimeter or track segments in the muon 
      chambers.  The L1 calorimeter trigger requires a single CEM or CHA 
      trigger tower with transverse energy greater than 8 or 12 GeV, 
      respectively (these  thresholds were set at 6 and 8 GeV during Run 1A). 
      The L1 muon trigger infers the track momentum from  the deflection 
      of the track segment in the muon chambers due to the  magnetic field;
      it requires a minimum transverse momentum of 6 and 10 \gevc in the CMU 
      and CMX chambers, respectively. A minimum energy of 300 MeV is required 
      in the hadron calorimeter tower associated with the track segment.

      The second-level trigger (L2) uses the calorimetry information with 
      greater sophistication. The L2 trigger is a fastbus based
      processor~\cite{L2_FB} with a decision time of approximately 20 $\mu$s. 
      It combines calorimetry towers forming electromagnetic and jet-like 
      clusters. An electromagnetic cluster is constructed as a set of 
      contiguous CEM (PEM)  towers each with $E_T \geq$ 7 GeV (4 GeV),
      including at least one seed tower with $E_T \geq$ 8 GeV (6 GeV).

      The L2 jet clusters are formed starting with a seed tower with 
      $E_T \geq3$ GeV and summing all contiguous towers with 
      $E_T \geq 1$ GeV.
      A crude estimate of $\met$ is also available at this trigger level.
      The  L2 trigger utilizes the list of $r-\phi$ tracks provided
      by the central fast tracker (CFT), a hardware processor which uses 
      fast timing information from the CTC as input. The events
      used in this analysis were collected using the L2 
      high-$p_T$ electron trigger, which 
      requires an electromagnetic cluster of transverse 
      energy greater than 16 GeV matched by a CFT track with transverse 
      momentum $p_T \geq 12$ GeV/c. A second trigger requires an 
      electromagnetic cluster of $E_T \geq 16$ GeV and $\met \geq 20$ GeV and 
      is used to recover losses due to the CFT inefficiency. The L2  
      high-$p_T$ muon  
      trigger requires a CFT track with $p_T \geq 12$ GeV/c pointing within 
      $5^{\circ}$ to a L1 track segment in the muon detectors. To ensure good
      efficiency, additional L2 muon triggers require only a L1 track 
      segment accompanied by at least one jet cluster with $E_T \geq$ 15 GeV 
      or $\met \geq$ 35 GeV.

      The L3 trigger decision is made after the full event reconstruction.
      Events accepted by the L2 trigger are processed by a farm of SGI
      processors running the full off-line reconstruction package.
	The Level~3 electron trigger requires a CEM cluster with 
      $E_T\geq 18$ GeV and a reconstructed track with $p_T \geq 13$ GeV/c
      pointing to it. The ratio of hadronic to electromagnetic energy in 
      the cluster is required to be less than 0.125. The Level~3 muon trigger
      requires a match within 10 cm in the $r-\phi$ plane between a
      reconstructed track with $p_T>18$ GeV/c extrapolated to the radius 
      of the muon detectors and a track segment in the muon chambers.

      Trigger efficiencies have been measured directly using events 
      with overlapping triggers. The electron trigger efficiency is found 
      to be larger than 99.6\% for electrons inside the detector
      fiducial volume. Likewise, the muon trigger efficiency 
      is ($70\pm 2$)\%; this includes an inefficiency due to the fact
      that the muon trigger does not cover the entire detector fiducial volume. 
      The measured trigger efficiencies have been included in the detector 
      simulation described in Section~\ref{sec:Simul}. 
      A check of the muon trigger simulation was performed by comparing 
      the rate of $W \rightarrow \mu \nu$ events in the data to that of a
      simulation of this process using the Herwig generator
      (see Section~\ref{sec:degrade}) normalized to the same number of
      $W \rightarrow e \nu$ events. We observe  agreement between data 
      and simulation within 10\%, and this difference is taken as
      the systematic error on the muon trigger simulation.
 \subsection{Electron reconstruction}\label{dataselec}
      The $\W+$ jet sample is selected requiring electrons reconstructed in 
      the central pseudo-rapidity region $|\eta| \leq$ 1. Stricter cuts,
      described in detail in Ref.~\cite{top_prd}, are applied to central 
      electron candidates which passed the trigger prerequisites.      
      The following variables are used to discriminate against charged hadrons:
      (1) the ratio  of hadronic to electromagnetic energy of the cluster,
      $E_{had}/E_{em}$; (2) the ratio of cluster energy to track
      momentum, $E/P$; (3) a comparison of the lateral shower profile
      in the calorimeter cluster with that of test-beam electrons, $L_{shr}$;
      (4) the distance between the extrapolated track-position and the 
      CES measurement in the $r-\phi$ and $z$ views, $\Delta x$ and $\Delta z$;
      (5) a $\chi^{2}$ comparison of the CES shower profile with those of 
      test-beam electrons, $\chi^{2}_{strip}$; (6) the interaction vertex 
      position, $z_{ver}$ and the distance between the interaction vertex and 
      the reconstructed track in the $z$-direction, $z$-vertex match; and 
      (7) the isolation, $I$, defined as the ratio of the
      additional transverse energy in a cone of radius 
      $R=0.4$ around the cluster axis  to the transverse energy of the 
      electron  cluster. The  electron selection criteria
      are listed in Table~\ref{tab:tab_3_0}.

      Fiducial cuts on the electromagnetic shower position, as measured 
      in the CES,
      are applied to insure that the electron candidate is away from the
      calorimeter boundaries and the energy is well measured.

      Electrons from  photon conversions are removed using an algorithm based 
      on tracking information.  Electron tracks close to a companion track 
      with opposite charge are considered conversion candidates. The following
      variables are  used to identify and remove photon conversions:
	(1) the difference of the  polar angles,  $\delta \cot \theta$; 
      (2) the distance between the two tracks in the $r-\phi$ plane at the 
      radius $R_{conv}$ where the tracks are parallel, $\Delta_{sep}$; and 
	(3) the conversion radial position, $R_{conv}$. If a companion
	track is not found, we identify conversion candidates using 
      $f_{VTX}$ which is the ratio
      of the measured to expected number of VTX hits 
      associated to the electron candidate. Table~\ref{tab:tab_3_1} summarizes
      the criteria used to identify and remove electrons from photon 
      conversions. The efficiency of the conversion algorithm is measured with
      a sample of photon conversions selected using the CPR detector.
      The efficiency of the conversion removal algorithm is 90.7 $\pm$ 3.8\%. 
      The fraction of electrons erroneously removed is estimated using a 
      sample of $Z \rightarrow e^+ e^-$ events to be 2.2 $\pm$ 0.6\% and is 
      properly accounted for by the simulation.

	The total primary electron identification efficiency has been measured 
      using a sample of $Z \rightarrow e^+ e^-$ decays and is listed in 
	Table~\ref{tab:tab_3_4}.

      When an electron candidate is found, the calorimeter towers belonging to the 
      electron cluster are not used by the jet clustering algorithm.
\subsection{Muon reconstruction}\label{muselec}
      Muons are identified in the $|\eta| \leq 1.0$ region by extrapolating CTC 
      tracks to the muon detectors and matching them to track segments 
      reconstructed in the muon chambers. The following variables, described 
      in detail  in Ref.~\cite{top_prd}, are used to separate muon candidates
      from cosmic rays and from hadrons not contained by  the calorimeter:
      (1) an energy deposition in the electromagnetic and hadronic calorimeter
      characteristic of minimum ionizing particles, $E_{em}$ and $E_{had}$; 
      (2) the distance of closest approach of the reconstructed track to the 
      beam line, $d$; (3) the $z$-vertex match; (4) the matching distance 
      between the extrapolated track and the track segment in the muon chamber, 
      $\Delta x = r \Delta \phi$; and (5) the isolation $I$, the ratio of
      additional transverse energy in a cone of radius $R=0.4$ around the 
      track direction to the muon transverse momentum.

      The muon selection criteria are listed in Table~\ref{tab:tab_3_3}.
      The muon identification efficiency has been measured using a sample
      of $Z \rightarrow \mu^+ \mu^-$ decays and is listed in 
      Table~\ref{tab:tab_3_4}.

      Leptons passing the requirements listed in Table~\ref{tab:tab_3_0}
	and~\ref{tab:tab_3_3} are labeled primary leptons. As a consequence of the
      high luminosity of the collider, approximately 50\% of the events with a 
      primary lepton contain multiple interactions which result in more than 
      one primary vertex in the event. The ambiguity is resolved by selecting 
      the vertex associated with the primary lepton track to evaluate  jet
      pseudorapidities and the missing transverse energy.

\newpage
\narrowtext
\begin{table}[p]
\def\arraystretch{0.8}
\begin{center}
\caption[]{Selection requirements for primary electrons.} 
 \begin{tabular}{lc}
  Variable         &  Cut \\
\hline
$E/P$            &   $\leq$ 1.5 \\
$E_{had}/E_{em}$ &   $\leq$ 0.05   \\
$L_{shr}$        &  $\leq$ 0.2 \\
$|\Delta x|$     & $\leq$ 1.5 cm \\
$|\Delta z|$     & $\leq$ 3.0 cm \\
$\chi^{2}_{strip}$& $\leq$ 10.0 \\
$z_{ver}$	   & $\leq$ 60.0 cm\\
$z$-vertex match & $\leq$ 5.0 cm \\
$I$              & $\leq$ 0.1\\
 \end{tabular}
\label{tab:tab_3_0}
\end{center}
\end{table}
\widetext

\narrowtext
\begin{table}[p]
\def\arraystretch{0.8}
\begin{center}
 \caption[]{Criteria used to identify electrons  from  photon conversions.}
 \begin{tabular}{lc}
  Variable         &  Cut \\
\hline
 $|\Delta_{sep}|$	&	$\leq$ ~0.3 cm  \\ 
 $|\delta \cot \theta |$	&	$\leq$ 0.06   \\ 
 $ R_{conv}$			&     -20 $\leq$ cm and $\leq$ 50 cm     \\
 $f_{VTX} $			& 	$\leq$ 20\%    \\ 
\end{tabular}
\label{tab:tab_3_1}
\end{center}
\end{table}
\widetext


\narrowtext
\begin{table}[p]
\def\arraystretch{0.8}
\begin{center}
\caption[]{Lepton identification efficiencies, including the isolation 
         requirement, measured using a sample of $Z\rightarrow \ell \ell$
         events collected during Run 1B. In Run 1A the muon efficiency is
 ($7.8 \pm 2.8$)\% lower.}
\begin{tabular}{lc}                     
{Lepton type}       & {Efficiency}      \\
\hline
Electrons           & 0.81 $\pm$ 0.02   \\
Muons               & 0.93 $\pm$ 0.03   \\ 
\end{tabular}
\label{tab:tab_3_4}
\end{center}
\end{table} 
\widetext

\narrowtext
\begin{table}[p]
\def\arraystretch{0.8}
\begin{center}
\caption[]{Selection requirements for primary muons.}
\begin{tabular}{lc}
  Variable     		&  Cut \\
\hline
$E_{em}$		&   $\leq$ 2 GeV \\
$E_{had}$     		&   $\leq$ 6 GeV \\
$E_{em}+E_{had}$	&   $\geq$ 0.1 GeV\\
$|d|$                   &   $\leq$ 0.3 cm \\
$|\Delta x|$		&   $\leq$ 2.0 cm (CMU) \\
			&   $\leq$ 5.0 cm (CMP,CMX) \\
$|z_{ver}|$		&   $\leq$ 60.0 cm         \\
$z$-vertex match 	&   $\leq$ 5.0 cm \\
$I$                     & $\leq$ 0.1 \\
\end{tabular}
\label{tab:tab_3_3}
\end{center}
\end{table}
\widetext

\subsection{jet reconstruction}~\label{sec:jetcor}
      The CDF jet reconstruction algorithm uses a cone of fixed radius
      in the $\eta -\phi$ space. In this analysis we use a cone of radius 0.4
      which has been shown to contain approximately  70\% of the jet
      energy~\cite{rob_prd}. A detailed description of the jet reconstruction 
      algorithm can be found in Ref.~\cite{rob_prd}. 

      The jet energy resolution can be parametrized as 
      $\sigma(E_T)/E_T \approx 1/\sqrt{E_T}$, where $E_T$ is measured in GeV. 
      Effects which contribute to the resolution are the lower calorimeter 
      response at the boundaries of different towers and
      of different calorimeter detectors, the loss  
      of low momentum particles inside the magnetic field, the energy 
      deposition in towers  outside the clustering cone, the contribution of 
      the underlying-event and energy losses due to minimum ionizing particles
      or neutrinos present in the jet. Corrections meant to reproduce the 
      average jet $E_T$ correctly (without improving the energy resolution) 
      are often used~\cite{rob_prd,jetcorr}. The average jet energy correction
      factor ranges from approximately 1.7 to 1.1 as the jet transverse energy 
      increases from 15 to 100 GeV.

      Checks of the jet energy corrections have been performed in 
      Ref.~\cite{top_prd} by studying the momentum balance in $\gamma+$ jet 
      and $Z+$jet events. The  energy imbalance is measured to be within
      3\% of the $Z$ or photon energy. However, the uncertainty in the
      modeling of the large-angle gluon emission results in a 10\% systematic 
      uncertainty of the jet energy scale. 
\subsection{$\met$ and neutrino reconstruction}~\label{sec:met_descr}
      The presence of neutrinos is inferred from transverse energy imbalance 
      in the detector. The transverse missing energy is defined as
 \begin{displaymath}
     \met = - |\sum_{i} E^i_T  \vec{n}_i | 
 \end{displaymath}
      where, $E^i_T$ is the magnitude of the transverse energy contained in 
      each calorimeter tower $i$ in the pseudo-rapidity region $|\eta|<3.5$ and
      $\vec{n}_{i}$ is the direction of the tower in the plane transverse to 
      the beam direction. When a muon is present in the event, $\met$ is 
      calculated as:
\begin{displaymath}
      \met = - |\sum_{i} E^i_T  \vec{n}_i 
      + \vec{p}_T^{~\mu} \left(1- \frac {E_T^{\mu}}{p_T^{\mu}} \right) |
\end{displaymath}
      where $E_T^{\mu}$ is the transverse energy deposited by the muon
      in the calorimeter and $p_T^{\mu}$ is the muon transverse momentum.
\section{Selection of the $W$ and $Z+$ jet samples}~\label{sec:W-sel}
      The $\W+$ jet sample, which contains the $t\bar{t}$ signal,
      is selected from the high-$p_T$  inclusive lepton data set by requiring
      at least one primary electron with 
      $E_T  \geq$ 20 GeV  or one primary muon with  $p_T \geq$ 20 $\gevc$,
      $\met \geq$ 20 GeV and at least one jet with uncorrected transverse
      energy $E_T \geq$ 15 GeV and pseudo-rapidity $|\eta | \leq$ 2.
      An appreciable fraction  of these events is due
      to $Z+$ jet production. Some $Z$ events can be identifyied and removed
      when the second lepton from the $Z$ decay falls into the detector acceptance.
      Because $W+$ jet and $Z+$ jet events have similar production
      mechanisms, we will use the $Z+$ jet sample to check our evaluation
      of the backgrounds to $t\bar{t}$ production.
      It is also interesting to study this sample because
      events in which one of the two leptons is not identified
      (unidentified $Z$'s)  are a background to $t\bar{t}$ production.
      The following subsection explains the removal of dilepton events. The events
      surviving dilepton removal constitute the $W+$ jet sample which is 
      described in the last subsection.

\subsection{Selection of the $Z+$ jet sample}~\label{sec:zremoval}
      $Z$ candidates are selected from the high-$p_T$ lepton data set by
      requiring a primary electron with $E_T  \geq$ 20 GeV  and by
      searching for a 
      second lepton of opposite charge which satisfies the criteria listed
      in Table~\ref{tab:tab_3_2}.

      Searching for additional electrons we relax the 
      isolation and $E_{had}/E_{em}$ cuts.
      We also search  in the PEM and FEM detectors.
      Additional muons are searched for by relaxing all selection cuts defining 
      primary muons. As shown in Table~\ref{tab:tab_3_2}, CTC tracks without a
      match to a track segment in the muon chambers but pointing to a 
      calorimeter tower with a small energy deposition are also considered 
      muon candidates.

\begin{figure}
\begin{center}
\leavevmode
\epsffile{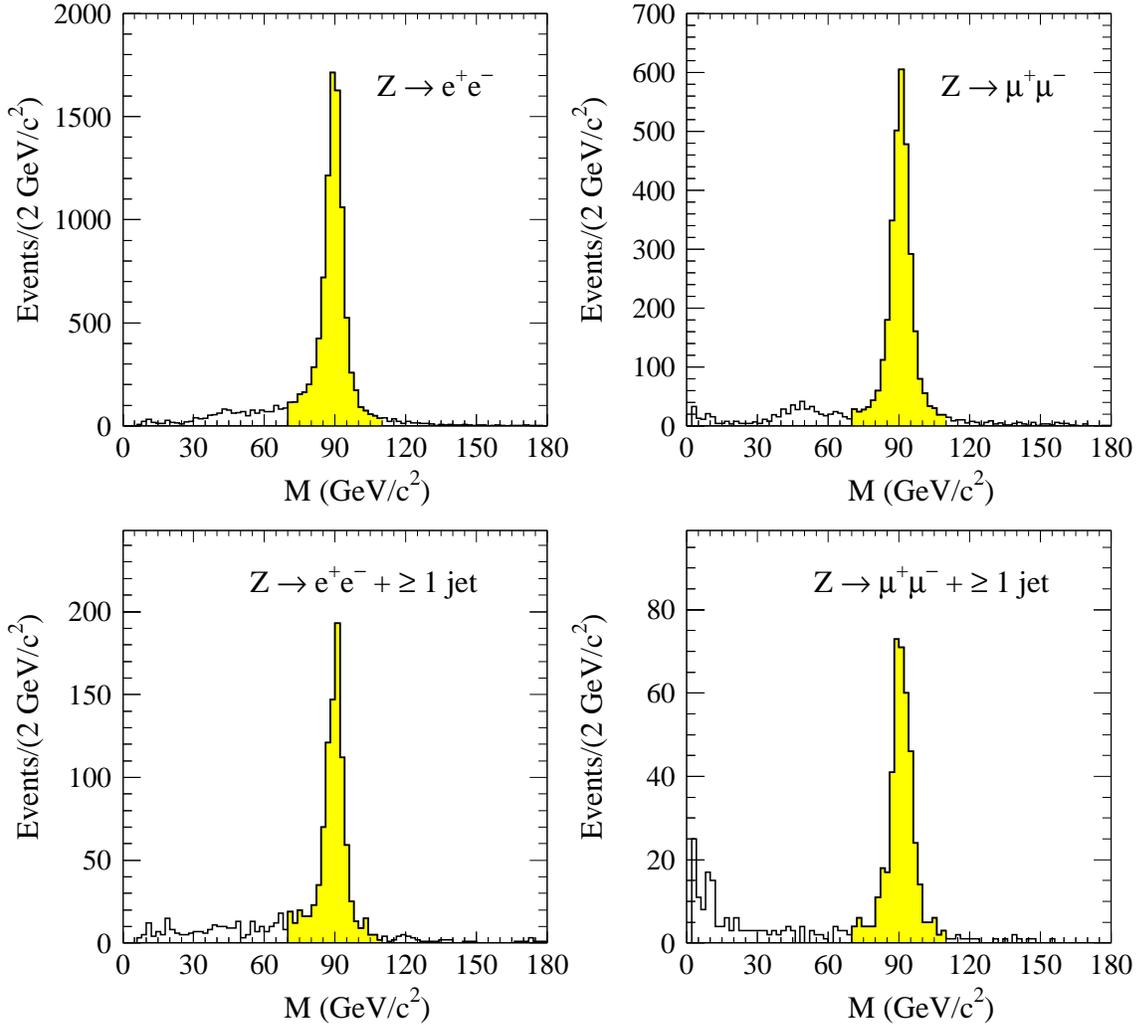}
\caption{Invariant mass distributions of electron and muon pairs before and 
         after requiring the presence of at least one jet with $E_T\geq 15$ 
         GeV and $|\eta|\leq 2$. The shaded area indicates the mass window used
         to select $Z$ candidate events.}
\label{fig:fig_3_0}
\end{center}
\end{figure}

      Events are flagged as $Z$ candidates if the invariant mass of the lepton
      pair falls in the range $70 \leq M_{\ell\ell} \leq 110~\gevcc$ 
      (see Figure~\ref{fig:fig_3_0}). The number of $Z$ candidate events as 
      a function of the jet multiplicity is shown in Table~\ref{tab:tab_5_3}.
\newpage
\narrowtext
\begin{table}[p]
\def\arraystretch{0.8}
\begin{center}
\caption[]{Selection requirements for loose leptons.}
 \begin{tabular}{cc}
  Variable         &  Cut \\
\hline
\multicolumn{2}{c} {Electrons }   \\
          $|\eta|$	& $\leq$ 4.2 \\
			$E_T$		  & $\geq$ 10 GeV \\
			$E_{had}/E_{em}$&   $\leq$ 0.12   \\
			$I$              & $\leq$ 0.15\\
\multicolumn{2}{c} { Muons  with a track segment in the muon chambers} \\		
         $p_T$ & $\geq 10$~\gevc        \\
		 $|d|$			 & $\leq$ 0.5 cm \\
		 $z$-vertex match	& $\leq$ 10 cm\\
		 $|\Delta x |$		& $\leq$ 10 cm\\
		 $E_{em}$		  & $\leq$ 5 GeV \\
		 $E_{had}$		& $\leq$ 10 GeV \\
   $I$                     & $\leq$ 0.15   \\
\multicolumn{2}{c} { Muons  without a track segment in the muon chambers} \\
         $p_T$ 		  & $\geq 10$~\gevc \\
		 $|d|$			& $\leq$ 0.5 cm \\
		 $z$-vertex match	& $\leq$ 10 cm\\
		 $E_{em}+E_{had}$	& $\leq$ 10 GeV \\
		                        & ($E_{em} \leq $ 2  or
                                         $E_{had} \leq $ 6 GeV)\\
   $I$                     & $\leq$ 0.15   \\
\end{tabular}
\label{tab:tab_3_2}
\end{center}
\end{table}
\widetext

\narrowtext
\begin{table}[p] 
\def\arraystretch{0.8}
\begin{center}
\caption[]{Number of $Z$ candidate events as a function of the observed jet
multiplicity.}
\begin{tabular}{cccc}
{Jet multiplicity} & $Z \rightarrow e^+e^-$ & 
$Z \rightarrow \mu^+\mu^-$ & {Total} \\
\hline
{$1$ jet}       &    791  &    357  &    1148 \\ 
{$2$ jets}      &    107  &     52  &     159 \\ 
{$3$ jets}      &      9  &      7  &      16 \\ 
{$\geq~4$  jets} &     3  &      1  &       4 \\
\end{tabular}
\label{tab:tab_5_3}
\end{center}
\end{table} 
\widetext

\subsection{Dilepton removal}~\label{sec:dilepremoval}
	All events containing a primary lepton and at least one additional 
      lepton selected using the criteria listed in Table~\ref{tab:tab_3_2} are 
      removed from the $\W+$ jet sample. These events arise 
      from $Z \rightarrow \tau^-\tau^+$, di-bosons, Drell-Yan and $\ttbar$ 
      production. The $\sigma_{\ttbar}$ production cross section using 
      dilepton events has been measured in Ref.~\cite{topdil} and we want
      to avoid obvious correlations.

      We also remove  events containing an isolated track with 
      $p_T \geq 10$~GeV/c with charge opposite to the primary lepton~
      \cite{isoltr}. The majority of these events originates from 
      genuine dilepton events in which one lepton is outside the region covered by
      the calorimeters or the muon detectors.
      
      Finally, to remove dileptons missed due to inefficiencies of the 
      tracking system, we remove events in which a jet with $E_T \geq 15$ GeV
      and $|\eta| \leq 2$ has a large electromagnetic fraction 
      ($E_{em}/E_{em+had} \geq  0.95$) and less than three tracks. These types of
      events are mostly produced by  $Z\rightarrow e^+ e^-$ decays. 

      The dilepton removal reduces the acceptance for $\ttbar$ events
      by  17.2\%.
\subsection{The $W+$ jet sample}
\label{sec:W-finsel}
\begin{figure}
\begin{center}
\leavevmode
\epsffile{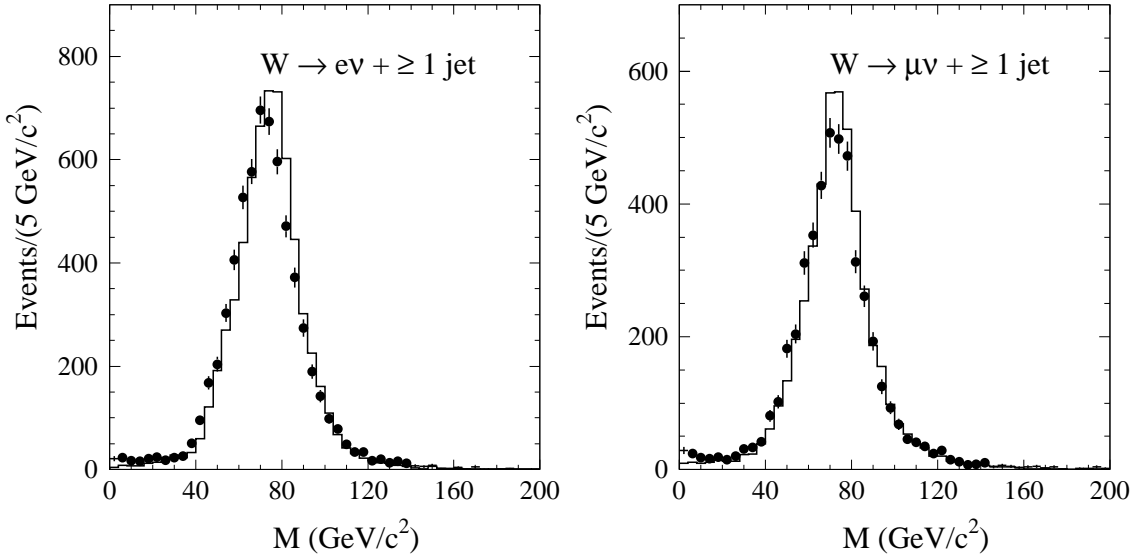}
\caption{ Distribution of the transverse mass $M$ of $\W$ candidates in the
          data ($\bullet$) and in a simulation using the Herwig generator
          (solid histogram). We utilize measured quantities 
          without the full set of
          corrections used to determine the $\W$ mass.}
\label{fig:fig_4_0}
\end{center}
\end{figure}

      The number of $W$ events surviving the $Z$ and dilepton removal is listed
      in Table~\ref{tab:tab_4_2} as a function of the jet multiplicity.
      The transverse mass distribution of the $\W$ candidates is shown in
      Figure~\ref{fig:fig_4_0}.
\narrowtext
\begin{table}[p]
\def\arraystretch{0.8}
\begin{center}
\caption[]{Number of $W$ candidate events as a function of the observed jet
multiplicity.}
\begin{tabular}{cccc}
{Jet multiplicity} & $\W \rightarrow e \nu$ & $\W \rightarrow \mu \nu$  & {Total} \\
\hline
{$1$ jet}       &   5472   &   3982  &   9454 \\ 
{$2$ jets}      &    744   &    626  &   1370 \\ 
{$3$ jets}      &    111   &    87   &    198 \\
{$\geq~4$ jets} &    26    &    28   &     54 \\ 
\end{tabular}
\label{tab:tab_4_2}
\end{center}
\end{table}

\section{Additional data samples}~\label{sec:other_samples}
      In addition to the $Z+$ jet sample, we use a number of independent data 
      sets for the purpose of calibrating the Monte Carlo generators and the 
      detector simulation. The generic-jet samples are described in 
      Subsection~A.
      We will use these sample to derive the new parametrization of the 
      mistag rate, to check our evaluation of the efficiency of the tagging 
      algorithms, and to calibrate the calculation of the fraction of $W+$ jet 
      events with heavy flavor. Subsection~B describes the low-$p_T$ inclusive
      lepton sample which will be used to determine the efficiency of the 
      tagging algorithms. Finally, Subsection~C details the selection of 
      the isolated photon sample. We will use this sample to check the
      parametrization of the mistag rate of our tagging algorithms.
\subsection{Generic-jet  samples}~\label{sec:jet_sample}
      The samples JET~20, JET~50, JET~70 and JET~100 are data collected 
      requiring the presence of a L2 calorimeter cluster with transverse 
      energy $E_T \geq$ 20, 50, 70 and 100 GeV, respectively.
     
      The samples $\sum E_T$~175 and $\sum E_T$~300 are data collected 
      requiring the scalar sum of the transverse energy of all calorimeter 
      towers, as evaluated by the L2 trigger, to be larger than 175 and 300~GeV, 
      respectively.
     
      The last generic-jet sample, $\sum E_T$~125~4CL, was collected requiring the 
      presence of four L2 calorimeter clusters with $E_T\geq 15$ GeV and the 
      scalar sum of the transverse energy of all calorimeter towers to be 
      larger than  125~GeV.

      The L2 triggers calculate the above quantities with respect to the 
      nominal interaction point. Off-line we take as the event vertex the one with
      the largest $ {\displaystyle \sum_i} p_T^i$ using all tracks $i$ 
      associated with the vertex. We retain the events in which 
      the L2 requirements are also matched after the event is reconstructed
      using this vertex. 
      In these events, we inspect all jets with $E_T \geq$ 15 GeV and which 
      contain at least two SVX tracks (taggable jets).
\subsection{The low-$p_T$ inclusive lepton sample}~\label{sec:bb_sample}
      The efficiency of the $b$-tagging algorithms needs to be measured in
      a sample enriched in $\bbbar$ production. The 
      low-$p_T$ electron sample is collected with 
      the L2 requirement that a CFT track with $p_T \geq$ 7.5\gevc is matched 
      by an electromagnetic L2 cluster with $E_T \geq$ 8 GeV. The fraction of 
      electrons coming from semileptonic $b$-decays is enhanced with the 
      selection criteria listed in Table~\ref{tab:tab_5_1}. We use electrons 
      in the CEM fiducial region
      and remove photon conversion candidates. We require 
      the lepton to be in a cone of radius 0.4 around the direction of a
      taggable  jet.
      We require also the presence of at least one additional taggable jet. 
      The $b$-purity of this sample is approximately 50\%.

      We check  the results obtained using the low-$p_T$ electron sample using
      a lower statistics low-$p_T$ muon sample collected using the inclusive
      muon trigger. In this case, a CFT track with $p_T \geq$ 7.5 \gevc must
      be matched to a reconstructed track-segment  in both sets of
      central muon detector (CMU+CMP). Central muons which passed the trigger
      prerequisite are selected with the same criteria used for the high-$p_T$
      muons listed in Table~\ref{tab:tab_3_2} (we require $I \geq$ 0.1).
\newpage
\narrowtext
\begin{table}[p]
\def\arraystretch{0.8} 
\begin{center}
\caption[]{Criteria used to select 
          the low $p_t$ inclusive electron sample.}
\begin{tabular}{lc}
Variable	&	Cut\\
\hline
$E_T$	&	$\geq$ 10 GeV \\
$E/P$            &   $\leq$ 1.5 \\
$E_{had}/E_{em}$ &   $\leq$ 0.05   \\
$L_{shr}$        &  $\leq$ 0.2 \\
$|\Delta x|$     & $\leq$ 1.5 cm \\
$|\Delta z|$     & $\leq$ 3.0 cm \\
$\chi^{2}_{strip}$& $\leq$ 10 \\
$z$-vertex match & $\leq$ 5.0 cm \\
$I$              & $\geq$ 0.1\\
\end{tabular} 
\label{tab:tab_5_1}
\end{center}
\end{table}
\widetext

\subsection{The isolated photon sample}~\label{sec:photon_sample}
      The isolated photon sample was collected requiring a L2 isolated
      electromagnetic cluster
      with $E_T \geq$ 16 GeV and with less than 5 GeV of additional energy in a
      5$\times$10 grid of calorimeter towers centered on the photon direction~
      \cite{cdf_gamma}. Photon candidates which pass the L3 trigger must be in 
      the good fiducial region of the calorimeter and there must be less than 
      4 GeV in a cone of radius 0.7 around the photon direction.
      Table~\ref{tab:tab_5_2} summarizes the off-line criteria used for the 
      selection of the photon sample. After requiring the presence of an 
      additional  jet with $E_T \geq$ 15 GeV and $|\eta|\leq 2$, the final 
      sample consists of $3000~\gamma +\geq 1$ jet events. The expected 
      background contamination of the sample due to $\pi^0$ and $\eta$ decays 
      is estimated to be ($45.0\pm 4.5$)\%~\cite{cdf_gamma,rharris}.
\narrowtext
\begin{table}[p]
\def\arraystretch{0.8}
\begin{center}
\caption[]{Criteria used to select the isolated photon sample.} 
\begin{tabular}{lc}

  Variable         &  cut-value \\
\hline
$E_T$		 &   $\geq$ 23 GeV \\
$\eta$            &   $\leq$ 1.0 \\
$E_{had}/E_{em}$ &   $\leq 0.055+0.00045 \cdot E_T$   \\
Transverse energy  deposited in a cone of  & \\
radius 0.7 around the $\gamma$       &  $\leq$ 2 GeV \\
CTC tracks pointing to the $\gamma$ cluster & None \\
$\chi^{2}_{strip}$ & $\leq$ 20 \\
\end{tabular}
\label{tab:tab_5_2} 
\end{center}
\end{table}
\widetext

\clearpage
\section{Monte Carlo generators}~\label{sec:degrade}
      In this analysis we use three Monte Carlo generators,
      Herwig~\cite{herwig}, Pythia~\cite{pythia} and Vecbos~\cite{vecbos}.  
      The acceptance for $\ttbar$ events is calculated using Pythia version~{\sc 5.7}.
      The  $\ttbar$ acceptance has been also evaluated using the version 
      {\sc 5.6} of Pythia and  Herwig~{\sc 5.6}. 
      Herwig is used to estimate the fraction of $W+ \geq$ 1 jet events with 
      heavy flavor (using the process 2100). We first calibrate the Herwig 
      simulation using generic-jet data, as described
      in Section~\ref{sec:herwig_norm}.

      Both Herwig and Pythia generators use tree-level matrix element 
      calculations for the parton hard scattering, convoluted with parametrizations 
      of the parton distribution functions. The outgoing initial and final 
      state partons are converted into a cascade of gluons and $q\bar{q}$ 
      pairs with energy and angular distributions determined by the 
      Altarelli-Parisi equations~\cite{Altarelli}. The strength of these 
      generators is the modeling of the parton shower which accounts for the 
      color correlation between the initial and final state partons. 
      The parton shower terminates when the invariant mass of the parton falls
      below the perturbative QCD scale. At this level the partons are turned 
      into colorless hadrons according to phenomenological models (the process
      is called hadronization or fragmentation).  For  $b$ and $c$-quarks the 
      fragmentation is modeled in Pythia with the Peterson
      parametrization~\cite{Peterson}. We use the fragmentation parameter 
      $\epsilon =0.006$ for $b$-quarks and $\epsilon=0.05$ for $c$-quarks.
      Herwig uses its own hadronization model, the settings for which are listed
      in Ref.~\cite{herw_setup}. Both generators include a model of the 
      underlying event which describes the hadronization products of the 
      beam remnants.

      The Vecbos Monte Carlo is used to study the part of the phase-space in 
      the $W+\geq$ 1 jet production that is not treated correctly by parton 
      shower Monte Carlos, specifically $Wb\bar{b}$ and $Wc\bar{c}$ events in which 
      the two $b$ or $c$-partons produce two well separated jets. The Vecbos
      Monte Carlo generator provides a parton level calculation of the $W+$
      n jet cross section based on the leading order matrix elements of the 
      hard scattering. Infrared and collinear singularities are regulated by 
      requiring that the final-state partons have a transverse momentum 
      exceeding a cutoff value $p_{T}^{\rm min}$ and are separated by more 
      than $R_{\rm min}$ ($R=\sqrt{(\Delta \phi)^{2}+(\Delta \eta)^{2}}$). 
      We use $p_{T}^{\rm min}$= 8 GeV/c and $R_{\rm min}$ =0.4.
      We use the renormalization scale $Q^{2}$=$<p_T>^{2}$, where
      $<p_T>$ is the average transverse momentum of the outgoing partons.
      We have verified  that after our selection cuts the fraction of jets with
      heavy flavor calculated with Herwig matches the Vecbos prediction at the 
      $R_{\rm min}$ threshold. We transform the partons produced by Vecbos 
      into hadrons and jets using the Herwig program  adapted to perform the coherent 
      shower evolution of both initial and final-state partons~\cite{matrix}.

      In summary, we use Herwig to predict the fraction of $W+\geq$ 1 jet 
      events where only one jet clustered in a cone of radius 0.4 contain $b$ 
      or $c$-hadrons while we rely on Vecbos to extend the prediction to the 
      cases where two different jets contain both heavy-flavored hadrons.

      We use the MRS~D$_0^{\prime}$ set of parton distribution
      functions~\cite{mrsd0p} 
      to generate $W+$ jet events because it has  been  shown to reproduce
      the results of the $W$ asymmetry measured by CDF~\cite{w_assym}.

      The decay of hadrons with heavy flavor produced by the Monte Carlo 
      generators is modeled using the {\sc CLEO} Monte Carlo generator  
      (QQ)~\cite{cleomc}. We use the QQ table of branching ratios for 
      each decay but our own lifetime table because decay lengths are modeled
      inside the detector simulation. 
%
\section{Detector Simulation}~\label{sec:Simul}
      The QFL detector simulation is used to decay all generated particles
      and model their interactions with the various elements
      of the CDF detector.
      The  detector response is based upon parametrizations
      and simple models which depend on the particle kinematics.
      The calorimeter simulation is based upon a parametrization
      of the calorimeter response to single particles parametrized as 
      a function of the pseudo-rapidity and azimuthal angle (to account for 
      cracks in the calorimetry ) and of the transverse momentum using 
      test-beam data.
      After the simulation of the CDF detector, the Monte Carlo events are 
      treated as if they were real data.
\subsection{CTC track simulation}~\label{sec:ctceff}
      The CTC simulation is not a hit-level simulation. It converts each 
      particle's momentum vector at generator level into a reconstructed track 
      using covariance matrices derived from the data. Not surprisingly,
      the track-reconstruction efficiency in the detector simulations
      is higher than that measured in the data. The major factor
      influencing the track reconstruction efficiency is the density of hits 
      in the tracking detector. In this respect, the problem is aggravated by 
      the fact that the Monte Carlo generators do not contain multiple 
      interactions.

      To adjust the tracking reconstruction efficiency in the simulation,
      CTC hits of Monte Carlo generated tracks have been embedded  in 
      generic-jet data. The efficiency is determined by counting the fraction 
      of times the embedded track is reconstructed. The tracking efficiency is
      measured as a function of the hit density around the track for low 
      luminosity runs, and then for runs of typical luminosities.
      Table~\ref{tab:tab_7_1} compares the track reconstruction efficiency in 
      the detector simulation to the efficiency for reconstructing simulated 
      tracks embedded in the data. 
      The degradation of the track reconstruction efficiency is parametrized  
      in the detector simulation as a function of the number of hits around the
      tracks and of the average luminosity of the data. This procedure
      accounts well for the dependence of the tracking efficiency on the jet 
      transverse energy. 

\newpage
\widetext
 \begin{table}[p]
 \begin{center}
\def\arraystretch{0.8}
 \caption[]{Track reconstruction efficiency
for charged particles   in the detector simulation (QFL)
 and for Monte Carlo tracks embedded in generic-jet data
 acquired in low luminosity running. The effect of the average luminosity
 of the data is shown separately.}
 \begin{tabular}{ccccc}
          & CTC track & SVX track      & Luminosity Effect \\
 Embedded-track  & 0.94$\pm$0.02     & 0.87$\pm$0.03 & 0.95$\pm$0.02\\
  QFL simulation     & 0.993    & 0.983         &   1       \\
 \end{tabular}
 \label{tab:tab_7_1}
 \end{center}
 \end{table}
\narrowtext

\subsection{Lepton identification efficiencies}
      Aside from the efficiency for reconstructing a track, the primary lepton
      identification efficiency in the simulation depends also on how well the
      Monte Carlo generators reproduce the isolation distribution and how well
      the calorimeter response has been parametrized.
      In the simulation, the primary lepton identification efficiencies are 
      measured as the ratio of the number of leptons passing the selection cuts
      listed in Tables~\ref{tab:tab_3_0} and~\ref{tab:tab_3_3} to the number 
      of leptons generated  in the kinematical acceptance.
      The identification efficiencies in the simulation are 97 $\pm$ 2\% for 
      muons and 87.5 $\pm$ 2.0\% for electrons. The identification efficiencies
      for primary leptons are degraded in the detector simulation to match the
      ones measured in the data (see Table~\ref{tab:tab_3_4}). 
      Altogether, we degrade the rates of simulated  primary leptons by the 
      factor of 0.936$\pm$0.125 (the error includes a 10\% uncertainty on the
      muon trigger simulation).

      The efficiency for identifying soft lepton tags is a far more complicated
      problem because some detector responses, such as $dE/dx$ in the CTC and the CPR 
      chambers, have not been parametrized in the detector simulation.
      The SLT simulation weights tracks corresponding  to leptons from
      $b$ and $c$-quark decays at generator level with a parametrization of 
      the efficiency of each selection cut measured using the data, as
      described in Section~\ref{sec:slt_Tagger}.
\subsection{SVX track simulation}
      The detector simulation becomes unwieldy when simulating tracks 
      that are measured by both the  CTC and SVX tracking detectors as  is the
      case for input tracks to the SECVTX and jet-probability algorithms. 
      The SVX track reconstruction is performed by assigning hits on the silicon
      vertex detector to previously reconstructed CTC tracks. In the data hits are 
      assigned if they are contained in a road around the reconstructed CTC 
      track determined by its uncertainty (4 $\sigma$ in the $r-\phi$ plane). 
      A CTC track with at least two associated SVX 
      hits is defined to be a SVX track and is refitted using the SVX hits and 
      the CTC track parameters and covariance matrix. The simulation of the 
      SVX is a hit-level simulation in which the hit resolution is taken from the
      data.
      Simulated SVX tracks are reconstructed as in the data. However, 
      in the data we must multiply all the elements of the covariance
      matrix by a factor of 2 so that the CTC-SVX matching uncertainty
      agrees with the measured resolution~\cite{bsmass}
      while there is no such need in the simulation.

      The efficiency for finding SVX tracks in the detector simulation 
      also needs 
      to be degraded, by a factor determined by measuring the efficiency for 
      reconstructing Monte Carlo generated tracks embedded at hit-level in 
      generic-jet data (see Table~\ref{tab:tab_7_1}).

      Having done this, the simulation is still not a perfect 
      reflection of the data. For example, as shown in  
      Section~\ref{sec:jpb_Tagger}, the distribution of the impact parameter 
      significance of SVX tracks in the data and in the detector simulation 
      are slightly different. We conclude that it is neccessary
      to measure the tagging efficiencies of each algorithm in the data and in 
      the simulation and correct the detector simulation for any observed 
      difference. This is done in Section~\ref{sec:simu_mod}.
\section{Description of the tagging algorithms}~\label{sec:taggers}
      The presence of jets originating from $b$ quarks is one of the 
      characteristic signatures of $\ttbar$ events. Following previous 
      work~\cite{cdf_xsec,top_prd}, we tag $b$-quarks using two of their
      distinctive properties: the relatively long lifetime and the presence 
      of semileptonic decays. 
      Two tagging techniques based on tracking 
      information using the SVX detector have been developed to identify jets 
      containing heavy flavor. The Secondary Vertex Tagging algorithm (SECVTX)
      is described in Subsection~A. The jet-probability algorithm, used to check
      SECVTX results, is described in Subsection~B. The Soft Lepton Tagging
      algorithm (SLT) is discussed in Subsection~C, which also includes the 
      evaluation of the SLT fake rate and a description of the simulation of this
      algorithm.
\subsection{SECVTX algorithm}~\label{SVX_Tagger}
      The SECVTX algorithm is described in more
      detail in Refs.~\cite{topxs,top_prd}. SECVTX is based on the 
      determination of the primary event vertex and the reconstruction of 
      additional secondary vertices using displaced tracks associated with
      jets.

      The positions of the $p\bar{p}$ interactions (primary vertices) are 
      distributed along the beam direction according to a Gaussian with a 
      width of approximately 28 cm. In the plane transverse to the beam axis, 
      these interactions follow a distribution  that is a Gaussian with a 
      width of ~25 $\mu$m in both the $x$ and $y$ dimensions. To 
      reconstruct the primary event vertex, we first identify its $z$-position
      using the tracks reconstructed in the VTX detector. When  projected back
      to the beam axis, these tracks determine the longitudinal position with 
      a precision of about 0.2 cm.

      The transverse position of the primary vertex is determined for each 
      event by a weighted fit of all SVX tracks which have a $z$ coordinate 
      within 5 cm of the $z$-vertex position of the primary vertex associated
      with the trigger lepton. First, all tracks are constrained to originate
      from a common vertex. The position of this vertex is constrained by the 
      transverse beam envelope described above. Tracks that have impact 
      parameter significance
       $|d|/\sigma_{d}$, where $\sigma_{d}$
      is the estimate of the uncertainty on the impact parameter $d$,
      larger than three with respect to this vertex 
      are removed and the fit is repeated. This procedure is iterated until 
      all used tracks satisfy the impact parameter requirement. At least five 
      tracks must be used in the determination of the transverse position of 
      the primary vertex or we use the nominal beam-line position. 
      The primary vertex coordinates transverse to the beam direction have 
      uncertainties in the range of 10$-$25 $\mu$m, depending on the number of
      tracks and the event topology.

      The search for a secondary vertex in a jet is a two stage process. 
      In both stages, tracks in the jet are selected 
      based on the significance of their impact parameter 
      with respect to the primary vertex.
      The first stage (see Table~\ref{tab:tab_8_1}) requires at
      least three candidate tracks for the reconstruction of the secondary 
      vertex. Tracks consistent with coming from the decay 
      $K_s \rightarrow \pi^+ \pi^-$ or $\Lambda \rightarrow \pi^- p$ 
      are not used as candidate tracks. Two candidate tracks are constrained 
      to pass through the same space point to form a seed vertex. If at least 
      one additional candidate track is consistent with intersecting this seed
      vertex, then the seed vertex is used as the secondary vertex. If the 
      first stage is not successful in finding a secondary vertex, a second 
      pass is attempted. More stringent track requirements (on $|d|/\sigma_{d}$ 
      and $p_T$, for example) are imposed on the candidate tracks. All 
      candidate tracks satisfying these stricter criteria are constrained to 
      pass through the same space point to form a seed vertex. This vertex has
      an associated $\chi^2$. Candidate tracks that contribute too much to the
      $\chi^2$ are removed and a new seed vertex is formed. This procedure is
      iterated until a seed vertex remains that has at least two associated 
      tracks and an acceptable value of $\chi^2$. Table~\ref{tab:tab_8_1} 
      lists the selection criteria used for the determination of the secondary
      vertex candidates.

      The decay length of the secondary vertex $L_{xy}$ is the projection
      of the two-dimensional vector pointing from the primary vertex to the 
      secondary vertex on the jet axis; if the cosine of the angle between 
      these two vectors is positive (negative), then $L_{xy}$ is positive 
      (negative). Most of secondary vertices from the decay of $b$ and 
      $c$-hadrons are expected to have positive $L_{xy}$. 
      Secondary vertices from random
      combination of mismeasured tracks are expected to have a symmetric 
      distribution around $L_{xy}$=0~\cite{wenzel}. To reduce the background 
      from false secondary vertices (mistags), a jet is considered tagged by 
      SECVTX  if it contains a secondary vertex with 
      $\frac{L_{xy}}{\sigma_{L_{xy}}} \geq 3.0$, where $\sigma_{L_{xy}}$ is 
      the estimated uncertainty on $L_{xy}$ ($\sim$ 130 $\mu$m).  The mistag 
      contribution to positive  SECVTX tags is evaluated starting from the 
      rate of negative SECVTX tags and detailed in Section~\ref{sec:mistags}.
\newpage
\narrowtext
 \begin{table}[htp]
 \begin{center}
 \def\arraystretch{0.8}
 \caption[]{Selection criteria for CTC and SVX tracks used in the SECVTX 
            $b$-tagging algorithm. A good SVX hit is defined as a hit in 
            the SVX linked to only one CTC track.}
\begin{tabular}{llc}
Variable & \multicolumn{2}{r} {Cut} \\
\hline
\multicolumn{3}{c}{ CTC track selection criteria}\\
  No. of axial superlayers & \multicolumn{2}{r} {$\geq$ 2}\\
  No. of hits in each axial superlayer  &  \multicolumn{2}{r} {$\geq$ 2}\\
  No. of stereo superlayers &  \multicolumn{2}{r} {$\geq$ 2}\\
  No. of hits in each stereo superlayer  &  \multicolumn{2}{r} {$\geq$ 2}\\
  $\chi^2$/d.o.f. of the track fit	&  \multicolumn{2}{r} {$\leq$ 6}\\
  $z$-vertex match	&  \multicolumn{2}{r} {$\leq$ 5 cm}\\
\hline
\multicolumn{3}{c} { SVX track selection criteria - Pass 1}  \\
 if $N_{\rm SVX-hits} \geq$ 3 &  $ \left\{ \begin{array}{l}
			  N_{\rm SVX-hits}^{\rm Good} \\
			  p_T
			  \end{array}  \right.$
			      & $ \begin{array}{c}
			  \geq 1 \\
			  \geq 0.5\hspace{0.15cm} {\rm GeV/c}
			  \end{array} $ \\ 
 if $N_{\rm SVX-hits} =$ 2 & $ \left\{ \begin{array}{l}
			  N_{\rm SVX-hits}^{\rm Good} \\
			  p_T
			  \end{array}  \right.$
			      & $ \begin{array}{c}
		          \geq  2 \\
			  \geq 1.5\hspace{0.15cm} {\rm GeV/c}
			  \end{array} $ \\
  $|d|$	      &		& $\leq$ 0.1 cm\\ 
  $|d|/\sigma_d$ &	& $\geq$ 2.5  \\
\hline
\multicolumn{3}{c} { SVX track selection criteria - Pass 2}  \\
  if $N_{\rm SVX-hits} =$ 4 &  $ \left\{ \begin{array}{l}
			  N_{\rm SVX-hits}^{\rm Good} \\
			  p_T
			  \end{array}  \right.$
			      & $ \begin{array}{c}
			  \geq 1 \\
			  \geq 1.0\hspace{0.15cm} {\rm GeV/c}
			  \end{array} $ \\ 
  if $N_{\rm SVX-hits} =$ 3 & $ \left\{ \begin{array}{l}
			  N_{\rm SVX-hits}^{\rm Good} \\
			  p_T
			  \end{array}  \right.$
			      & $ \begin{array}{c}
			  \geq 2 \\
			  \geq 1.0\hspace{0.15cm} {\rm GeV/c}
			  \end{array} $ \\ 
  $|d|$	      &		& $\leq$ 0.1 cm\\
  $|d|/\sigma_d$& 	& $\geq$ 3.0  \\
\end{tabular}
 \label{tab:tab_8_1}
 \end{center}
 \end{table}
\widetext

\clearpage
\subsection{Jet-probability algorithm}~\label{sec:jpb_Tagger}
      The jet-probability $b$-tagging algorithm~\cite{aleph} is used to 
      cross-check the SECVTX results. The jet-probability algorithm compares 
      track impact 
	parameters to measured resolution functions in order to calculate for
	each jet a probability that there are no long lived particles in the 
      jet cone.  This probability is uniformly distributed between 0 and 1 for light 
	quark or gluon jets, but is very small for jets containing displaced
	vertices from heavy flavor decays. We briefly describe the 
      transformation from the track impact parameters to the
      jet-probability measure.

\begin{figure}
 \begin{center}
 \leavevmode
 \epsffile{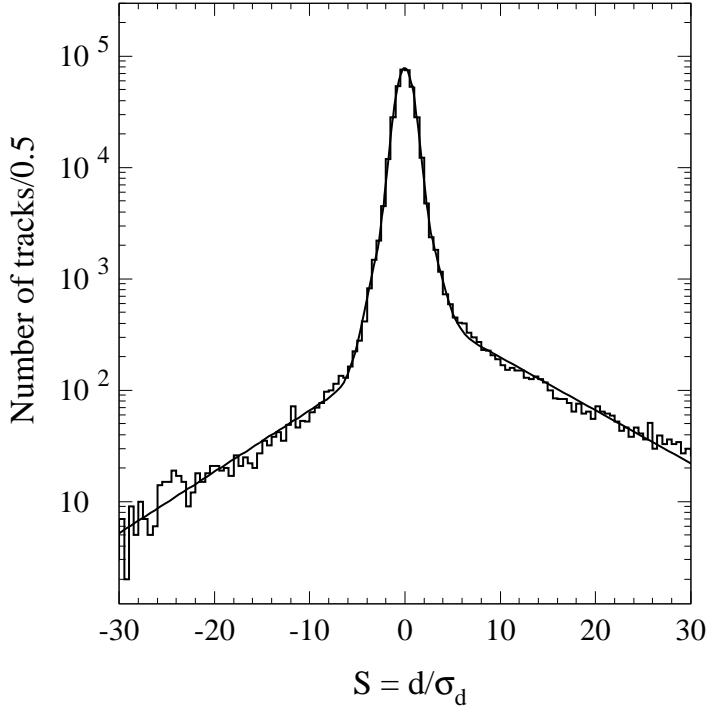}
 \caption[]{Distribution of the signed impact parameter significance of tracks
            in the JET 50 sample. The resolution function ${\cal R}(S)$
            is the result of a fit using two Gaussians plus an exponential 
            function, separately for the positive and negative sides.}
 \label{fig:fig_8_0}
 \end{center}
 \end{figure}

	The track impact parameter significance $S$ is defined as the
      value of the impact parameter $d$ divided by its uncertainty
      $\sigma_{d}$. Tracks used in the calculation of jet-probability 
      are required to 
      satisfy the quality criteria listed in Table~\ref{tab:tab_8_2}.
      The sign of the impact parameter significance is defined to be 
      positive if the point of closest approach to the  primary vertex lies 
      in the same hemisphere as the jet direction, and negative otherwise. 
      Figure~\ref{fig:fig_8_0} shows the 	distribution of the impact 
      parameter significance of tracks in the JET 50 sample. This 
      distribution is fitted with the resolution function ${\cal R}(S)$.

\begin{figure}
 \begin{center}
 \leavevmode
 \epsffile{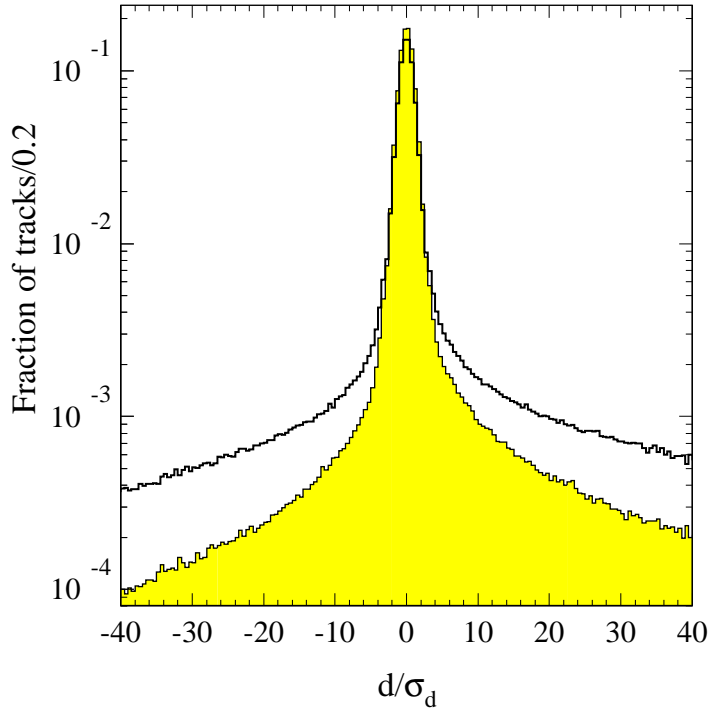}
 \caption{ Distribution of the impact parameter significance $d/\sigma_d$
           of tracks in the JET 50 data (histogram) and the corresponding 
           Herwig simulation (shaded histogram). The tracks are required to 
           satisfy the criteria listed in Table~\protect\ref{tab:tab_8_2}.}
 \label{fig:fig_8_1}
 \end{center}
 \end{figure}

      The negative side of the resolution function ${\cal R}(S)$ derived 
      using JET 50 data is used to determine the probability $P(S_0)$ that
      the impact parameter significance $S_0$ of a given track
      is due to the detector resolution:
  \begin{displaymath}
   P(S_0) = \frac{ \int_{-\infty}^{-|S_{0}|} {\cal R}(S) dS}
               {\int_{-\infty}^0 {\cal R}(S) dS} 
  \end{displaymath}
      Figure~\ref{fig:fig_8_1} shows that the impact parameter significance 
      distribution of tracks in the JET 50 data and in the corresponding 
      simulation are slightly different. The resolution functions
	${\cal R}(S)$ are therefore defined separately for the data and the 
      simulation in order to account for the differences in the resolution  
      between the true and the simulated detector performance.

      The probability that a jet is consistent with a zero lifetime 
      hypothesis is calculated as:
 \begin{displaymath}
   \mbox{JPB} = \prod \sum_{k=0}^{N-1}\frac{(-\ln\prod)^k}{k!}
 \end{displaymath}
      where $\prod$ is the product of the individual probabilities of the
      $N$  SVX tracks in a jet which satisfy the criteria listed in 
      Table~\ref{tab:tab_8_2}. Jet-probability is defined using tracks with 
      positive impact parameter and requiring $N \geq$ 2.
      We also define a negative jet-probability 
      in which we select only tracks with negative impact parameter in the 
      calculation.  This is used as a control sample and a check of our
      method.
\newpage
\widetext
 \begin{table}[p]
 \begin{center}
 \def\arraystretch{0.8}
 \caption[]{Selection criteria for tracks used by the
 jet-probability algorithm.}
 \begin{tabular}{llc}
Variable & \multicolumn{2}{r} {Cut} \\
\hline
\multicolumn{3}{c} { SVX track selection criteria }  \\
  $|d|$			&  \multicolumn{2}{r} {$\leq$ 0.15 cm}\\
  $p_T$			&  \multicolumn{2}{r} {$\geq$ 1.5 GeV/c}\\
 $N_{\rm SVX-hits} $	&  \multicolumn{2}{r} {$\geq$ 2}\\
\end{tabular}
 \label{tab:tab_8_2}
 \end{center}
 \end{table}
\widetext

\clearpage

\begin{figure}
 \begin{center}
 \leavevmode
 \epsffile{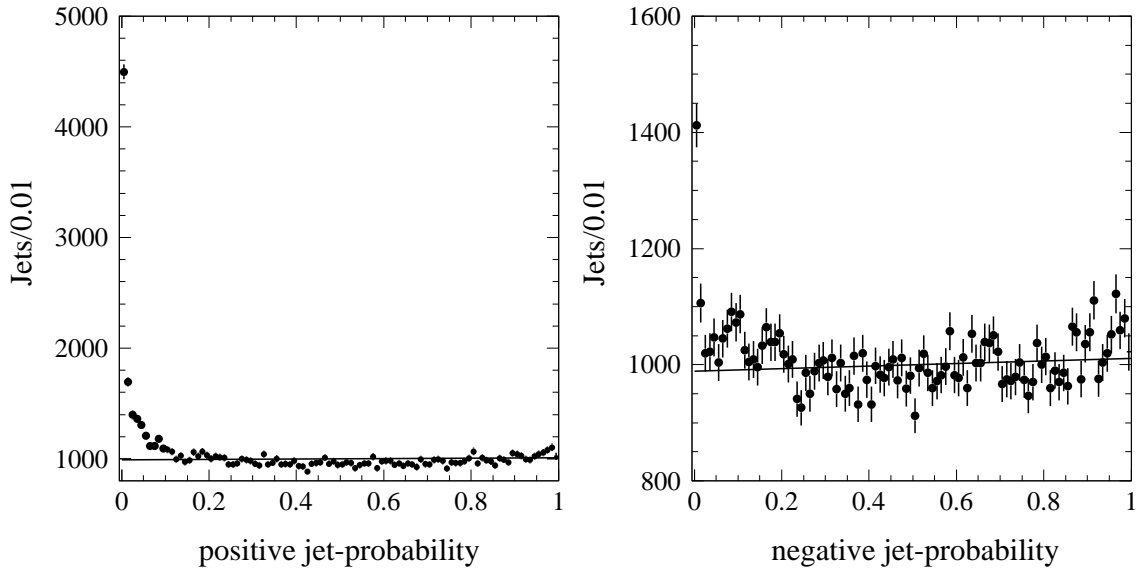}
 \caption[]{Distributions of positive  and negative  jet-probability 
            in a mixture of  JET 50 and JET 140 data. The lines represent a 
            fit to the negative  distribution  with a first order polynomial. 
            The slope of the fit corresponds to a 1.6\% change of the 
            distribution over the entire jet-probability range. }
 \label{fig:fig_8_2}
 \end{center}
 \end{figure}

      Figure~\ref{fig:fig_8_2} shows the positive and negative 
      jet-probability distributions in a sample of JET 50 and JET 140 data.
      The positive jet-probability distribution shows jets containing 
      hadrons with heavy flavor as a large excess 
      at jet-probabilities smaller than 0.05 over a  flat distribution.
      A jet has a positive  JPB tag if the jet-probability value
      is smaller than 0.05.

\begin{figure}
 \begin{center}
 \leavevmode
 \epsffile{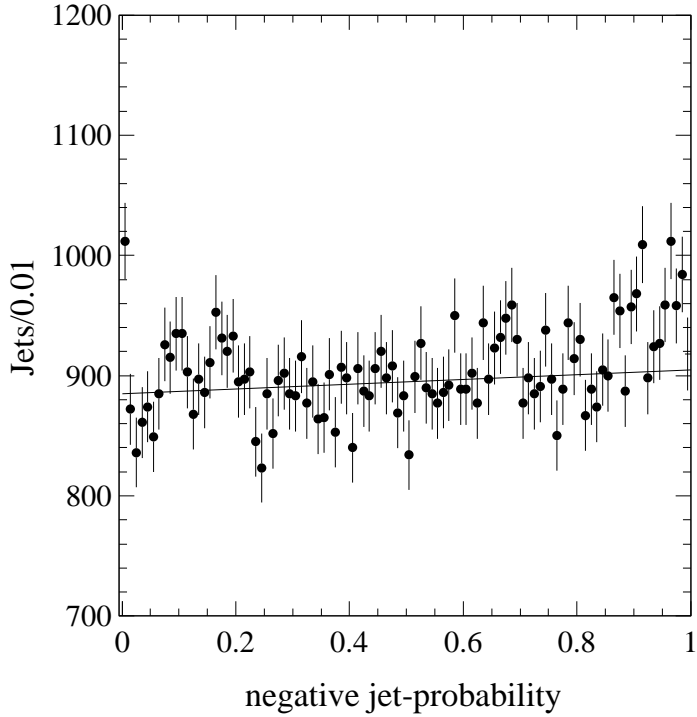}
 \caption[]{ Negative jet-probability distribution for jets with positive 
             jet-probability greater than 0.1. This selection requirement 
             removes most of the jets with heavy flavor. The line corresponds
             to the  fit to the negative JPB distribution shown in 
             Figure~\protect\ref{fig:fig_8_2}.}
 \label{fig:fig_8_3}
 \end{center}
 \end{figure}

        The negative jet-probability distribution is quite flat, as expected, 
        since  the resolution files were constructed using tracks with negative
        impact parameter. The small excess at negative jet-probability
        smaller then 0.05 (negative JPB tags)
        is due to the increase of the fraction of jets from heavy flavor in 
        the JET 140 with respect  to the JET 50 data. This excess
        largely disappears, as 
        shown in Figure~\ref{fig:fig_8_3}, when plotting the negative 
        probability of jets which have a large positive jet-probability 
        (0.1$-$1).
        Since tracks with negative signed impact parameter in JET 50 data are 
        used to define the resolution function, the small contribution
        to negative tags from jets
        with heavy flavor  is uncorrectly attributed to the 
        detector resolution by this procedure. It will be accounted for in 
        the evaluation of the JPB mistags in  Section~\ref{sec:mistags}.

        Ideally JPB tags corresponding to jet-probability values smaller than
        0.05 should contain a 5\% mistag rate. This expectation is tested in 
        Figure~\ref{fig:fig_8_3} fitting a first order polynomial function  
        to the jet-probability distribution  in the interval 0.1$-$1.0.
       The extrapolation of the fitted function predicts 4441$\pm$34
        negative JPB tags while 4455 are observed; this corresponds to 4.94\% 
        of the total number (101050) of jets in the sample.
%
\subsection{SLT algorithm}~\label{sec:slt_Tagger}
        The SLT algorithm tags $b$ quarks by searching for an electron or muon from 
        their decay. Low momentum or soft leptons can also result
        from $b$-hadron decay through  
        sequential $c$-decays, or $\tau$ and $J/\psi$ cascade decays. This 
        analysis follows the guidelines for the identification of soft 
        electrons or soft muons documented in Refs.~\cite{top_prd,kestenbaum}.
       While previous measurements of the $t\bar{t}$ cross section used rates
        of events with  SLT tags, in this analysis we search for soft lepton 
        candidates only in a cone of radius 0.4 around the axis of a jet with 
        $E_T \geq 15$ GeV and $|\eta| \leq 2$. 

        To search for soft electrons, every CTC track with $p_T \geq 2$ GeV/c,
        which is associated to a jet, is extrapolated into the fiducial region of 
        the calorimeter and is matched to a CES cluster. The matched CES 
        cluster is required to be consistent in shape and position to the 
        expectations for an electron. In addition, we require 
        $0.7 \leq E/P \leq 1.5$ and $E_{had}/E_{em} \leq 0.1$. 
        The energy deposited by the track in the preradiator (CPR) is required
        to be consistent with an electron shower. The track ionization rate 
        ($dE/dx$), derived from the charge deposition of the CTC hits associated with 
        the track, is also required to be consistent with the electron 
        hypothesis. Electrons from photon conversions are removed.
        Photon conversions are identified as combinations of the electron 
        candidate and an additional track with opposite charge passing the 
        criteria listed in Table~\ref{tab:tab_3_3} with the additional 
        requirement that the invariant mass  be smaller than
        500 MeV/c$^2$. The selection criteria 
        used to define the soft electron are described 
        in more detail in Ref.~\cite{kestenbaum}.  The efficiency of each
         criteria
        used to select soft electron candidates has been measured using a 
        sample of electrons produced by photon conversions~\cite{top_prd}
        (the efficiency of the $E/P$ and $E_{had}/E_{em}$ cuts is
         calculated using the simulation).

        Soft muons are identified by matching CTC tracks with $p_T \geq$ 2 
        $\gevc$ to track segments in the CMU, CMP and CMX muon chambers. 
        Muons candidate tracks with $p_T \geq$ 3 $\gevc$ that are extrapolated
        to
        the fiducial volume of both the CMU and CMP system  are required to be
        matched to track segments in both muon detectors. To maintain high 
        efficiency for non-isolated muons, we do not impose minimum-ionization
        requirements on the calorimeter deposition. 
        However, in order to reduce hadronic punch-through
        in the region not covered by the CMP system, we check that the energy, 
        $E_{had}$, in the tower traversed by  muon candidates with $p_T \geq$
        6 $\gevc$  is consistent with the muon hypothesis; we require 
        $E_{had} \leq 6 +\sum p$, where $\sum p$ is the scalar sum of the 
        momenta of all tracks contained in a cone of radius 0.2 around the 
        muon direction.
        The efficiency of each selection cut has been measured using a sample 
        of $J/\psi \rightarrow \mu^+ \mu^-$ and $Z \rightarrow \mu^+ \mu^-$ 
        decays~\cite{top_prd,kestenbaum}.

\begin{figure}
 \begin{center}
 \leavevmode
 \epsffile{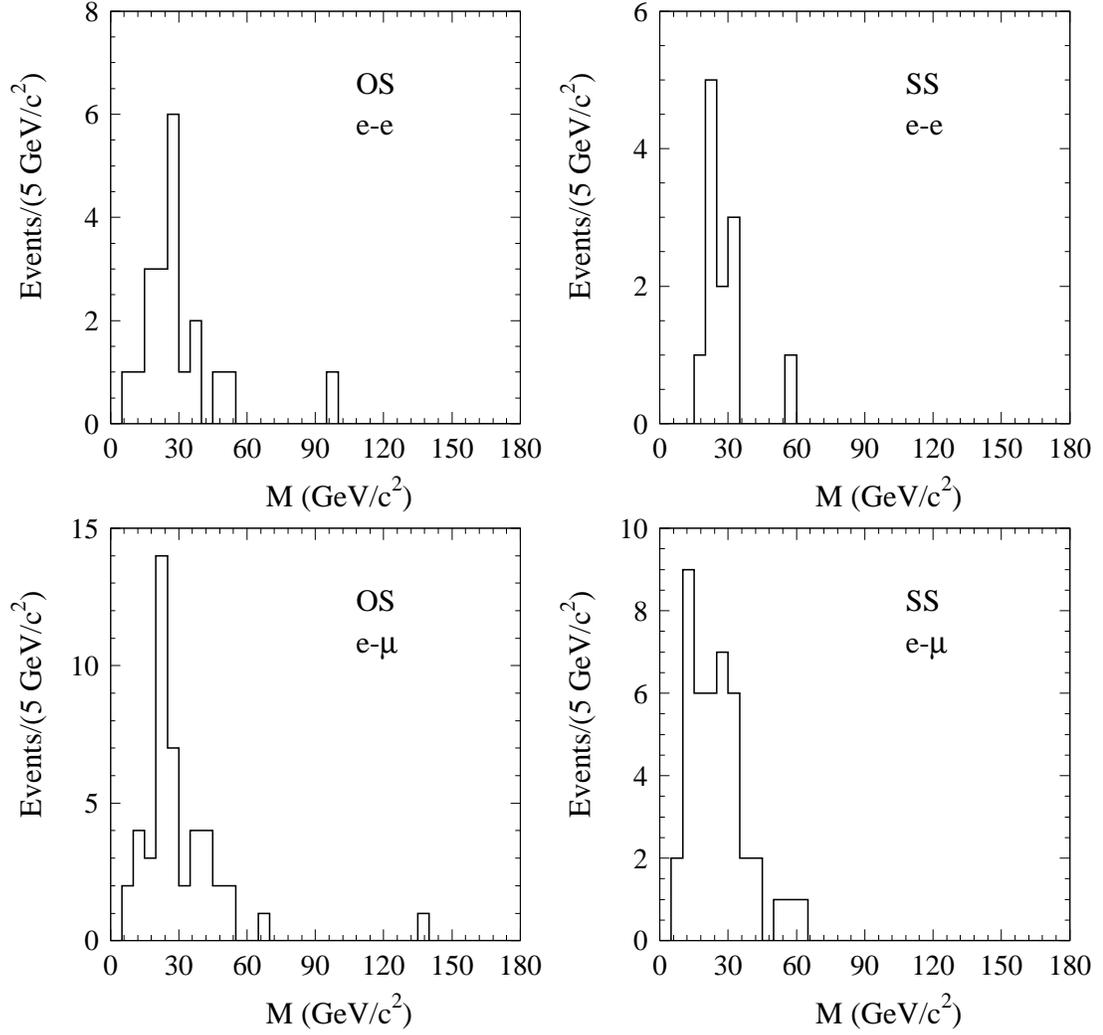}
 \caption[]{ Invariant mass distributions between the primary electron
             and the soft lepton candidates in $W+\geq$1 events.
             OS and SS refer to lepton pairs with opposite and same charge, 
             respectively.}
 \label{fig:fig_8_4}
 \end{center}
 \end{figure}

\begin{figure}
 \begin{center}
 \leavevmode
 \epsffile{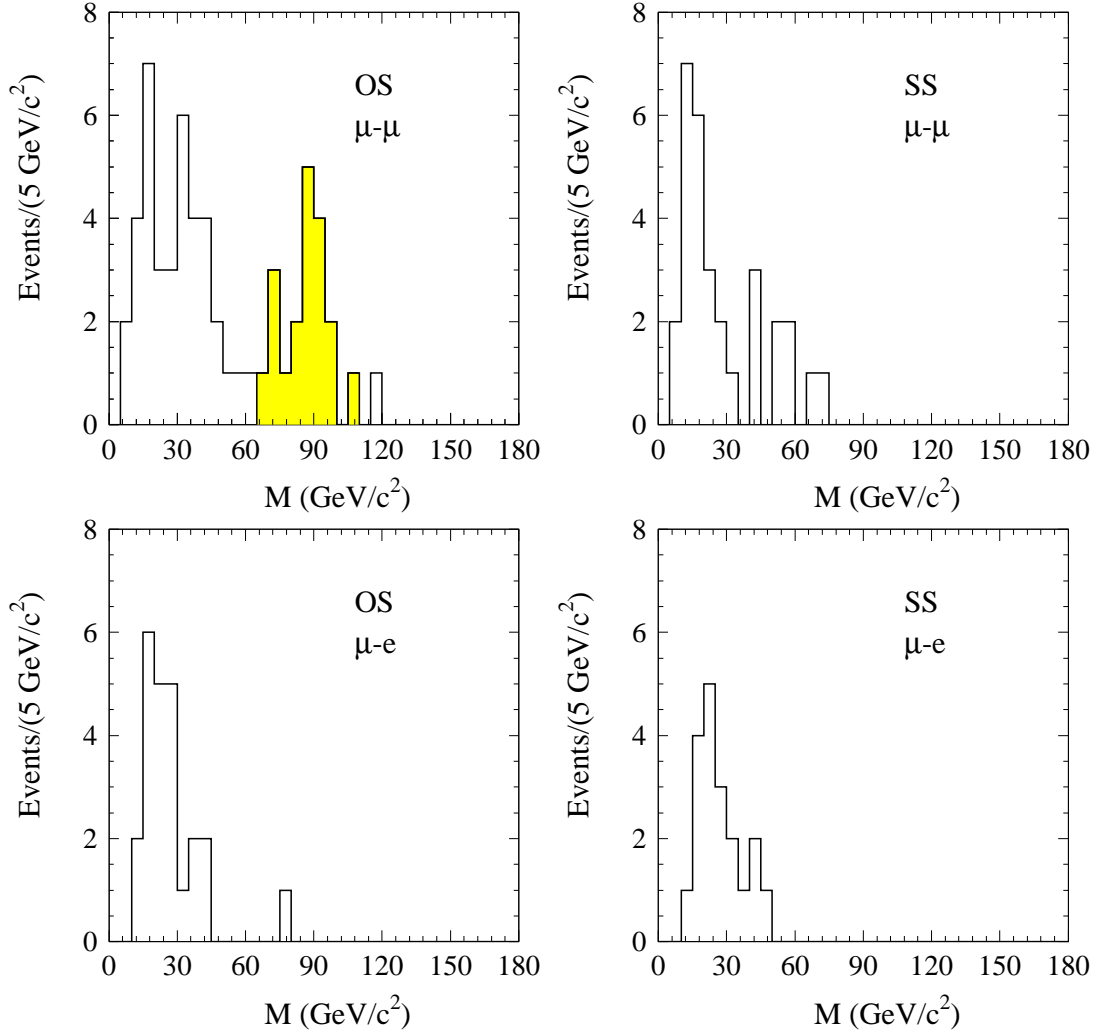}
 \caption[]{ Invariant mass distributions between the primary muon and the soft 
             lepton candidates in $W+\geq$1 events. OS and SS refer to lepton 
             pairs with opposite and same charge, respectively. The shaded 
             area indicates soft muons not considered tags.}
 \label{fig:fig_8_5}
 \end{center}
 \end{figure}

        Figures~\ref{fig:fig_8_4} and~\ref{fig:fig_8_5} show distributions of
        the invariant mass between  primary and soft leptons in all 
        $W+\geq$1 jet events.
        As shown in Figure~\ref{fig:fig_8_5}, there is a handful of events
        where the soft muon is consistent with being the second leg of a
        $Z$ boson decay embedded in a jet. Soft muons which,
        when combined with the primary muons of opposite charge, 
        yield an invariant mass 
        $70 \leq M_{\mu\mu} \leq 110$ GeV/c$^2$ are not considered tags.
\subsubsection{ Fake soft lepton tags}
~\label{sec:slt_fake}
        This background includes hadrons which pass the lepton selection cuts 
        (such as pions which fake an electron or a muon) as well as electrons from 
        conversions or muons from pions or kaons which decay in the detector.
        Because of the complication of accurately modeling these processes in
        the QFL detector simulation, this background is estimated using the 
        data.

        The SLT fake rate is  measured starting from the ratio of the number 
        of tracks passing the soft lepton selection criteria to the total 
        number of tracks which satisfy the soft lepton fiducial requirements
        in generic-jet data~\cite{top_prd,kestenbaum}. In the JET 20, JET 50 
        and JET 70 samples the probability $P$ that a track produces a SLT tag
        is computed separately for electrons and for different types of muon 
        detectors (CMU,CMP and CMX). This probability is parametrized as 
        a function of the track $p_T$ and isolation~\cite{top_prd,kestenbaum}.
        Since in this analysis we search a jet for SLT candidates in a cone of 
        radius of 0.4 around its axis, we define a SLT probability per 
        jet $P_{SLT}^{jet}(N) = \sum_{i=1}^{N} P_{SLT}^{jet}(i-1)+
        (1-P_{SLT}^{jet}(i-1)) \times P_i $ where $N$ is the number of tracks
        contained in a cone of radius 0.4 around the jet axis.

        In Table~\ref{tab:tab_10_4} the observed rates of SLT tags
        in various generic-jet samples  are compared to the rates predicted
        by the probability $P_{SLT}^{jet}$ described above.
        Since in generic-jet data the trigger jet is biased toward a lower
        yield of soft muons (a jet containing a muon  has a lower energy 
        deposition in the calorimeter and therefore is less likely to be the 
        trigger jet) the comparison is performed with and without the trigger 
        jet. However, when more than one jet is above the trigger threshold, 
        all jets are considered. Excluding trigger jets from the comparison 
        one observes agreement between the observed and predicted rates of 
        tagged jets. The last seven samples shown in Table~\ref{tab:tab_10_4} 
        were not used to determine the SLT probability per track. Predicted 
        and observed yields of SLT tags in all samples agree within
        15\%. As the amount and type of heavy flavor changes appreciably in 
        different QCD samples (see Section~\ref{sec:mistags}) the apparent 
        agreement  suggests that the rate of SLT tags in 
        generic-jet data is dominated by fakes.

        The SLT fake probability is obtained by removing the contribution of SLT 
        tags due to heavy flavor decays in the generic-jet data used to 
        construct the SLT probability per track. For this purpose, we use the 
        signed impact parameter significance distribution of the soft lepton 
        tracks. The distribution observed in the data is fitted with the shape
        expected for leptons coming from the decay of $b$ and $c$-hadrons, 
        derived in simulated events, in addition to the shape of fake SLT tags.
        The shape for fake SLT tags is derived using all tracks taggable by 
        the SLT algorithm in events which do not contain any SECVTX, JPB or 
        SLT tags.
 \begin{figure}
 \begin{center}
 \leavevmode
 \epsffile{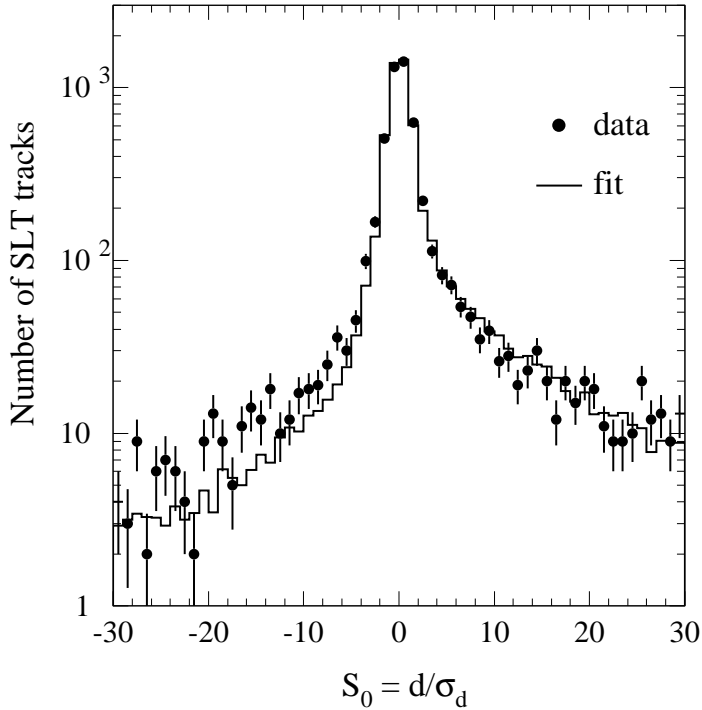}
 \caption[]{Distribution of the signed impact parameter significance of SLT 
            tracks contained in the JET 50 data ($\bullet$). The solid 
            histogram represents a fit using the shapes expected for $b$ and 
            $c$ semileptonic decays and for fake tags.}
 \label{fig:fig_10_15}
 \end{center}
\end{figure}
      Figure~\ref{fig:fig_10_15} shows the signed impact significance
      distribution of SLT tags in  JET 50 data  along with the fit result. 
      The composition of the SLT tags determined from these fits  is 
      ($74.0\pm3.2)\%$ fakes, ($10.5\pm2.3)\%$ $b$'s and ($14.5\pm4.3)\%$ $c$'s
      for all three generic-jet samples used to evaluate the SLT tagging 
      probability. Due to the systematic associated with using simulated shapes
      for the impact parameter distribution of lepton coming from $b$ and 
      $c$-decays, we estimate a 10\% error on the fraction of
      fake SLT tags  determined by the fits.
      Based on this result, the SLT mistag probability 
      per jet is obtained by rescaling the SLT tagging probability in
      generic-jet data by ($74.0 \pm 7.4$)\%.
\newpage
\narrowtext
 \begin{table}[p]
 \begin{center}
\def\arraystretch{0.8}
 \caption[]{Comparison of the observed and predicted  yields 
        of jets with SLT tags.}
\begin{tabular}{lccc}
\multicolumn{4}{c}{Samples used in the fake parametrization} \\ \hline
 Sample              & Predicted (P)   & Observed (O) &  (P-O)/O \\ 
 JET 20             &   5353.9       &   4994      &  $  7.2\%$ \\
 JET 20 without leading jet  &   3392.4       &   3383      &  $ 0.3\%$ \\
 JET 50             &   7082.9       &   6408      &  $ 10.5\%$ \\
 JET 50 without leading jet  &   4947.4       &   4988      &  $ -0.8\%$ \\
 JET 70             &   8089.2       &   7277      &  $ 11.2\%$ \\
 JET 70 without leading jet  &   5724.9       &   5678      &  $ 0.8\%$ \\
 \hline
 \multicolumn{4}{c}{Independent samples} \\ \hline
 JET 100             &  8603.6       &   7483      & $  15.0\%$ \\
 JET 100 without leading jet & 6109.8   &   5909      & $   3.4\%$ \\
 JET 140             &   1324.1      &   1196      & $  10.7\%$ \\
 $\sum E_T$ 175      &   3392.6      &   3392      & $   0.02\%$ \\
 $\sum E_T$ 125 4CL &   9651.9      &  10095      & $  -4.4\%$ \\
 $\sum E_T$ 300      &   1627.1      &   1401      & $  16.1\%$ \\
 Isolated $\gamma$      &    365.8      &    352      & $   3.9\%$ \\
 \end{tabular}
 \label{tab:tab_10_4}
 \end{center}
 \end{table}
\widetext

\subsubsection{Simulation of the SLT algorithm}
\label{sec:simu_slt}
        A main feature of the soft lepton tagger is that it was
        developed from studying real  leptons from photon conversions
        and $J/\psi$ mesons. The efficiency  of each selection criteria
        is well measured in the data.
        Therefore, the simulation of the soft lepton 
        tagger does not need to rely on the QFL modeling of the detector 
        response in order to estimate the tagging efficiency. 
        The SLT simulation matches tracks produced by QFL to
        electrons and muons at generator level. The electrons or muons are 
        required to come from $b$ or $c$ decay or any of their cascade decays.
        Electron tracks  are extrapolated to the CPR and CES detectors,
        and required to pass fiducial cuts.
        Electron candidates are eliminated if they are consistent with arising from
        photon conversions.
        Muon tracks are extrapolated to the muon detectors, required to pass 
        the fiducial cuts and classified according to the muon detector type.
        Finally tracks are weighted with the measured efficiencies of the selection
        criteria, which are  functions of the 
        track transverse momentum~\cite{top_prd,kestenbaum}.
        This procedure ensures that the simulation accurately models
        the soft lepton tagging efficiency.

        In Section~\ref{sec:herwig_norm} we compare rates of SLT tags in
        generic-jet data to the corresponding simulation to verify that
        the procedure has been implemented correctly.
        By construction, the SLT simulation does not produce mistags. However,
        when applying the SLT mistag probability to tracks in simulated 
        generic-jet data, we predict the same rate of fake SLT tags per jet 
        that is observed in the data. This implies that tracks in simulated 
        jets have the same multiplicity and transverse momentum distributions 
        as in the data.
\section{Efficiency of the SECVTX and JPB taggers}
\label{sec:simu_mod}
        We first describe  the calibration of the efficiency of the tagging
        algorithms in the simulation. For this purpose, we use
        the low-$p_T$ inclusive electron sample described
        in Section~\ref{sec:bb_sample} and the corresponding
        simulation. A large fraction of the events in 
        this sample is expected to originate from $b\bar{b}$ production
        in which a jet containing an electron from a semileptonic $b$-decay, 
        called e-jet, recoils against a jet from the other $b$, called 
        the away-jet or a-jet.
        The  tagging efficiency in the simulation, $\varepsilon_{b}^{\rm MC}$,
        is adjusted to the  value $\varepsilon_{b}$ of the tagging efficiency
        in the data using the scale factor
  \[ SF = \frac{ \varepsilon_{b} } {\varepsilon_{b}^{\rm MC}}. \]  
        Following the derivation of the scale factor, 
        Subsections A-I discuss the various sources of systematic
        uncertainty and also present cross-checks. In Subsection~J
        we provide an explanation for the deviation  of the scale factor
        from unity.

        The data sample consists of 55248 events. The simulated sample is 
	  generated with Herwig~\cite{herw_setup} (process 1500, $2 \rightarrow 2$
        hard scattering with transverse momentum threshold 
        $p_T^{\rm min} \geq 13$ GeV/c).
        Using the generic hard parton scattering,  $b\bar{b}$ and $c\bar{c}$ 
        pairs are produced through processes of order $\alpha_s^{2}$ as
        $gg \rightarrow b\bar{b}$ (direct production). Processes of order
        $\alpha_s^{3}$  are implemented in the generator
        through flavor excitation processes
        such as $gb \rightarrow g b$ or gluon splitting, where the process
        $gg \rightarrow gg$ is followed by $g \rightarrow b\bar{b}$.
        We use the MRS(G) set of parton distribution functions~\cite{mrsg}. 
        Apart from the parton distribution functions, the simulation package
        is the same as that used to generate $W+$ jet events. The generated 
        hard scattering sample corresponds to an integrated luminosity of 
        83.5 pb$^{-1}$. In this sample we select 
        events with an e-jet containing hadrons with heavy flavor. After 
        applying the same selection used for the data, the simulated sample
        low-$p_T$ electron sample contains 16547 events.

        Table~\ref{tab:tab_9_1} shows the composition of the simulated 
        inclusive electron sample. One notices that 80\% of the e-jet are due
        to $b\bar{b}$ production and that only 33\% of the away-jets contain 
        heavy flavor. 

        In the simulated events  where the away jet is tagged by SECVTX
        ($ST_{a-jet}$),
        94\% of the electron-jets are due to $b\bar{b}$ production. 
        It is therefore convenient to measure the $b$-tagging efficiency 
        as the fraction of these events in which the electron-jet is tagged 
        by SECVTX or JPB
\begin{eqnarray}
       \varepsilon_{b}^{MC} = \frac{SDT}{ST_{a-jet}} 
\end{eqnarray}
        where $SDT$ is the number of events where both the electron
        and away-jet contain heavy flavor and are tagged. The $b\bar{b}$ production 
        accounts for 99\% of the simulated events with a double tag.

        Table~\ref{tab:tab_9_2} lists rates of tags in the data and in the 
        simulation. In the simulation there are very few mistags and they are 
        easily identified because the jet does not contain $b$ or $c$-hadrons in a
        cone of radius 0.4 around its axis. In the data,
        the rate of mistags is evaluated 
        using the parametrization described in Section~\ref{sec:mistags}.

        We use the simulation to describe $F_{hf}$, the fraction  of  data 
        in which electron-jets contain hadrons with heavy flavor. The data 
        contain also a relevant number of e-jets in which the electron is not 
        associated with the production of hadrons with heavy flavor (mostly
        from photon conversions in jets due to light quarks or gluons).
        In these events, the electron-jet contributes only mistags. To describe 
        the remaining fraction ($1-F_{hf}$) of the data, we make the additional 
        assumption that away-jets in these events contain the same fraction of
        heavy flavor as generic-jets. The parametrization of the probability 
        of tagging jets with heavy flavor in generic-jet data is derived in 
        Section~\ref{sec:mistags}. The 10\% uncertainty associated with this 
        parametrization is discussed in Section~\ref{sec:ss-qcdtag}.
\newpage
\narrowtext
 \begin{table}
 \begin{center}
 \def\arraystretch{0.8}
 \caption[]{Fractions of electron and away-jets  before and after tagging
            in the  low-$p_T$ inclusive electron simulation. 
             $SN_{a-jet}^{h.f.}$ and $SN_{a-jet}^{prompt}$
    are the fractions  of away-jets with and without heavy flavor.}
\begin{tabular}{lcccccc} 
 {~} & \multicolumn{2}{c}{direct production} & \multicolumn{2}{c}{flavor
excitation} & \multicolumn{2}{c}{gluon splitting} \\
 {~} & {$b$ (\%)} & {$c$ (\%)}& {$b$ (\%)} & {$c$ (\%)} & {$b$ (\%)} 
 & {$c$ (\%)}\\
 \hline
{$SN_{e-jet}$}        & 20.90 & 3.49 & 39.72 & 10.26 & 19.39 & 6.22 \\
{$SN_{a-jet}^{h.f.}$} & 19.93 & 3.31 &  5.91 &  1.35 &  2.61 & 0.53 \\
{$SN_{a-jet}^{\rm prompt}$}& 1.64 & 0.29 & 35.65 &  9.38 & 19.60 & 6.38 \\
{$ST_{e-jet}^{SEC}$} & 24.51 & 0.68 & 47.58 & 2.55  & 22.74 & 1.93 \\
{$ST_{e-jet}^{JPB}$} & 23.57 & 1.60 & 44.64 & 5.93  & 20.75 & 3.51 \\ 
{$ST_{a-jet}^{SEC}$} & 70.50 & 3.07 & 16.17 & 2.29  & 7.47  & 0.51 \\
{$ST_{a-jet}^{JPB}$} & 67.59 & 5.23 & 15.06 & 3.51  & 7.11  & 1.50 \\
{$SDT^{SEC}$}        & 73.46 & 0.54 & 17.01 & 0.43  & 8.45  & 0.11 \\
\end{tabular}
\label{tab:tab_9_1}
\end{center}
\end{table}
\widetext

\narrowtext
 \begin{table}[htpb]
 \begin{center}
 \def\arraystretch{0.8}
 \caption[]{Number of events before and after tagging electron and away-jets.
           $P_{QCD}$ is the probability of tagging away-jets if they contain 
           the same heavy flavor fraction as generic-jets (see text).}
\begin{tabular}{lcclc} 
  \multicolumn{3}{c}{ Data } & \multicolumn{2}{c}{ Simulation } \\
 Type & Observed$-$mistags   & $P_{QCD}$(\%) & Type & Observed-mistags \\
 \hline
 $N_{e-jet}$       & 55248            &        & $SN_{e-jet}$  & 16547    \\
 $T_{e-jet}^{SEC}$ & 8158  $ -$ 84.3     &       & $ST_{e-jet}^{SEC}$   &
   4549$ -$ 0 \\
 $T_{e-jet}^{JPB}$ & 9123  $ -$ 335.3    &       & $ST_{e-jet}^{JPB}$   &
   5990$ -$ 0 \\
 $T_{a-jet}^{SEC}$ & 3640  $ -$ 112.8    &   1.67 &$ST_{a-jet}^{SEC}$   & 
 1832$ -$ 7 \\
 $DT^{SEC}$       & 1126  $ - $23.8     &       &$SDT^{SEC}$    &   545$ -$ 1 \\
 $DT^{JPB}$       & 1225  $ -$ 35.3     &       &$SDT^{JPB}$     &   743$ -$ 1 \\
 \end{tabular}
 \label{tab:tab_9_2}
 \end{center}
 \end{table}
\widetext

\clearpage
        We use the following procedure to derive the tagging efficiency 
        scale factor separately for SECVTX and jet-probability,
        together with the heavy flavor purity $F_{hf}$ of the 
        data.  The data and the simulation are normalized to the same number 
        of tagged electron-jets that contain heavy flavor, $T_{e-jet}$ and 
        $ST_{e-jet}$, through the coefficient
 \[ \alpha = \frac{T_{e-jet}}{ST_{e-jet}}. \] 
        Before tagging, the heavy flavor purity of the data is therefore
        given by
\begin{eqnarray}
      F_{hf} = \frac { \alpha \cdot SN_{e-jet} } {SF \cdot N_{e-jet}  }
\end{eqnarray}
        where $N_{e-jet}$ and $SN_{e-jet}$ are the number of e-jets in the data
        and the simulation and $SF$ is the tagging efficiency scale factor. 
        We start assuming $SF=1$. 
        It follows that in  the data the number of events 
        in which a tagged away-jet with heavy 
        flavor is associated to an electron-jet without heavy flavor is
 \[ T^{QCD}_{a-jet} = (1- F_{hf}) \cdot N_{e-jet} \cdot P_{QCD} \]
        and that the number of events in which a tagged away-jet containing heavy 
        flavor is associated with an electron-jet also containing  heavy 
        flavor is
 \[ T_{a-jet}^{SEC} - T^{QCD}_{a-jet}. \]
        For the data the $b$-tagging efficiency,  analogy of equation (1),
        is then
  \[ \varepsilon_{b} = \frac{DT}{T_{a-jet}^{SEC} - T^{QCD}_{a-jet}}. \]
       where, as before, $DT$ is the number of events in which  the a-jet is 
       tagged by SECVTX and the e-jet has a SECVTX or JPB tag.

       The ratio of the tagging efficiencies in the data and simulation
       yields the scale factor 
       \[ SF = \frac{ \varepsilon_{b} } {\varepsilon_{b}^{MC}}. \]
       The value of the scale factor is inserted again in the equation (2) and
       we iterate until the scale factor value is stable to within 1\%. 
\narrowtext
 \begin{table}[htpb]
 \begin{center}
 \def\arraystretch{0.8}
 \caption[]{Data to Monte Carlo tagging efficiency scale factors.
             $F_{hf}$ is 
            the  fraction of e-jets containing heavy flavor in 
            the data.}
 \begin{tabular}{lcc} 
  Sample                     & SF             & $F_{hf}$ \\ \hline 
  SECVTX e-jet, SECVTX a-jet & 1.23$\pm$0.07  & 43.5$\pm$2.9 \% \\
  JPB e-jet, SECVTX a-jet    & 0.96$\pm$0.05  & 45.3$\pm$2.4 \% \\
 \end{tabular}
 \label{tab:tab_9_3}
 \end{center}
 \end{table}
\widetext

        Using the numbers of electron and away-jets listed in 
        Table~\ref{tab:tab_9_2}, we derive $SF$=1.23$\pm$0.07 for SECVTX and 
        0.96$\pm$0.05 for jet-probability. The error accounts for the sample 
        statistics (with the largest contribution coming from the simulation)
        and for 10\% uncertainties in the evaluation of the mistag rates and 
        in the prediction of the rate of tags in generic-jets with heavy flavor.

        The $b$-purity of the e-jets before tagging, $F_{hf}$ = (43.5 $\pm$ 2.9)\%, 
        is in agreement with the measurement in Ref.~\cite{top_prd}, 37$\pm$8\%,
        using the fraction of tagged electron-jets that also contain a muon 
        of opposite charge.

        The average SECVTX tagging efficiency is 
        36.7$\pm$1.9\% in the data and (29.8$\pm$1.1)\% in the simulation.  
        The corresponding numbers 
        for jet-probability are (39.2$\pm$2.1)\% and (40.7$\pm$1.1)\%, 
        respectively. 

\begin{figure}
\begin{center}
\leavevmode
\epsffile{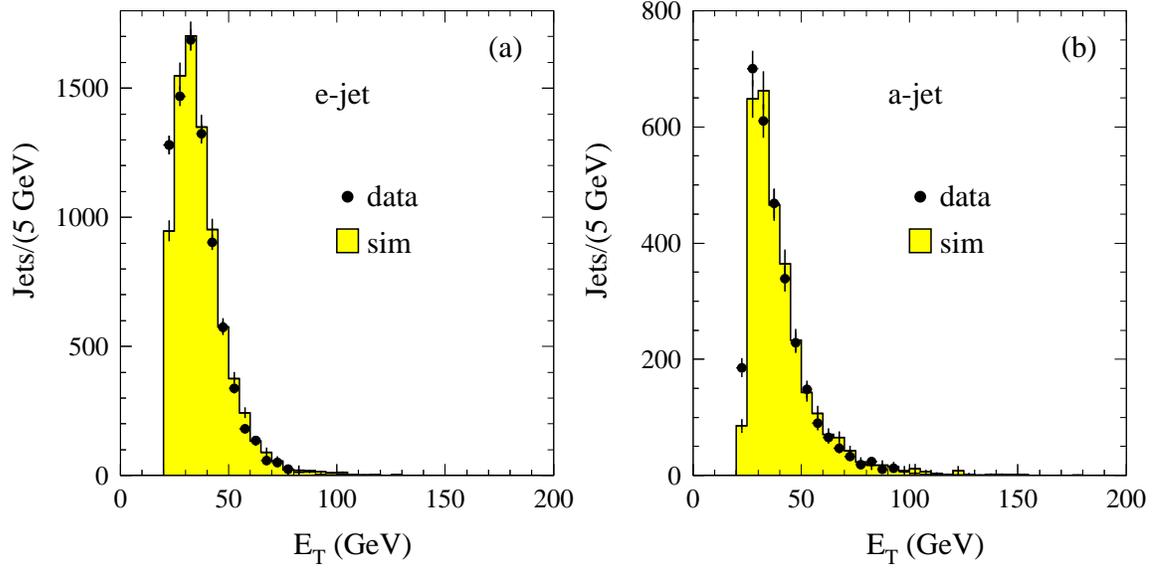}
\caption[]{ Distributions of the transverse energy of electron-jets (a) 
            and away-jets (b) tagged by SECVTX.}
 \label{fig:fig_9_0}
 \end{center}
 \end{figure}

\begin{figure}
\begin{center}
\leavevmode
\epsffile{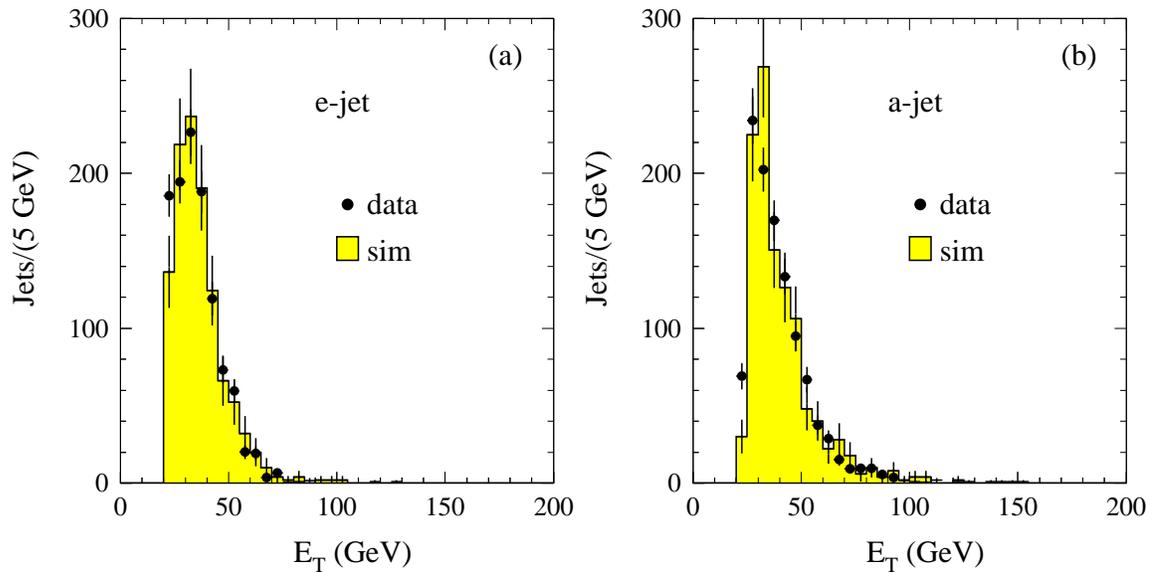}
\caption[]{ Distributions of the transverse energy of electron-jets
            (a) and away-jets (b) in events with double SECVTX tags.}
 \label{fig:fig_9_1}
 \end{center}
 \end{figure}

\begin{figure}
\begin{center}
\leavevmode
\epsffile{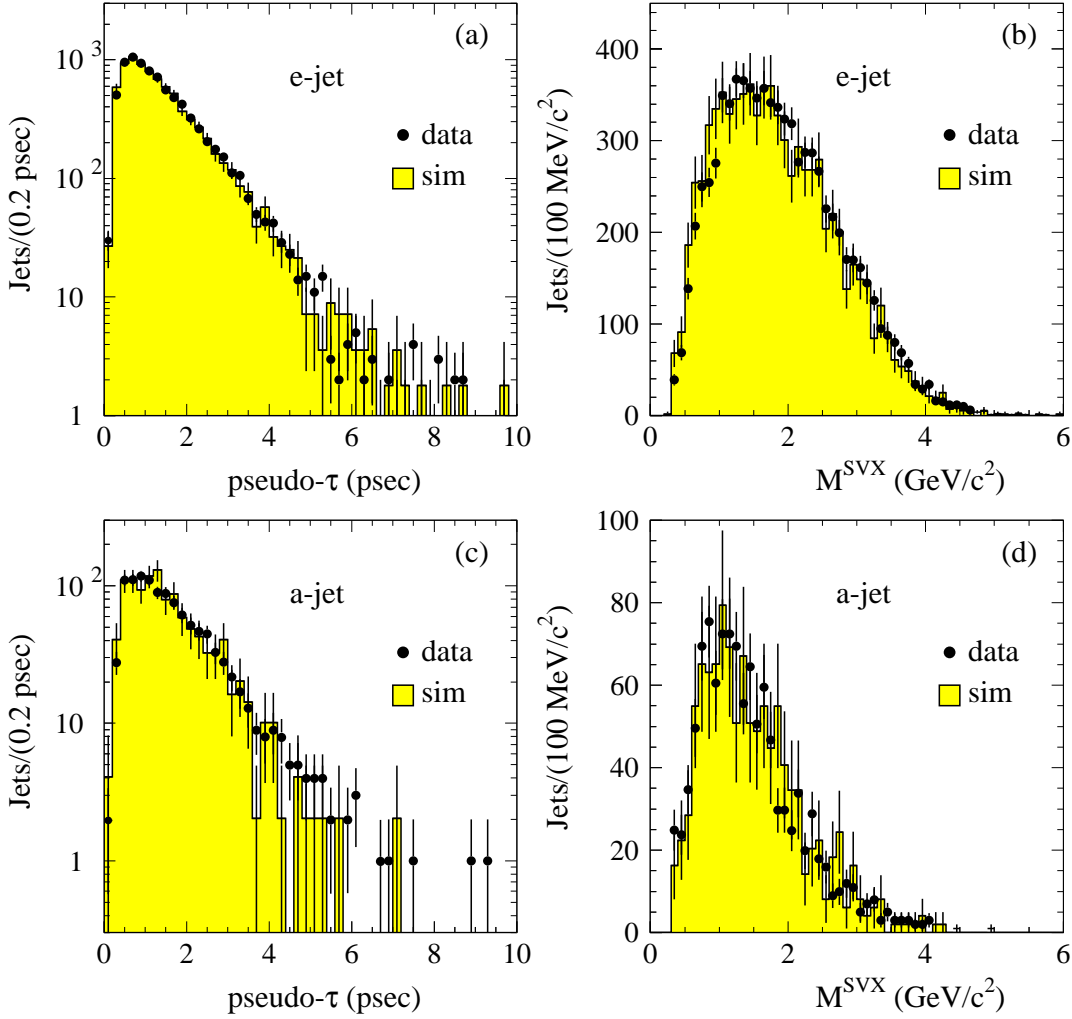}
\caption[]{Distributions of pseudo-$\tau$ (a) and of the invariant mass 
          (b) of SECVTX tags in electron-jets; (c) and (d) are the analogous
          distributions for away-jets in events with double tags.}
 \label{fig:fig_9_2}
 \end{center}
 \end{figure}

        Since the tagging efficiencies depend on the jet energy, it is important
        to verify the agreement between the tagging efficiencies as a
        function of the transverse energy in the 
        data and the simulation as is done in Figures~\ref{fig:fig_9_0} 
        and~\ref{fig:fig_9_1}. The distributions of the lifetime and invariant
        mass of the SECVTX tags is given 
        in Figure~\ref{fig:fig_9_2} and  supports our 
        determination of the $b$-purity of the sample. The lifetime of a 
        SECVTX tag is defined as
 \[ {\rm pseudo}-\tau = \frac { L_{xy}  \cdot M^{SVX}} { c \cdot {p_{T}^{SVX}} } \]
        where $M^{SVX}$ and $p_{T}^{SVX}$ are the invariant mass and the 
        transverse momentum of all tracks forming the SECVTX tag.

\subsection{Check of the background parametrization using a photon conversion sample}
\label{sec:ss-qcdtag}
        In events where the e-jet does not contain heavy flavor, we predict
        the rate of tagged away-jets containing heavy flavor using the 
        probability of tagging jets with heavy flavor as  measured in generic-jet 
        data. We test this method in a sample of data where the electrons
        in the e-jet are due 
        to photon conversions. The  criteria used to identify photon conversions
        are listed in Table~\ref{tab:tab_3_1}. In this case we require that 
        an electron is matched by a second track consistent with a photon 
        conversion and that it is not matched by a track segment in the VTX detector.
        Otherwise, we select this sample as the inclusive electron sample 
        where in contrast conversions were removed.

        Following the procedure used in the previous section, we determine the 
        fraction of events with heavy flavor to be $F_{hf}$ = 8.7$\pm$0.9\% 
        from the number of e-jets with a SECVTX or JPB tag. Tagging rates
        in  events due to heavy flavor production
        are described using the Herwig simulation as used in the previous section.
        In the 
        remaining 91.3\% of the events, we describe the rates of tagged 
        away-jets using the  parametrization derived from generic jets.

        Table~\ref{tab:tab_9_4} shows that this procedure  correctly predicts 
        the rates of tags observed in the data. We take the 10\% statistical 
        error of this comparison as the systematic uncertainty of the method.
\newpage
\widetext
 \begin{table}[p]
 \begin{center}
\def\arraystretch{0.8}
 \caption[]{ Rates of  events in which the electron jet is
 due to a photon conversion before and after tagging.
      The heavy
             flavor purity of this sample is $F_{hf}$=8.7$\pm$0.9\%. 
  Events where the e-jet contains heavy flavor are described with the Herwig
simulation. In the remaining events, the rate of tagged away-jets (QCD)
 is predicted using the probability for tagging jets with heavy flavor
 in generic-jet data.
 Mistags have been removed from the data and simulation.}
 \begin{tabular}{lcccc}
   Type             & Data  & Simulation & QCD & Prediction \\
 $N_{e-jet}$       & 4027   & 350$\pm$37 & 3677$\pm$37& 4027  \\
 $T_{e-jet}^{SEC}$ &  108.3$\pm$10.6 & 114$\pm$12 & 0   & 114$\pm$12  \\
 $T_{e-jet}^{JPB}$ &  133.1$\pm$12.5& 126$\pm$13 & 0   & 126$\pm$13  \\
 $T_{a-jet}^{SEC}$   & 102.2$\pm$10.5&41.6$\pm$5.0& 60.2$\pm$6.0
& 101.8 $\pm$ 7.8 \\
 $T_{a-jet}^{JPB}$   & 135$\pm$13.7 &45.0$\pm$4.5& 86.7$\pm$8.7
& 131.7 $\pm$ 9.8 \\
 \end{tabular}
\label{tab:tab_9_4}
 \end{center}
 \end{table}
\narrowtext

\subsection{Sensitivity of the scale factor to the modeling of $c$-jets} 
        In the simulation the tagging efficiency is defined as the ratio of events 
        with double tags to all events where the away-jet is tagged by SECVTX.
        As shown in Table~\ref{tab:tab_9_1}, the Herwig 
        simulation predicts that 94\% of the a-jets with a SECVTX tag are 
        due to $b\bar{b}$ production. The remaining 6\% of the a-jets are due to 
        $c\bar{c}$ production and are accounted for by the simulation but in 
        principle this could be improperly modeled. In events where a-jets have
        a JPB tag, the fraction of $c\bar{c}$ production increases to 11\% (see 
        Table~\ref{tab:tab_9_1}). 
        If SECVTX and JPB scale factors are determined using a-jets tagged
        by JPB instead of a-jets tagged by SECVTX, the scale factors change each
        by less than 1\%. Therefore, we conclude  that the modeling of 
        $c$-jets is satisfactory
        for the determination of the $b$-tagging efficiency scale factor.
\subsection{Dependence of the scale factor on the gluon splitting cross section}
        As shown in Table~\ref{tab:tab_9_1}, a fraction of the events in the
        inclusive electron sample is due to gluon splitting to heavy flavor 
        quarks. The calibration of the Herwig simulation using generic-jet 
        data in Section~\ref{sec:herwig_norm} shows that direct production 
        and the heavy flavor excitation as implemented in Herwig provide a 
        fair description of the data, but the gluon splitting cross section
        requires a (40 $\pm$ 20)\% correction. We repeat the calculation of the
        scale factor using this larger gluon splitting cross section. We find 
        that the SECVTX scale factor increases from 1.23 to 1.25. The final
        scale factor we use will be this latter value.
\subsection{ $E_T$ dependence of the scale factor}
\label{sec:sf_et}
        Jets produced directly in association with a $\W$ boson have transverse
        energies comparable to the jets in the low-$p_T$ inclusive electron 
        sample. However, $b$-jets produced by top decay have substantially 
        higher transverse energies. In this section, we investigate a possible
        $E_T$ dependence of the scale factor using two methods.

\begin{figure}
 \begin{center}
 \leavevmode
 \epsffile{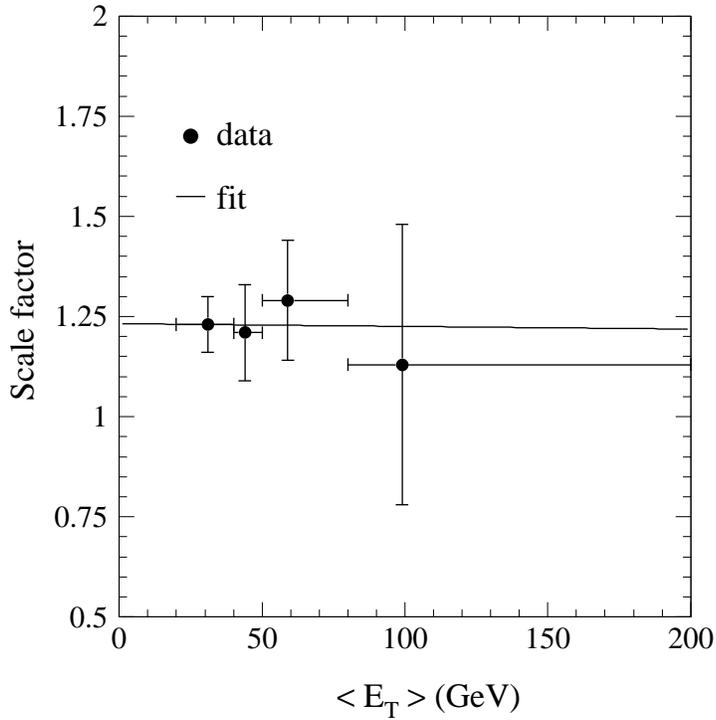}
 \caption[]{SECVTX tagging efficiency scale factor as a function of the 
            average transverse energy $<E_T>$ of the electron-jet. The line 
            represents a fit with a first degree polynomial.}
 \label{fig:fig_9_3}
 \end{center}
 \end{figure}

        First, we derive the value of the SECVTX scale factor in four
        different bins of the electron-jet transverse energy. In 
        each bin, we calculate the average e-jet transverse
        energy $<E_T>$ and the scale 
        factor using the iterative procedure previously described. The result
        of the study is shown in Figure~\ref{fig:fig_9_3}. A fit of the scale 
        factor as a function of the transverse energy with a first 
        order polynomial yields a $\chi^2$ of 0.3 for 2 d.o.f. and 
     \[ SF(E_T) = (1.23 \pm 0.17) -  
        (0.1 \pm 4.0 )\times 10^{-3} \times <E_T> ({\rm GeV)}  \]
        with a correlation $\rho= -0.95$ between the two fit parameters.
        The result of this fit is therefore consistent
        with a constant scale factor.
            
\begin{figure}
 \begin{center}
 \leavevmode
 \epsffile{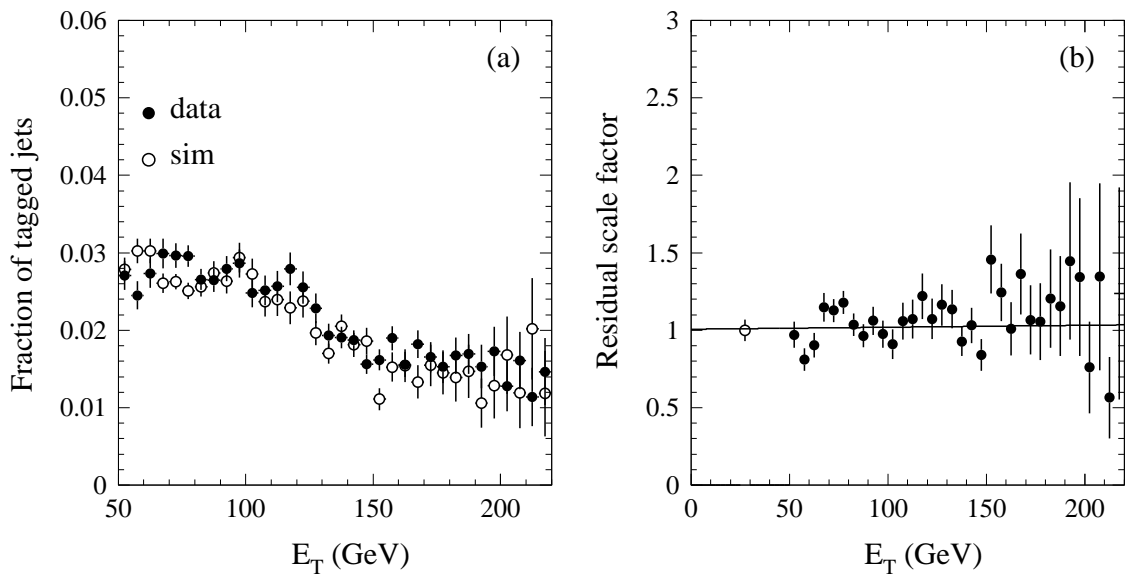}
 \caption[]{Fractions of tagged jets (a) as a function of the jet transverse 
            energy.
            The residual scale factor (b) is 
            defined as the ratio of these fractions in the data
            and the simulation. The open circle in (b) represents the 
            inclusive electron sample result.}
 \label{fig:fig_9_4}
 \end{center}
 \end{figure}

        In the second method, we compare the fraction of jets with heavy flavor 
        tagged by SECVTX in JET 50 and in JET 100 data and in the 
        corresponding Herwig simulation tuned as in 
        Section~\ref{sec:herwig_norm}. The $b$-tagging efficiency in the 
        detector simulation is increased by the factor 1.25 independently of 
        the jet transverse energy. The ratio $RSF$ of the fractions of
        tagged jets  in 
        the data and in the simulation is  sensitive to any residual 
        $E_T$ dependence of the scale factor. The result of this method is 
        shown in Figure~\ref{fig:fig_9_4}.  We fit the ratio $RSF$ of the 
        tagging efficiencies in the data to the simulation as a function of 
        the jet transverse energy with a first order polynomial. The fit 
        yields a $\chi^2$ of 51 for 49 d.o.f and  
  \[  RSF(E_T)= (1.01\pm 0.05) +  (1.3  \pm 4.6)\times 10^{-4}
        \times E_T ({\rm GeV)} \] 
        with a correlation $\rho=-0.92$ between the two fit parameters. The
        data are consistent with a constant scale factor.
%
\subsection{Uncertainty  of the scale factor}
\label{sec:sf_uncert}
        The SECVTX $b$-tagging efficiency scale factor measurement using 
        the inclusive electron sample 
        has a 5.6\% uncertainty. The uncertainty in the gluon splitting
        cross section predicted by Herwig  results in an additional 0.8\%
        uncertainty of the scale factor.
        By folding the $E_T$ spectrum of $b$-jets from
        top decays with the $E_T$ parametrization of the scale factor from the
        fit shown in Figure~\ref{fig:fig_9_3} (a variation of the fit parameters
	  by $\pm 1 \sigma$ yields a $\pm$4.2\% change in the efficiency for 
        tagging $b$-jets and $\pm$3.9\% change in the efficiency to tag top 
        events), we estimate a 4\%  uncertainty from any residual $E_T$ dependence.
        These errors are mostly systematic and in general highly correlated.
        Altogether, we assign a 10\%  error to the determination of the scale 
        factor after combining linearly the above contributions. 
        Our final estimate of the $b$-tagging efficiency scale factor for the
        SECVTX algorithm is $SF = 1.25 \pm 0.13$ and for
        the jet-probability algorithm is $SF = 0.96 \pm 0.10$. 
        The latter is consistent with unity.
\subsection{Check of the scale factor using a low-$p_T$ inclusive
           muon sample}
        The low-$p_T$ inclusive muon sample is  analogous to the electron 
        sample in that a muon with $p_T \geq$ 10 \gevc is required in place of
        an electron with $E_T\geq$ 10 GeV (see Section~\ref{sec:bb_sample}).
        It provides an independent sample 
        for checking the tagging efficiency scale factor.
        The low-$p_T$ muon sample consists of 
        10393 events. In these events muon-jets without heavy flavor are due 
        to fake muons arising from  non-interacting hadrons or 
        in-flight decays of $K$ and $\pi$  mesons. We compare to a simulated sample 
        also generated  using the option 1500 of Herwig which consists of 4280 
        events. The same procedure described above yields a SECVTX 
        tagging efficiency scale factor of 1.24$\pm$0.10, in agreement with 
        the value 1.23 $\pm$0.07 derived in the inclusive electron sample
        (before  correcting the gluon splitting cross section).
        At the same time the heavy flavor purity of the low-$p_T$ muon sample
        is measured to be $F_{hf}$=59.7$\pm$3.6\%.
\subsection{Check of the scale factor in jets containing inclusive
              $b$-decays}
        In this section we investigate whether the scale factor 
        is different in jets containing semileptonic $b$-decays and inclusive
        $b$-decays. We use the low-$p_T$ inclusive electron samples and 
        normalize the data and the  simulation to the same number of 
        electron-jets with a SECVTX tag after mistag removal.
        In the simulation, the rate of gluon splitting to $\bbbar$ and $\ccbar$
        pairs is corrected as in Section~\ref{sec:herwig_norm}. We compare 
        rates of away-jets which are taggable and which are tagged by SECVTX. 
        We find that the simulation predicts correctly the amount of 
        taggable away-jets but it underestimates by a factor 1.23 $\pm$ 0.08
        the rate of SECVTX tags with respect to the data.
\subsection{Check of the scale factor using rates of double tags in 
        generic-jet data}
        The studies of the $E_T$ dependence of the SECVTX scale factor 
        performed in Section~\ref{sec:sf_et} 
        depends upon the assumption that Herwig models correctly the 
        fractional yield of jets with heavy flavor as a function of their 
        transverse energy. We use the JET 50 and JET 100 data and simulation 
        for a test independent of this assumption.
        We select events with only two jets: one taggable jet with transverse 
        energy larger than the trigger threshold and one taggable jet with 
        $E_T \geq$ 15 GeV in the opposite hemisphere. We compare the number 
        of events with double JPB tags and double SECVTX tags in the data and 
        in the Herwig simulation after mistag removal. In the simulation, 
        92\% of these double tags are due to $b\bar{b}$ production.
        The ratio of double SECVTX to double JPB tags in the data and the 
        simulation is:
   \[ R_{\rm data}   = 0.92\pm 0.18 \mbox{~~and~~~} 
      R_{\rm sim}   = 0.61 \pm 0.05 \]
      This ratio does not depend on the absolute cross section for producing
      jets with heavy flavor. From the equivalence
   \[ \frac{ R_{data}}{R_{sim}} = \left(\frac {SF^{SEC}} {SF^{JPB}}\right)^2 \;,\]
        we measure  $\frac {SF^{SEC}} {SF^{JPB}}$=1.24$\pm$0.13
        using generic jets with high transverse energy,
        in agreement 
        with the value $\frac {SF^{SEC}} {SF^{JPB}}$=1.28$\pm$0.10 
        measured in the low-$p_T$ inclusive electron sample.
\subsection{SECVTX  efficiency for tagging $c$-jets}
\label{sec:sf_cjet}
        Since we need to apply a large correction to the simulated SECVTX 
        efficiency for tagging $b$-jets, it is worth investigating differences 
        between data and simulation for tagging $c$-jets. For this purpose, 
        we compare rates of tags in the JET 50 and JET 100 data to the 
        corresponding Herwig simulation, described in 
        Section~\ref{sec:herwig_norm}, normalized to the same number of events.

        We define $R$ as the ratio of the number of SECVTX to JPB tags 
        after mistag removal. In the data $R = 0.77 \pm 0.07$.
        Under the assumption that  the heavy flavor composition of the data
        is modeled correctly by Herwig, the SECVTX scale factor for $c$-jets, 
        $SF_c^{SEC}$, can then be derived solving the equivalence
  \[ R =  \frac
      { T_{b}^{SEC} \cdot SF_b^{SEC} + T_{c}^{SEC} \cdot SF^{SEC}_{c} } 
      {T^{JPB} \cdot SF^{JPB} }\]
        where $T_{b}^{SEC}$=5354 and $T_{c}^{SEC}$=2477 are the number of 
        simulated $b$ and $c$-jets tagged by SECVTX, and $T^{JPB}$=11958
        is the number of JPB tags. Using $SF_b^{SEC}$=1.23 $\pm$ 0.07 and 
        $SF^{JPB}$=0.96 $\pm$ 0.05, we derive that the SECVTX scale factor 
        for tagging $c$-jets is $SF_c$ = 0.92 $\pm$ 0.28. The error is 
        determined by the uncertainty of the heavy flavor composition 
        (see Section~\ref{sec:herwig_norm}) and by the errors of the scale 
        factors $SF_b^{SEC}$ and $SF^{JPB}$.
\subsection{Understanding  of the scale factor}
        In an effort to explain the 25\% difference of the SECVTX
        tagging efficiency in the data and the simulation
        we uncovered three oversights in 
        the simulation package used in this and in some previous CDF 
        analyses~\cite{cdf_xsec,topxs}. A significant
        fraction of the difference is due to the 
        use of an outdated version of the {\sc CLEO} decay tables and to outdated 
        $B$-lifetimes in the CDF particle database. The above two 
        inaccuracies account for $\sim 40$\% of the difference of the SECVTX scale 
        factor from unity. Small inconsistencies in the implementation of the SVX
        geometry in the simulation contribute an additional 16\% to this 
        difference. If we corrected for these effects, the new
        determination of the SECVTX scale factor would be 1.09$\pm$0.11;
        the uncertainty includes the error on the $b$-lifetime ($\sim$3\%)
	  and the uncertainty of the track degradation procedure described in 
        Section~\ref{sec:Simul} ($\sim$8\%). The  efficiency of 
        jet-probability is not affected by these changes in the QFL
        simulation.
\clearpage
\section{SECVTX and JPB mistags}
\label{sec:mistags}
        In this section we estimate the SECVTX and JPB mistag rate in a variety
        of control samples before applying it to $W+$ jet and $Z+$ jet events in
        Sections~XII and XIII.
        Tags in jets without heavy flavor, which we call mistags,
        are caused by detector 
        resolution effects. SECVTX mistags are poorly reproduced by our detector 
        simulation and traditionally CDF removed this background from the data
        using a parametrization of the probability of finding negative SECVTX 
        tags in JET 50 data~\cite{cdf_xsec,topxs,top_prd}.
        We derive a new parametrization of the mistag rate using the 
        JET 20, JET 50, JET 70, JET 100 and $\sum E_T$ 300  data
        described in Section~\ref{sec:jet_sample}. Even if JPB mistags
        are well reproduced by the detector simulation, we derive a mistag
        parametrization also for JPB tags because jet-probability has a higher 
        rate of mistags than SECVTX and provides a better check of the method.

        The method to evaluate the mistag probability starts with the measurement 
        of the number of positive and negative tags in generic-jet data and their 
        parametrization as a function of the jet $E_T$ and the jet track 
        multiplicity, $N^{SVX}_{TRK}$. The tagging probability is derived as a
        ratio of the number of tags to the number of taggable jets in bins of 
        transverse energy and track multiplicity.
        We use only jets which are far away from calorimeter cracks and 
        correct the jet energy for the detector response and out-of-cone 
        losses (see Section~\ref{sec:jetcor}).

        Negative tags are also produced in jets containing heavy flavor. In 
        particular, the probability of producing negative tags is different 
        for jets initiated by a heavy-quark or by  gluon splitting to a pair 
        of heavy quarks. Since this contribution to negative tags must be 
        accounted for and subtracted in order to obtain the mistag rate due 
        to jets without heavy flavor, it is important to parametrize the rate
        of  negative tags in a sample in which the composition of quark and 
        gluon jets is well understood and is not subject to 
        additional uncertainty due to the 
        simulation. For this reason, in each generic-jet sample, we use only
        jets with transverse energy above the trigger threshold (leading jets):
        jets with corrected $E_T \geq $ 30, 70, 90, 120 and 160 GeV in the 
        JET 20, JET 50, JET 70, JET 100 and $\sum E_T$ 300 data, respectively.
        In the generic-jet simulation, 95\% of the leading jets with a tag contain
        just one heavy-flavored hadron (a large fraction of these leading jets 
        is produced by heavy quarks from flavor excitation or 
        direct production).
        The $E_T$ region below 30 GeV is mapped selecting
        events containing  two leading jets, but using only the additional
        jets in the event ; in the simulation, 96\% of these non-leading jets
        which are tagged  contain two  hadrons with heavy flavor produced 
        by a gluon splitting process.

\begin{figure}
 \begin{center}
 \leavevmode
 \epsffile{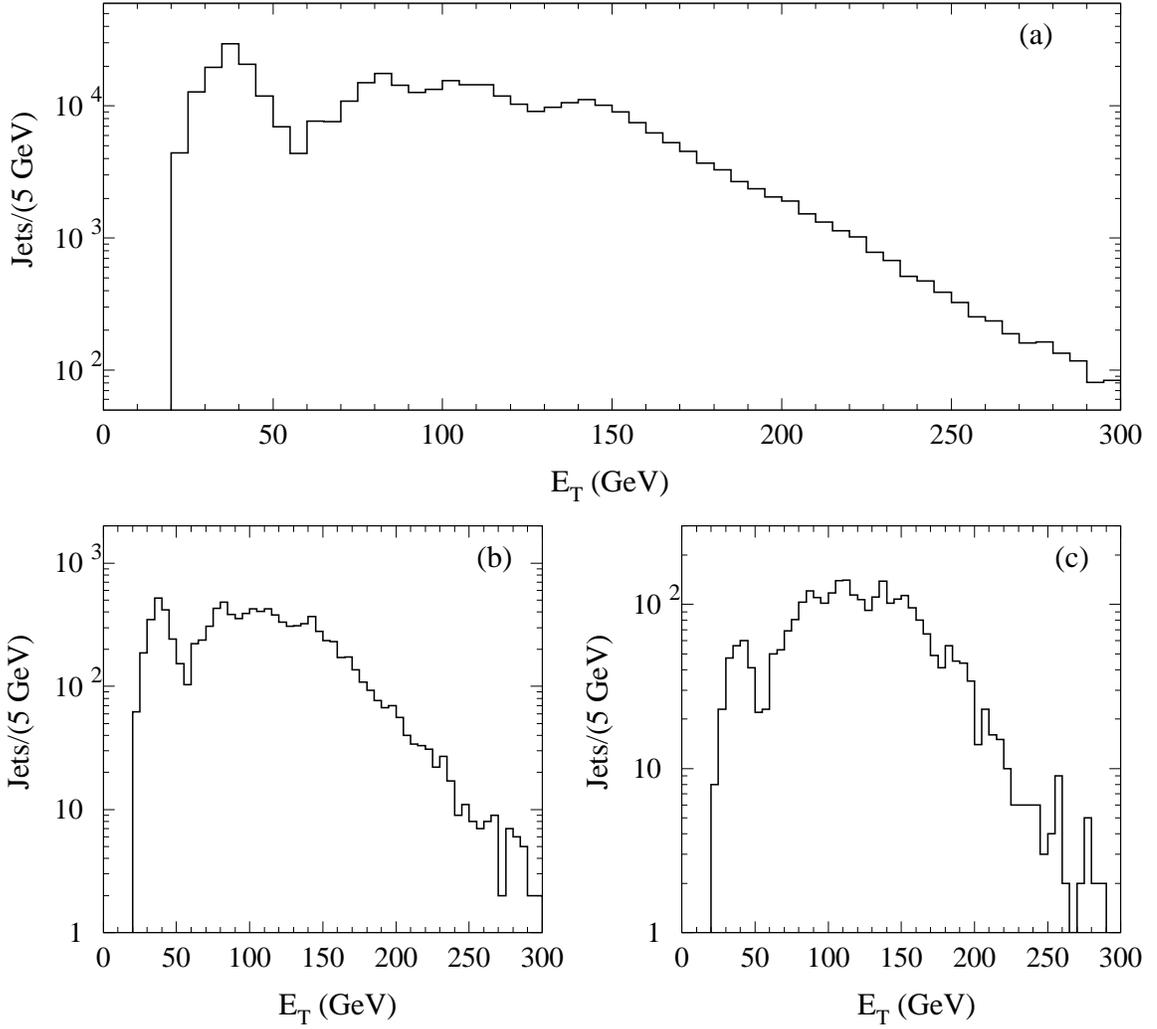}
 \caption[]{Transverse energy distributions  of (a) taggable jets, and jets 
            with positive (b) and (c) negative  SECVTX tags.}
 \label{fig:fig_10_0}
 \end{center}
 \end{figure}

\begin{figure}
 \begin{center}
 \leavevmode
 \epsffile{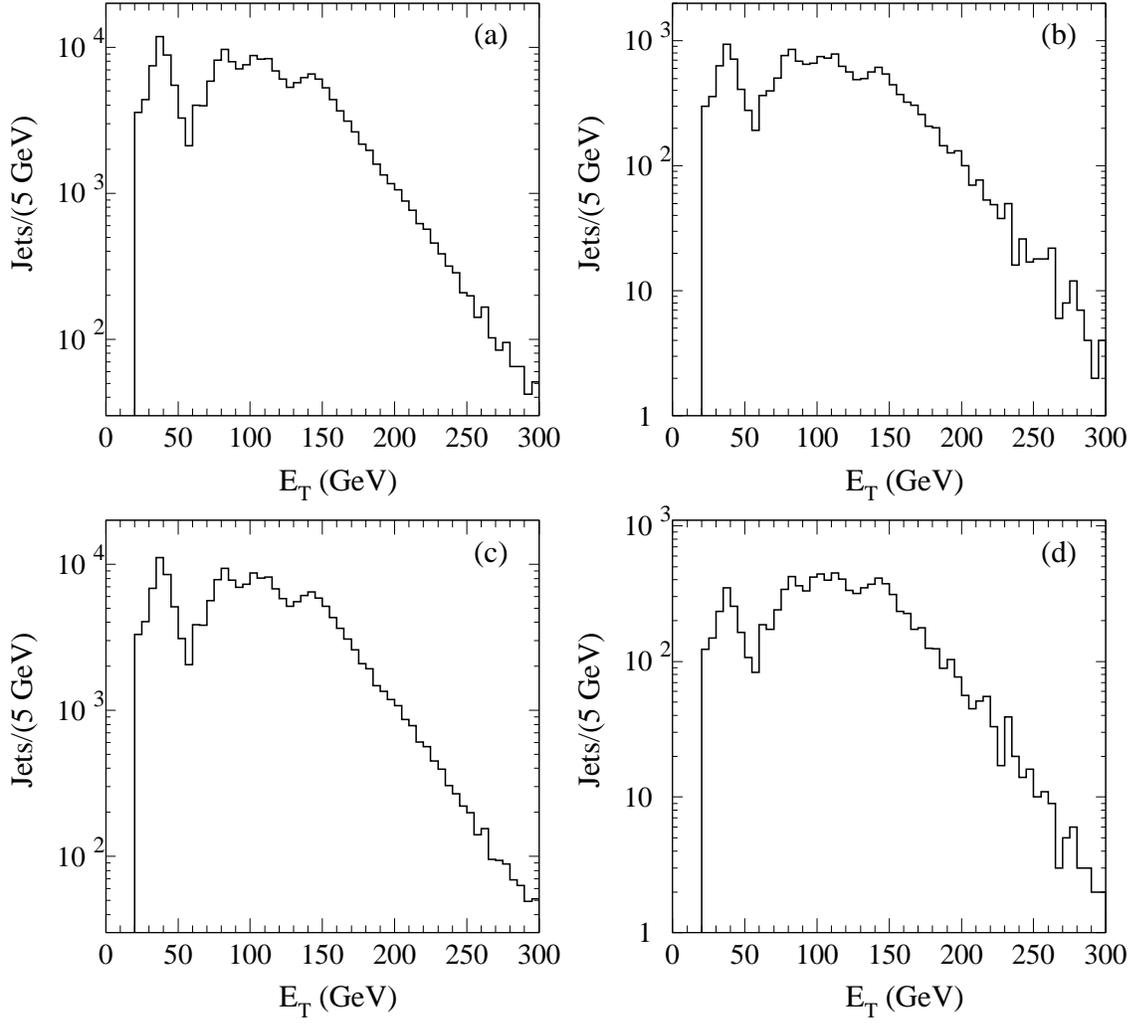}
 \caption[]{Transverse energy distributions of jets with (a) at least two JPB 
            tracks with positive impact parameter significance, (b) with 
            positive JPB tags, (c) with two or more JPB tracks with negative
            impact parameter significance, and  (d) with negative JPB tags.}
 \label{fig:fig_10_1}
 \end{center}
 \end{figure}

\begin{figure}
 \begin{center}
 \leavevmode
 \epsffile{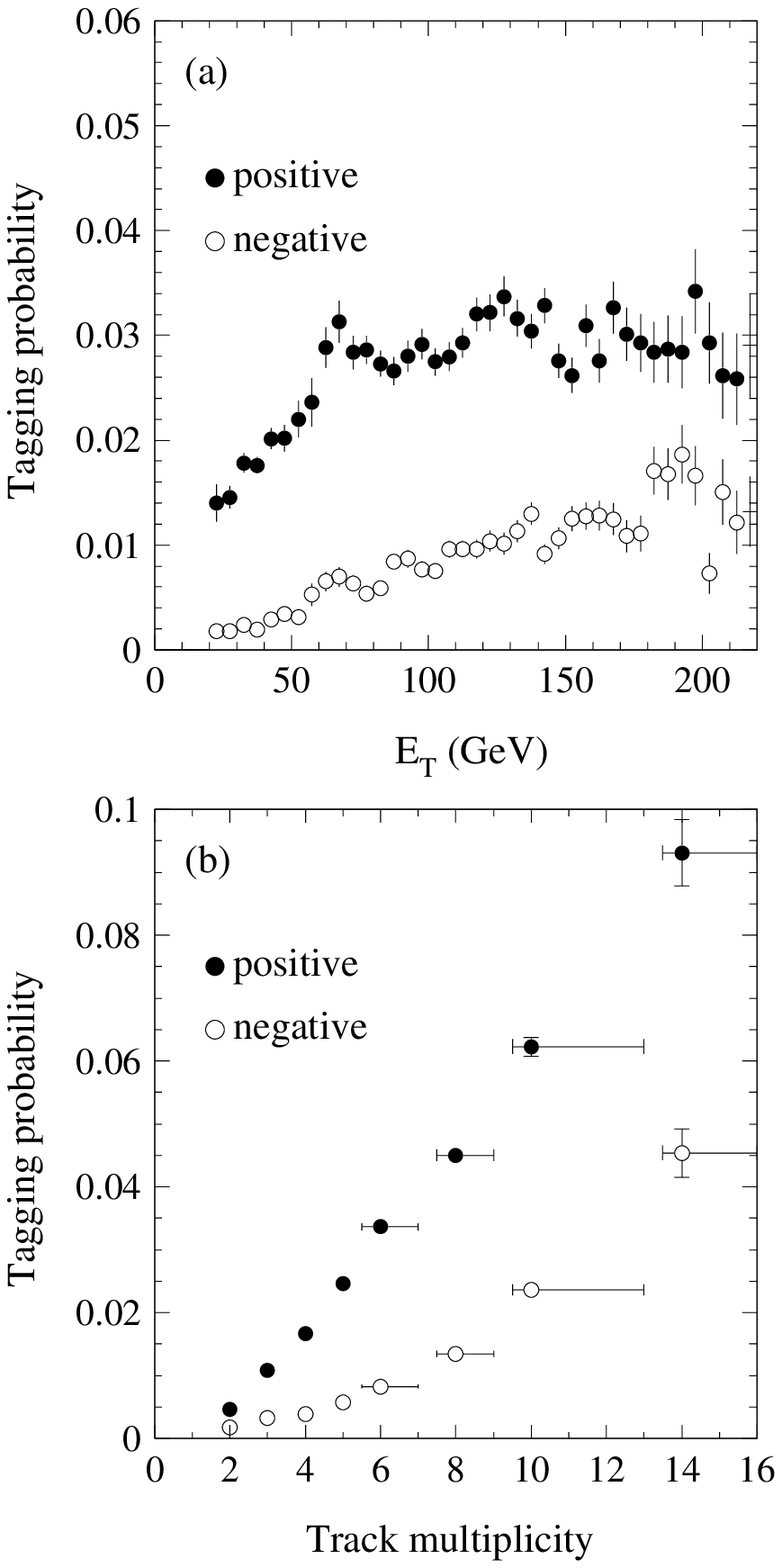}
 \caption[]{The positive and negative SECVTX tagging probability as a function 
            of (a) the jet $E_T$  and  (b) the number of SVX tracks in a jet.}
 \label{fig:fig_10_2}
 \end{center}
 \end{figure}

\begin{figure}
 \begin{center}
 \leavevmode
 \epsffile{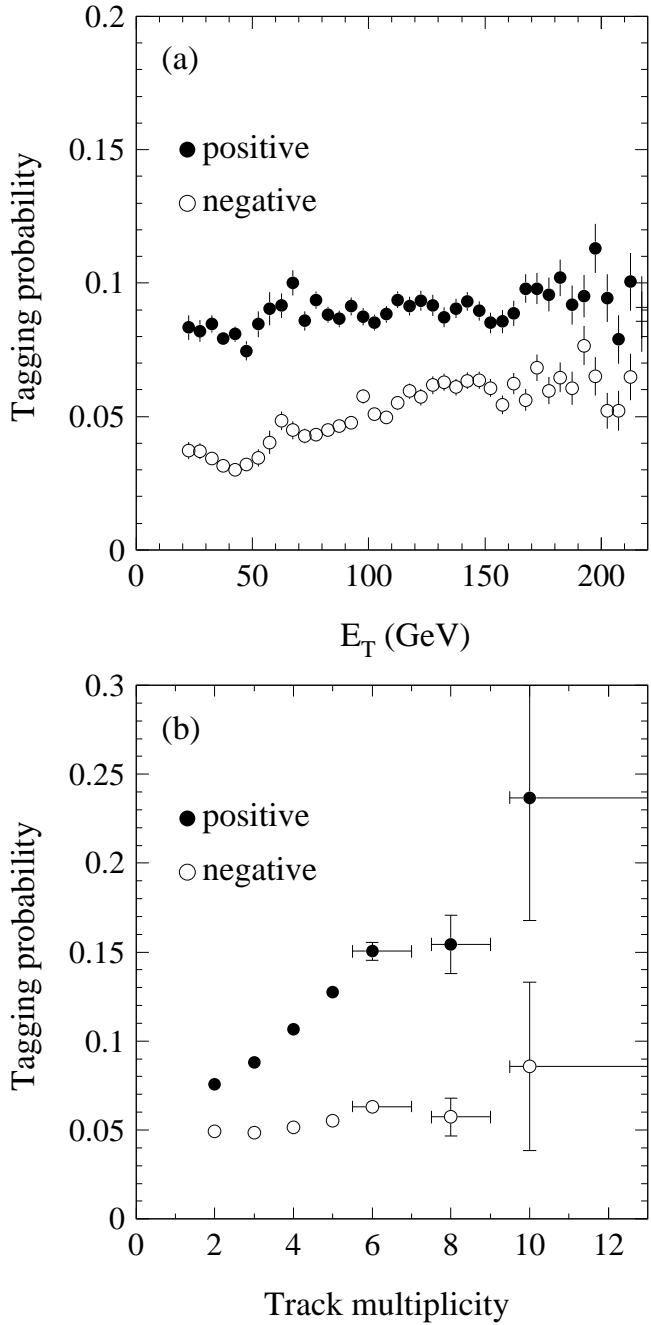}
 \caption[]{The positive and negative JPB tagging probability as a function of
            (a) the jet $E_T$  and (b) the number of SVX tracks in a jet.}
 \label{fig:fig_10_3}
 \end{center}
 \end{figure}

\begin{figure}
 \begin{center}
 \leavevmode
 \epsffile{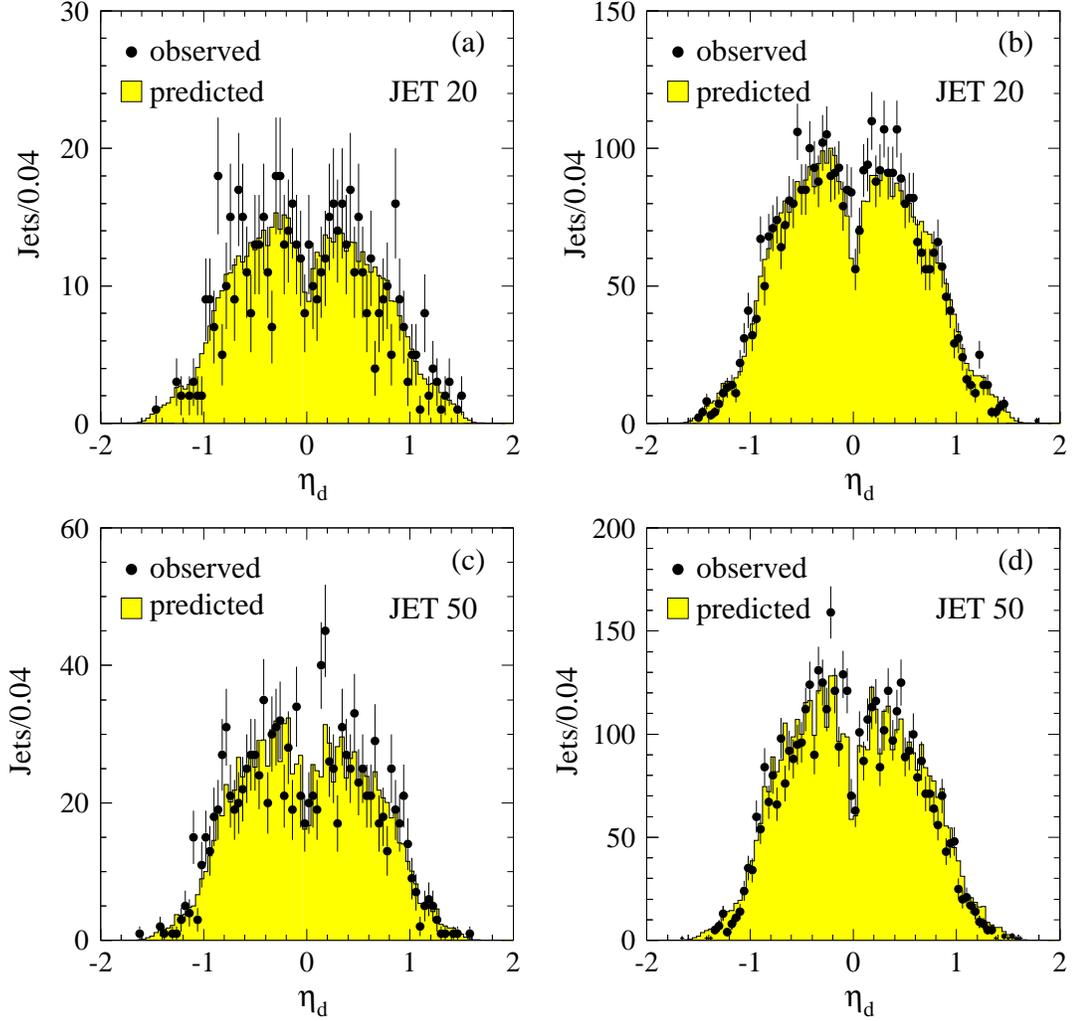}
 \caption[]{Pseudo-rapidity distributions of all jets tagged by SECVTX 
            ($\bullet$) are compared to the prediction derived using only jets
            away from calorimeter cracks (shaded histogram) in JET 20 and 
            JET 50 data. (a) and (c) are negative tags; (b) and (d) are
            positive tags.}
  \label{fig:fig_10_4}
  \end{center}
  \end{figure}

        Transverse energy distributions of the jets used to measure the
        tagging probability
        are shown in Figures~\ref{fig:fig_10_0} and~\ref{fig:fig_10_1}.
        Projections of the tagging probability matrices are shown in
        Figures~\ref{fig:fig_10_2} and~\ref{fig:fig_10_3}.
\newpage
\clearpage
          Figure~\ref{fig:fig_10_4} shows that the tagging probability 
          parametrization derived using jets with well measured energies 
          works well for all jets.
         
      Since the heavy flavor  contribution to negative tags is expected to be 
      small, the number of tags due to heavy flavor in a given $E_T$ bin of 
      the tagging probability matrix is estimated fairly well by the 
      difference, $P-N$, between the numbers of positive (P) and negative (N)
      tags in this bin. In simulated jets with heavy flavor, we measure the 
      ratio,  $R = \frac{N}{P-N}$, as a function of the jet transverse energy.
      We measure this ratio separately for jets which contain only one hadron
      with heavy flavor ($R_1$) and for jets which contain two hadrons with 
      heavy flavor ($R_2$). The following empirical parametrization provides
      a good description of $R$ for jets containing $b$ as well as $c$-hadrons:
 \begin{displaymath}
   R_1(E_T) = \left\{ \begin{array}{ll}
  \mbox{$0.0088 + 0.000158\cdot E_T$} & \mbox{for SECVTX} \\
              \mbox{$0.039 + 0.00117 \cdot E_T$}  & \mbox{for JPB}
                     \end{array}
                     \right.
 \end{displaymath}
 \begin{displaymath}
   R_2(E_T) = \left\{ \begin{array}{ll}
              \mbox{$0.075 + 0.000158\cdot E_T$} & \mbox{for SECVTX} \\
              \mbox{$0.14 + 0.00117 \cdot E_T$}  & \mbox{for JPB}
                    \end{array}
                    \right.
 \end{displaymath}

     With this parameterization we construct the mistag probability matrix 
     by  correcting each bin of the negative tagging probability matrix by 
     the factors:
\begin{eqnarray*}
 {\rm  N - (P-N)} \cdot R_1(E_T) & \mbox{~~~for jets with } E_T \geq 30\; \gev \\
 {\rm N - (P-N)} \cdot R_2(E_T) & \mbox{~~~for jets with } E_T \leq 30 \; \gev 
\end{eqnarray*}
        The fraction of negative tags contributed by heavy flavors is shown
        in Table~\ref{tab:tab_10_1}.
\newpage
\narrowtext
 \begin{table}[p]
 \begin{center}
\def\arraystretch{0.8}
 \caption[]{Fraction of negative tags (\%) due to heavy flavor as a
            function of the $E_T$ of the jet.}
 \begin{tabular}{lcc|lcc}
  Jet $E_T$ (GeV)      & SECVTX & JPB & Jet $E_T$ (GeV)        & SECVTX & JPB \\
 $~0\leq E_T \leq 20$  &        & 10  & $~80\leq E_T \leq 100$ & 6      & 12 \\
 $20\leq E_T \leq 35$  &  12    & 19  & $100\leq E_T \leq 120$ & 6      & 12 \\
 $35\leq E_T \leq 50$  &  10    & 15  & $120\leq E_T \leq 150$ & 6      & 10 \\
 $50\leq E_T \leq 65$  &   7    & 13  & $150\leq E_T \leq 180$ & 5      & 12 \\
 $65\leq E_T \leq 80$  &   8    & 15  & $180 \leq E_T$         & 5      & 12 \\
 \end{tabular}
 \label{tab:tab_10_1}
 \end{center}
 \end{table}
\widetext

        In the generic-jet samples used to derive the mistag matrices,
        approximately 70\% of the events contain additional interactions. 
        The rate of multiple interactions is different in other samples,
        e.g. $W+$ multi-jet events where we require an isolated primary lepton. 
        The negative tagging rate in the generic-jet data depends on the
        number of additional interactions.

\begin{figure}
 \begin{center}
 \leavevmode
 \epsffile{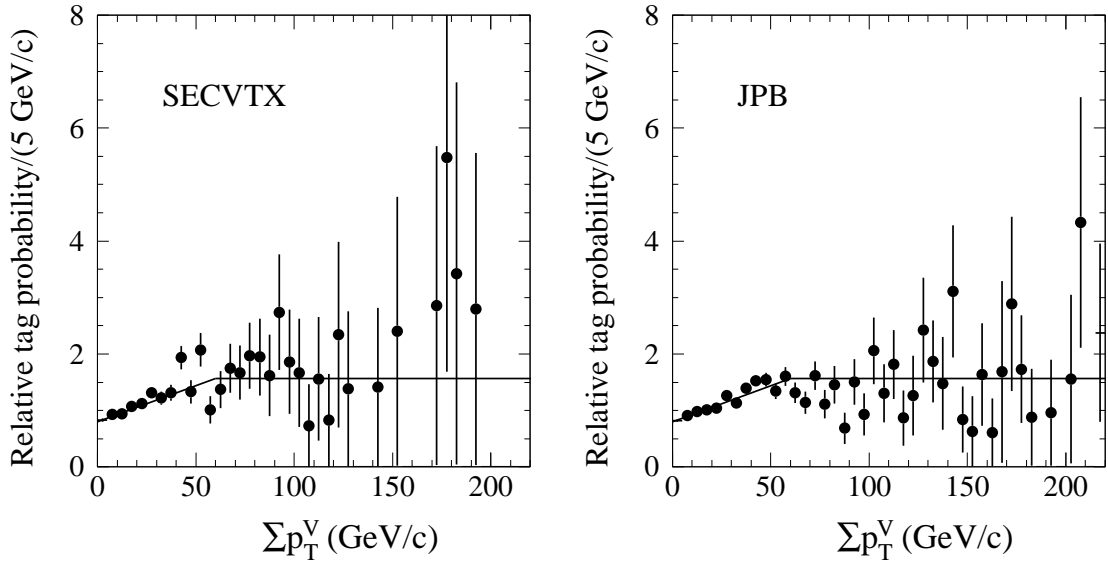}
 \caption[]{Yield of the negative tagging probability as a function of 
            $\sum p_T^V$ for (a) SECVTX  and (b) JPB. The solid line represents
            an empirical parametrization described in the text.}
 \label{fig:fig_10_5}
 \end{center}
 \end{figure}

        Figure~\ref{fig:fig_10_5} shows the relative  negative tagging 
        probability, normalized to the average, 
 as a function of the sum of the transverse 
        momenta of all tracks associated with additional vertices displaced 
        by more than 5 cm from the primary vertex $\sum p_T^V$.
        Accordingly, the mistag rate is parametrized with the additional
        empirical function for both SECVTX and JPB:
\begin{displaymath}
  F(\sum p_T^V) = \left\{ \begin{array}{ll}
  \mbox{$0.8 + 0.0128 \cdot \sum p_T^V$} & \mbox{for $\sum p_T^V < 60$ GeV/c}\\
  \mbox{$1.57$}                          & \mbox{for $\sum p_T^V \geq 60$ GeV/c}
                  \end{array}
                  \right.
\end{displaymath}
%
\subsection{Check of the SECVTX mistag parametrization}
~\label{sec:fake_Test_svx}
   In this section, we test the capability of our model to predict the rate of 
   negative tags in all available generic-jet samples.
 \begin{figure}
 \begin{center}
 \leavevmode
 \epsffile{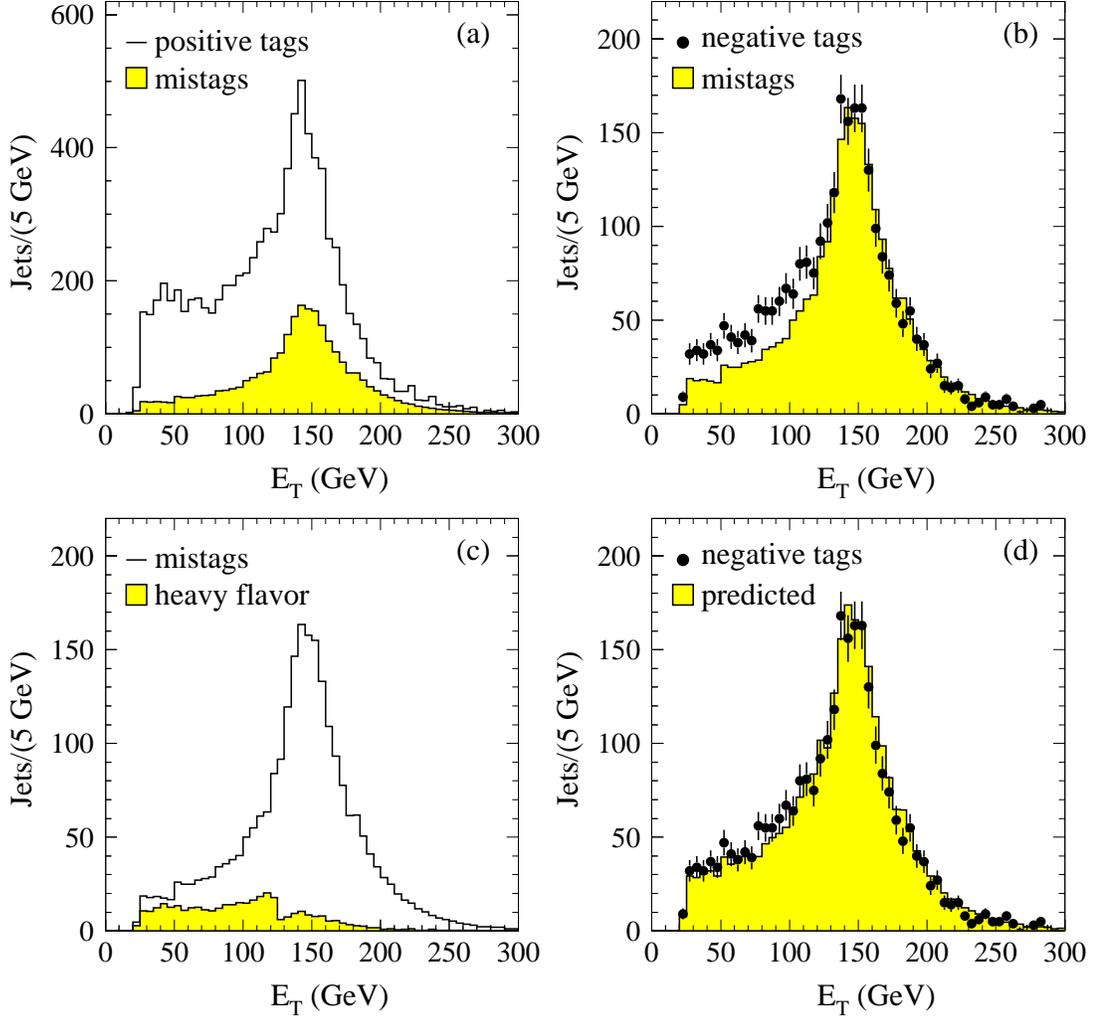}
 \caption[]{$E_T$ distributions of jets with SECVTX tags in the JET 100 sample.
            On (a), observed positive tags (histogram) are compared to the 
            predicted mistags (shaded histograms).
            On (b), observed negative tags ($\bullet$) are compared to the 
            predicted mistags (shaded histogram). On
            (c), predicted mistags (histogram) are compared to the predicted 
            heavy flavor contribution to the negative tags (shaded histograms).
            On (d), observed negative tags ($\bullet$) are compared to the sum 
            of the predicted mistags and heavy flavor contribution to the 
            negative tags (shaded histogram).}
 \label{fig:fig_10_6}
 \end{center}
 \end{figure}
        Figure~\ref{fig:fig_10_6} serves to illustrate the 
        procedure followed to predict the rates of negative tags.
        They are evaluated as the sum 
        of the mistags plus the heavy flavor contribution using the $R_1$
        and $R_2$ parametrizations derived in the previous section.
        This procedure requires the knowledge of the fraction of quark and 
        gluon jets as a function of jet-$E_T$ in each data sample (literally,
        we need to know the fraction of jets containing one or two hadrons
        with heavy flavor). In the JET 20, JET 50, JET 70, JET 100, 
        $\sum E_T$ 175 and $\sum E_T$ 300 samples we make the assumption, 
        corroborated by the corresponding simulations, that all jets below  
        trigger threshold are
        gluon jets and all jets above trigger threshold are quark jets.

        Figure~\ref{fig:fig_10_6}(a) shows the rate of observed positive tags 
        and  predicted mistags as a function of the
        jet $E_T$. Figure~\ref{fig:fig_10_6}(b) compares rates 
        of negative tags to the  predicted mistags. The mistag rate does not 
        include any heavy flavor contribution and is lower than the observed 
        rate of negative tags. Figure~\ref{fig:fig_10_6}(c) compares the rate
        of mistags and the heavy flavor contribution to the negative tags
        obtained by multiplying the difference between positive tags and 
        predicted mistags in Figure~\ref{fig:fig_10_6}(a) by $R_1$ ($R_2$) if 
        the jet-$E_T$  is above (below) the trigger threshold. 
        Figure~\ref{fig:fig_10_6}(d) compares the observed and predicted yield of 
        negative tags. The predicted yield is derived by adding the two 
        distributions shown in Figure~\ref{fig:fig_10_6}(c).

\begin{figure}
 \begin{center}
 \leavevmode
 \epsffile{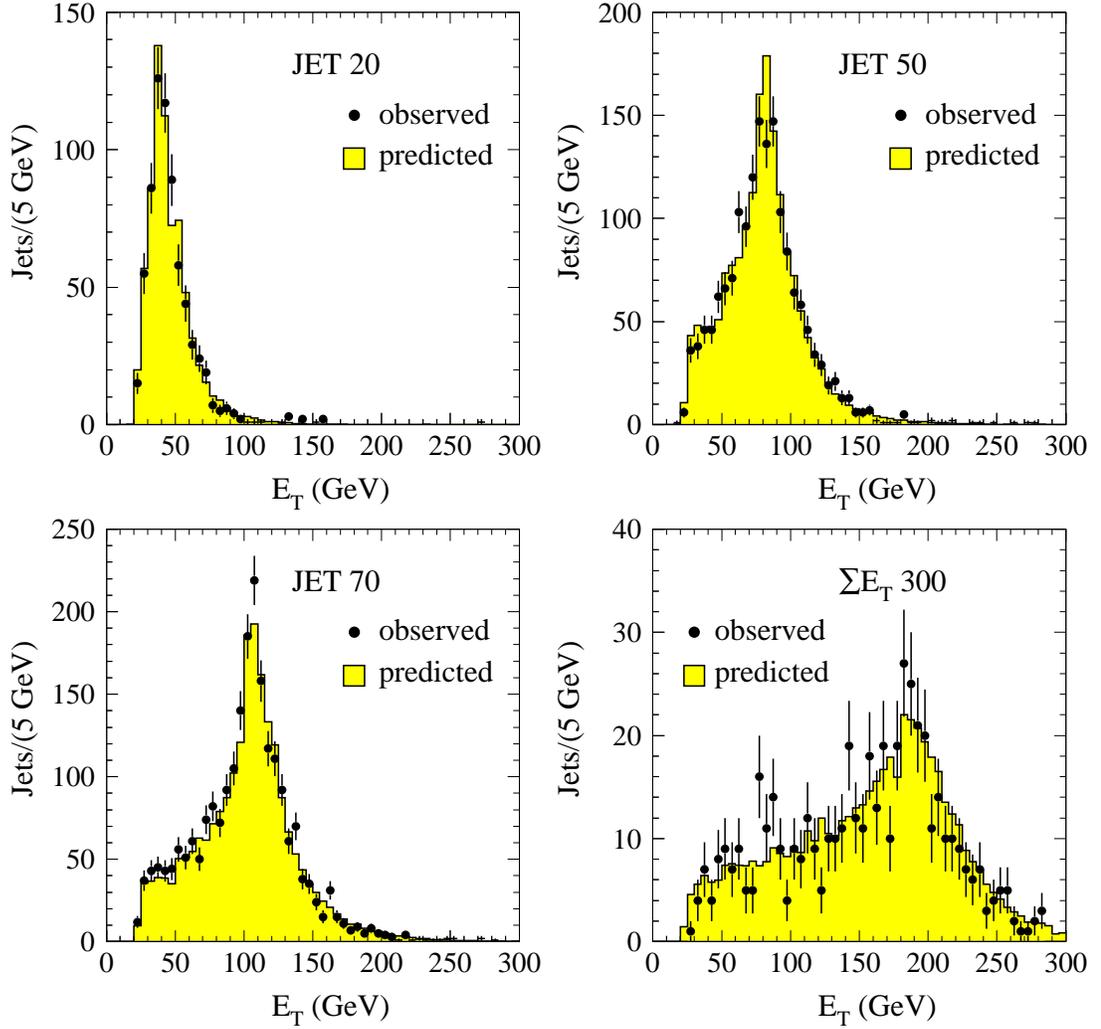}
 \caption[]{Transverse energy distributions of jets with negative SECVTX tags.
            The 4 data  samples were used for the construction of the mistag 
            probability matrix.}
 \label{fig:fig_10_7}
 \end{center}
 \end{figure}

\begin{figure}
 \begin{center}
 \leavevmode
 \epsffile{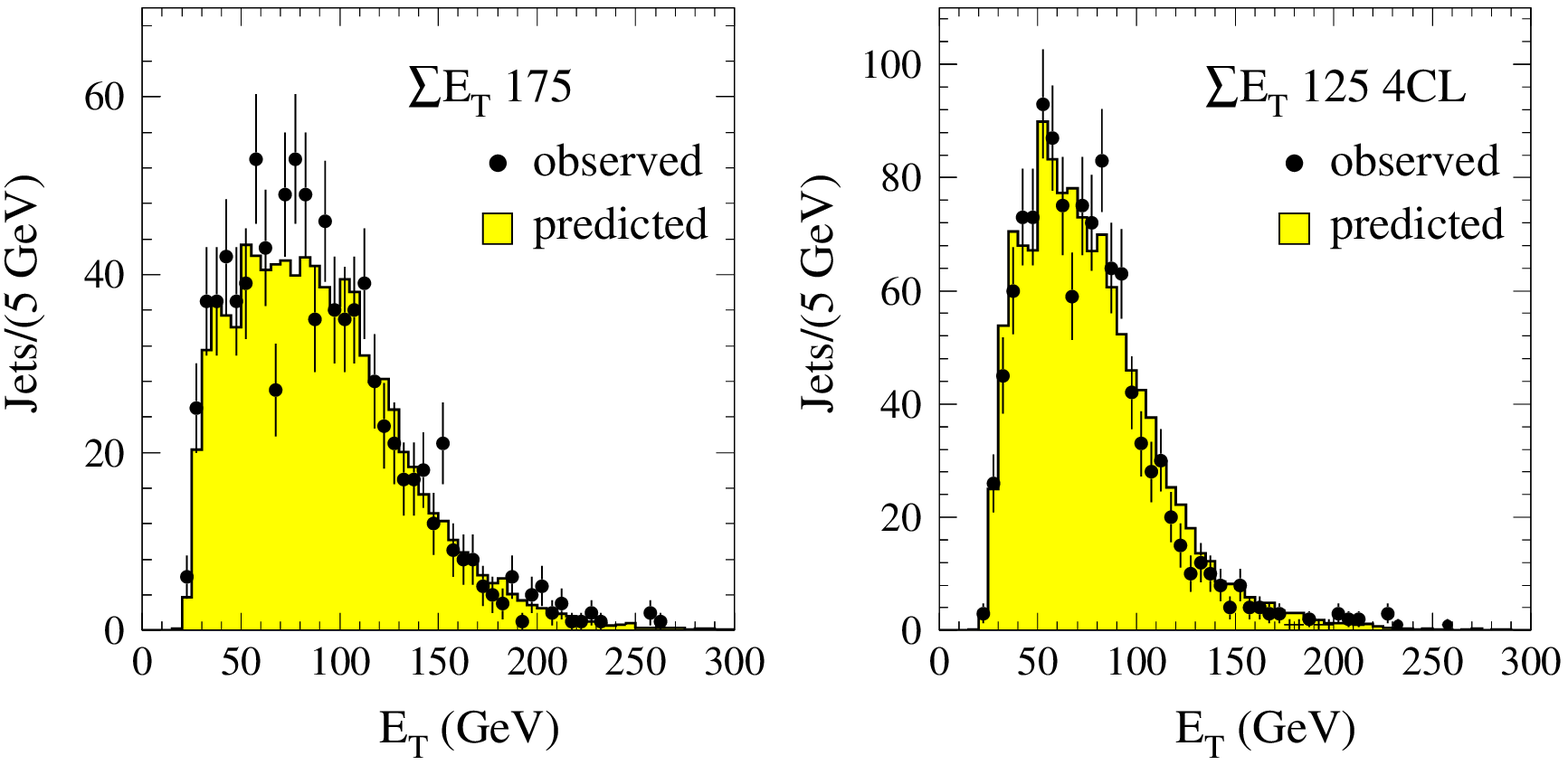}
 \caption[]{Transverse energy distributions of jets with negative SECVTX
            tags in the $\sum E_T$ 175 (a) and $\sum E_T$ 125 4CL (b) 
            samples, which were not used for the construction of the 
            mistag probability matrix.}
 \label{fig:fig_10_8}
 \end{center}
 \end{figure}

\begin{figure}
 \begin{center}
 \leavevmode
 \epsffile{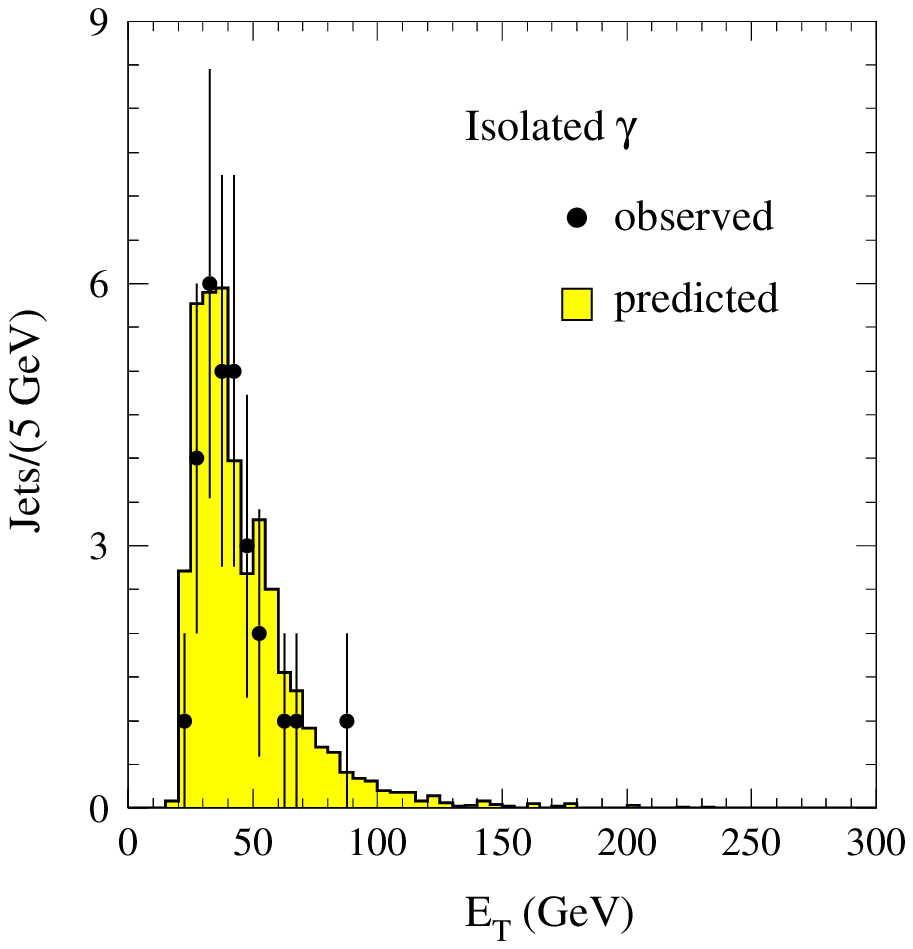}
 \caption[]{Transverse energy distributions of jets with negative SECVTX
            tags in the isolated photon sample.}
 \label{fig:fig_10_9}
 \end{center}
 \end{figure}

        Following the same procedure, comparisons between the corrected jet 
        $E_T$ distributions of observed and predicted negative SECVTX tags
        are shown in Figures~\ref{fig:fig_10_7} to~\ref{fig:fig_10_9}. 
        In the case of the $\sum E_T$ 125 4CL sample, the ratio of quark to 
        gluon jets (1/6, independent of $E_T$) is evaluated using  the
        corresponding Herwig simulation. In the inclusive photon sample,
        we use only the $R_1$ parametrization as the simulation shows that 
        tagged jets are mostly contributed by the $\gamma c$ Compton production.

\begin{figure}
 \begin{center}
 \leavevmode
 \epsffile{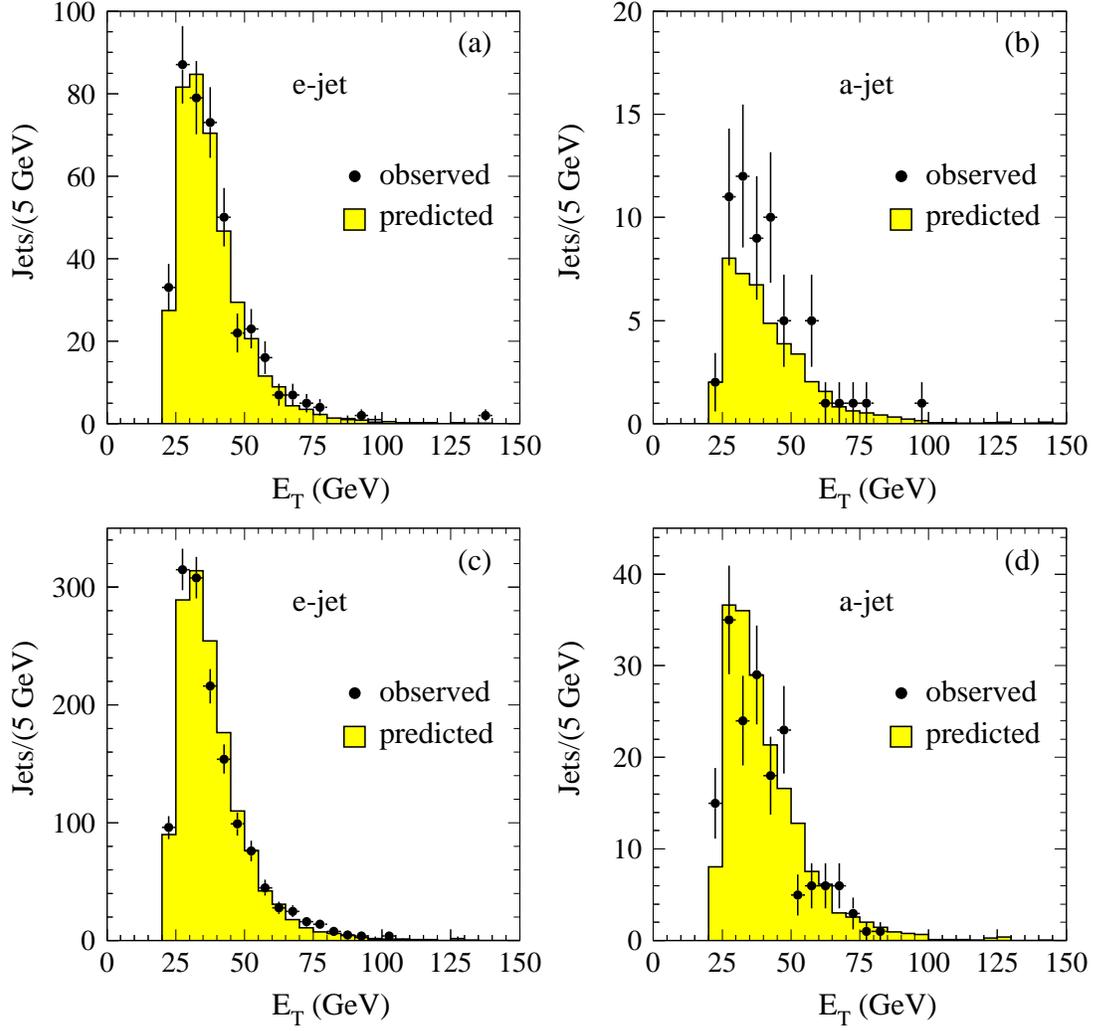}
 \caption[]{Transverse energy distributions in the inclusive electron data. 
            In (a) e-jets with a negative SECVTX tag; (b) a-jets with a 
            negative SECVTX tag in events where the e-jet is tagged by SECVTX;
            (c) e-jets with a negative JPB tag; (d) a-jets with a negative JPB
            tag in events where the e-jet is tagged by SECVTX.}
 \label{fig:fig_10_10}
 \end{center}
 \end{figure}

\begin{figure}
 \begin{center}
 \leavevmode
 \epsffile{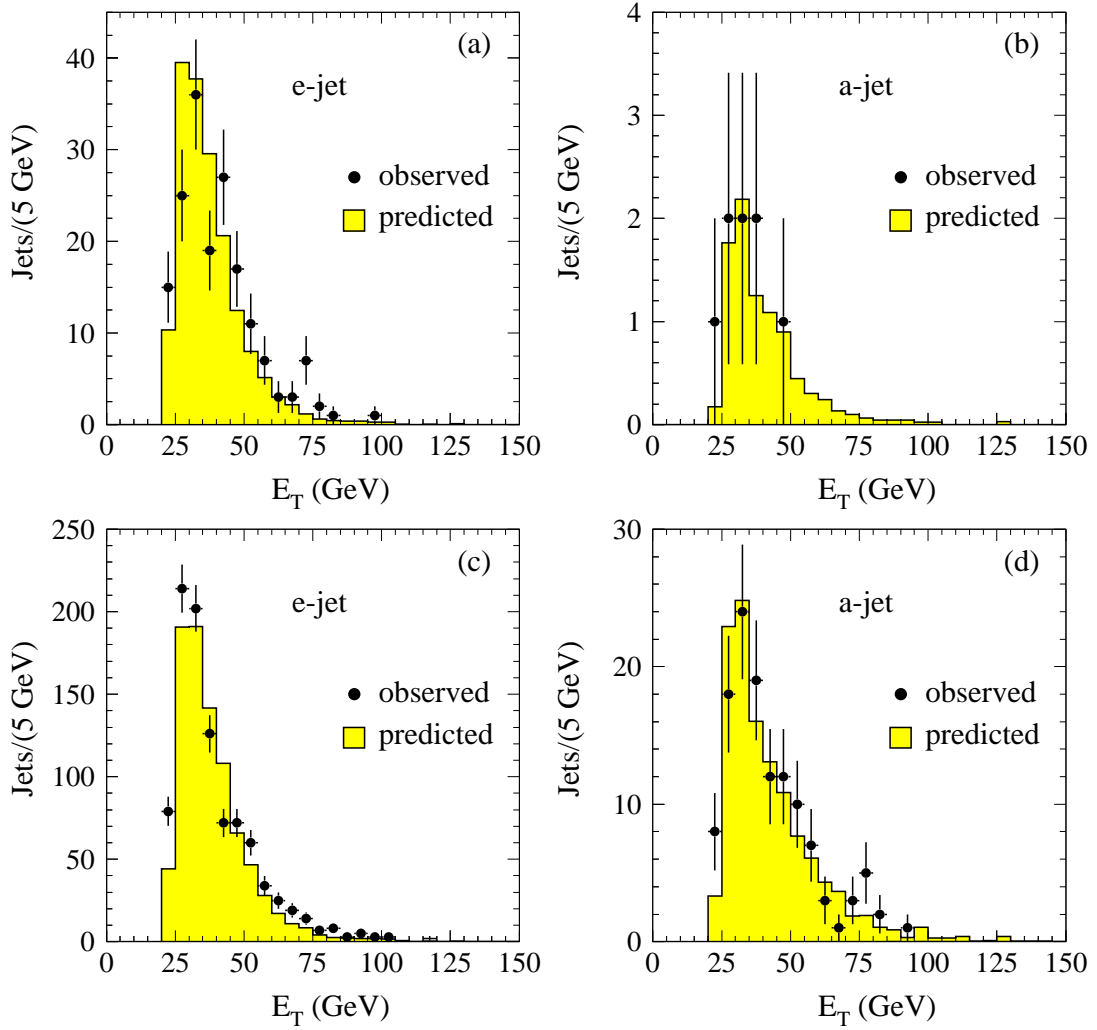}
 \caption[]{Transverse energy distributions in the inclusive electron 
            simulation. In (a) e-jets with negative SECVTX tags; (b) a-jets 
            with negative SECVTX tag; (c) e-jets with negative JPB tags; 
            (d) a-jets with negative JPB tag.}
 \label{fig:fig_10_11}
 \end{center}
 \end{figure}

        The inclusive low-$p_T$ electron sample, used to measure the tagging 
        efficiency scale factor, is also a good sample to test the validity of
        the $R_1$ and $R_2$ parametrizations because it is enriched in heavy 
        flavor content. We compare rates of observed and predicted negative 
        tags both in the data and the corresponding Herwig simulation. The 
        fraction of gluon jets in the simulation is taken from 
        Table~\ref{tab:tab_9_1}. In the data, this fraction is increased by 
        40\% according to the calibration of the Herwig simulation performed 
        in Section~\ref{sec:herwig_norm}. Comparisons between observed and 
        predicted rates of negative tags are shown in 
        Figure~\ref{fig:fig_10_10} for the data and Figure~\ref{fig:fig_10_11}
        for the simulation.

        Table~\ref{tab:tab_10_2} summarizes the rates of observed and 
        predicted negative SECVTX tags in all generic-jet samples.
        Based on the observed agreement a  10\% systematic error
        is assigned to the estimate of the SECVTX mistag probability.
\newpage
\narrowtext
 \begin{table}[p]
 \begin{center}
\def\arraystretch{0.8}
 \caption[]{Numbers  of observed positive and 
            negative SECVTX tags in all generic-jet samples. 
             The method for  predicting the number of negative
            tags, P$_{\rm N}$, is explained in the text.}
 \begin{tabular}{lcccc} 
 \multicolumn{5}{c}{Samples used in the mistag parametrization} \\ \hline 
 {Sample}          & {Pos. tags} & {Neg. tags} & {Mistags}  
                   & P$_{\rm N}$  \\
 {JET 20}           &   4731      &  699        &  652      &  722  \\
 {JET 50}           &   6874      & 1648        & 1426      & 1695  \\
 {JET 70}           &   7758      & 2248        & 1858      & 2192  \\
 {JET 100}          &   8335      & 2723        & 2385      & 2756  \\
 {$\sum E_T$ 300}   &   1507      &  501        &  438      &  521  \\ 
 \hline
 \multicolumn{5}{c}{Independent samples} \\ \hline 
 {$\sum E_T$ 175}    & 3790    &  947        &  675      & 908   \\
 {$\sum E_T$ 125 4CL}& 5637    & 1203        &  897      & 1249  \\
 {Isolated $\gamma$}&  284    &   29        &   35      & 40    \\
 \end{tabular}
 \label{tab:tab_10_2}
 \end{center}
 \end{table}
\widetext

\clearpage
\subsection{Check of the JPB mistag parametrization}
 ~\label{sec:fake_Test_jpb}
\begin{figure}
 \begin{center}
 \leavevmode
 \epsffile{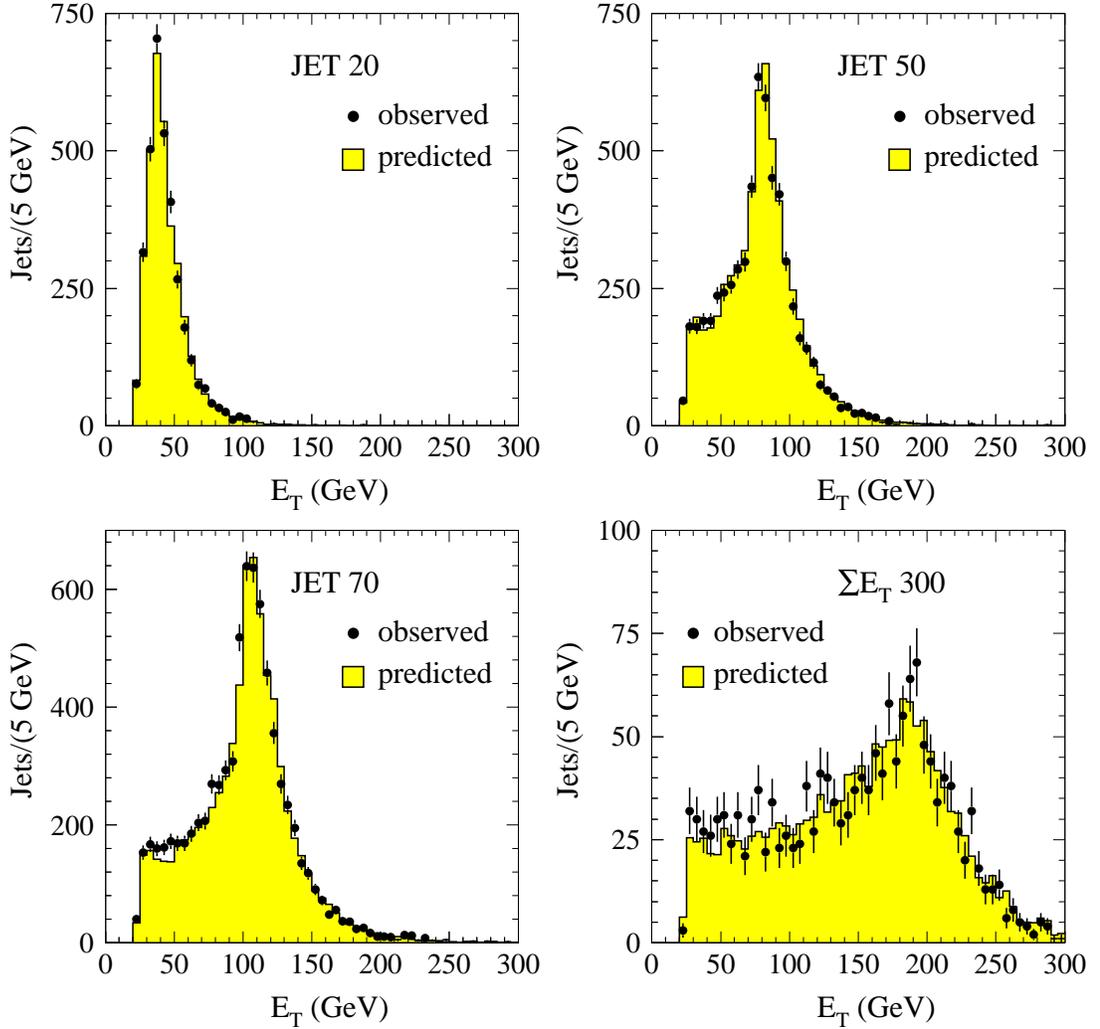}
 \caption[]{Transverse energy distributions of jets with a negative JPB tag.
            The 4 samples were used for the construction of the mistag 
            probability matrix.}
 \label{fig:fig_10_12}
 \end{center}
 \end{figure}

\begin{figure}
 \begin{center}
 \leavevmode
 \epsffile{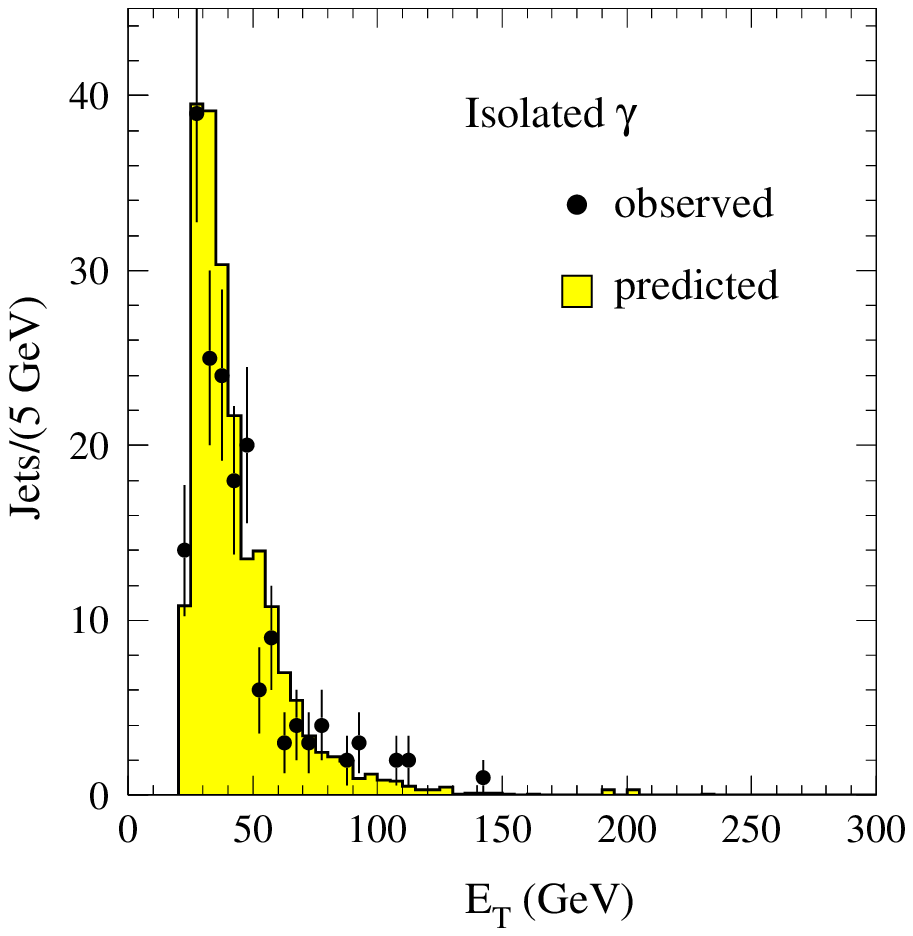}
 \caption[]{Transverse energy distributions of jets with negative JPB tags 
            in the isolated photon sample.}
 \label{fig:fig_10_14}
 \end{center}
 \end{figure}

        We follow the same procedure of the previous section to test the  
        parametrization of the mistag rate of jet-probability.
        Figures~\ref{fig:fig_10_12} trough~\ref{fig:fig_10_14} compare $E_T$ 
        distributions of observed and predicted jets with negative JPB tags 
        for all generic-jet samples. Rates of JPB tags are summarized in 
        Table~\ref{tab:tab_10_3}. As before, by comparing
        the observed and predicted number of negative tags, we assign
        a 10\% systematic error
        to the parametrization of the JPB mistags.
 \begin{figure}
 \begin{center}
 \leavevmode
 \epsffile{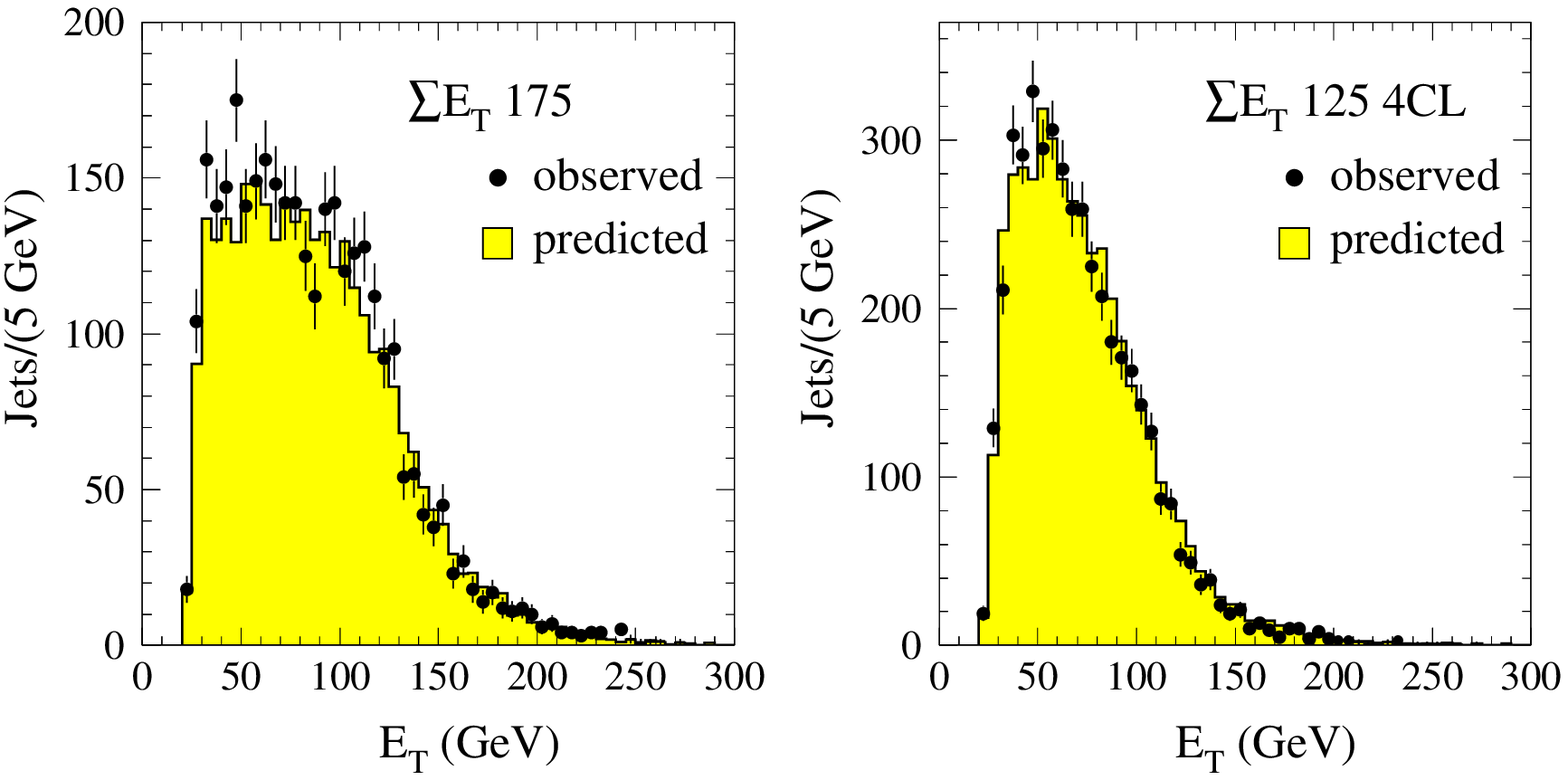}
 \caption[]{Transverse energy distributions of jets with negative JPB tags  
            in the $\sum E_T$ 175 (a) and $\sum E_T$ 125 4CL (b) samples,
            which were not used for the construction of the 
            mistag probability matrix.}
 \label{fig:fig_10_13}
 \end{center}
 \end{figure}

\newpage
\narrowtext
 \begin{table}[p]
 \begin{center}
\def\arraystretch{0.8}
 \caption[]{Numbers  of observed positive and
            negative JPB tags in all generic-jet samples.
             The method for  predicting the number of negative
            tags, P$_{\rm N}$, is explained in the text.}
 \begin{tabular}{lcccc} 
 \multicolumn{5}{c}{Samples used in the parametrization} \\ \hline
 {Sample}          & {Pos. tags} & {Neg. tags} & {Mistags} & P$_{\rm N}$  \\ 
 {JET 20}         &   8418      &  3414       & 2919      & 3421 \\
 {JET 50}         &  12124      &  5970       & 4948      & 6156 \\
 {JET 70}         &  13254      &  7567       & 6020      & 7437 \\
 {JET 100}        &  14528      &  8827       & 7010      & 8721 \\
 {$\sum E_T$ 300} &   2712      &  1581       & 1162      & 1566 \\
 \hline
 \multicolumn{5}{c}{Independent samples} \\ \hline
 {$\sum E_T$ 175} &   6217      &  3235       & 2227      & 3069 \\
 {$\sum E_T$ 125 4CL}& 9283    &  4407       & 3166      & 4481 \\
 {Isolated $\gamma$}&   537      &   179       &  176      &  209 \\
 \end{tabular}
 \label{tab:tab_10_3}
 \end{center}
 \end{table}
\widetext

\clearpage
\section{Calibration of the fraction of $\W+ jet$ events  with heavy flavor} 
         ~\label{sec:herwig_norm}

        $W\bbbar$ and $W\ccbar$ events are produced through the so-called
        gluon  splitting process, where a final state gluon branches into
        a heavy quark pair. In this analysis the fraction of $W +$ jet 
        events containing heavy flavor is estimated using the Herwig generator.
        The uncertainty in the rate 
        of gluons splitting into  heavy quarks  based
        on the parton shower approach is estimated to be approximately 40\% in 
        Ref.~\cite{mlm_wbb} and approximately  25\% in Ref.~\cite{seymour}.
        Because of this large uncertainty  we calibrate the gluon splitting cross 
        section modeled with Herwig using generic-jet data. Heavy flavor in 
        generic-jet data stems from three primary sources: (1) direct 
        production (e.g. $gg \rightarrow b\bar{b}$); (2) flavor excitation 
        (e.g. $gb \rightarrow g b$); and (3) gluon splitting. 
        The calibration of the simulation package is performed by tuning 
        the various cross sections calculated by Herwig to reproduce the tagging 
        rate observed in the  JET 50 and JET 100 data. In these samples,
        the gluon splitting contribution is comparable to the other
        production mechanisms. In the JET 20 simulation, the gluon splitting 
        contribution is negligible; we compare observed and predicted rates of 
        tags in this sample using the tuned simulation as a check that we 
        disentangled correctly the different heavy flavor production mechanisms.

        In each generic-jet sample we count the number of SECVTX tags in taggable 
        jets. Mistags are evaluated using the mistag probability 
        evaluated in Section~\ref{sec:mistags}.

        The simulated samples (corresponding to the JET 20, JET 50 and 
        JET 100 data) are generated using the option 1500 of Herwig
        and requiring  hard scattering partons with $|\eta| \leq$ 4.5 and 
        $p_T^{\rm min} \geq$ 10, 40 and 80 GeV/c respectively~\cite{herw_setup}.
        We use the MRS(G) set of structure functions~\cite{mrsg}. 
        Generated events are simulated with the standard package
        discussed in Section~\ref{sec:Simul}. As in the 
        data, we select events containing at least one jet above the trigger 
        threshold.

        In the simulation a jet is classified as a $b$ or a $c$-jet if it 
        contains a $b$ or a $c$-hadron in a cone of radius 0.4 around its 
        axis. Hadrons with heavy flavor resulting from the fragmentation 
        of one of the hard scattering partons are indicative of 
        direct production or flavor excitation (if one of the incoming 
        partons of the hard scattering has heavy flavor we attribute the 
        process to flavor excitation; in this case a second hadron of the
        same flavor is produced by the backward-evolution of the  
        structure functions). All pairs of hadrons with heavy flavor of 
        the same type which do not come from the hadronization of the 
        hard scattering partons are attributed to gluon splitting.
        Table~\ref{tab:tab_11_1} lists the  rate of jets containing heavy 
        flavor per event in the simulated JET 50 and JET 100 samples.

\widetext
 \begin{table}[p]
 \begin{center}
\def\arraystretch{0.8}
 \caption[]{Average  numbers of jets containing heavy flavor per event in the 
            JET 50 and JET 100 samples generated with Herwig, 
            split by flavor type and production mechanism.}
 \begin{tabular}{lccccc}
 {~}     & \multicolumn{2}{c}{\small {direct production + flavor excitation}}
         & \multicolumn{2}{c}{\small {gluon splitting}} & \\ 
{Sample} & ~~~~~$b$-jets~~~~ & ~~~~$c$-jets~~~~ & $g \rightarrow b\bar{b}$
         & $g \rightarrow c\bar{c}$ & {Total}\\
 \hline
 JET 50  & ~~~~~2.14 $\times~10^{-2}$~~~~  & ~~~~3.04 $\times~10^{-2}$~~~ 
          &      1.67 $\times~10^{-2}$      &     3.79 $\times~10^{-2}$ & 10.64 $\times~10^{-2}$ \\
 JET 100 & ~~~~~2.15 $\times~10^{-2}$~~~~  & ~~~~2.89 $\times~10^{-2}$~~~ 
          &      2.58 $\times~10^{-2}$      &     5.73 $\times~10^{-2}$ & 13.35 $\times~10^{-2}$ \\
\end{tabular}
\label{tab:tab_11_1}
\end{center}
\end{table}
\narrowtext

\begin{figure}
 \begin{center}
 \leavevmode
 \epsffile{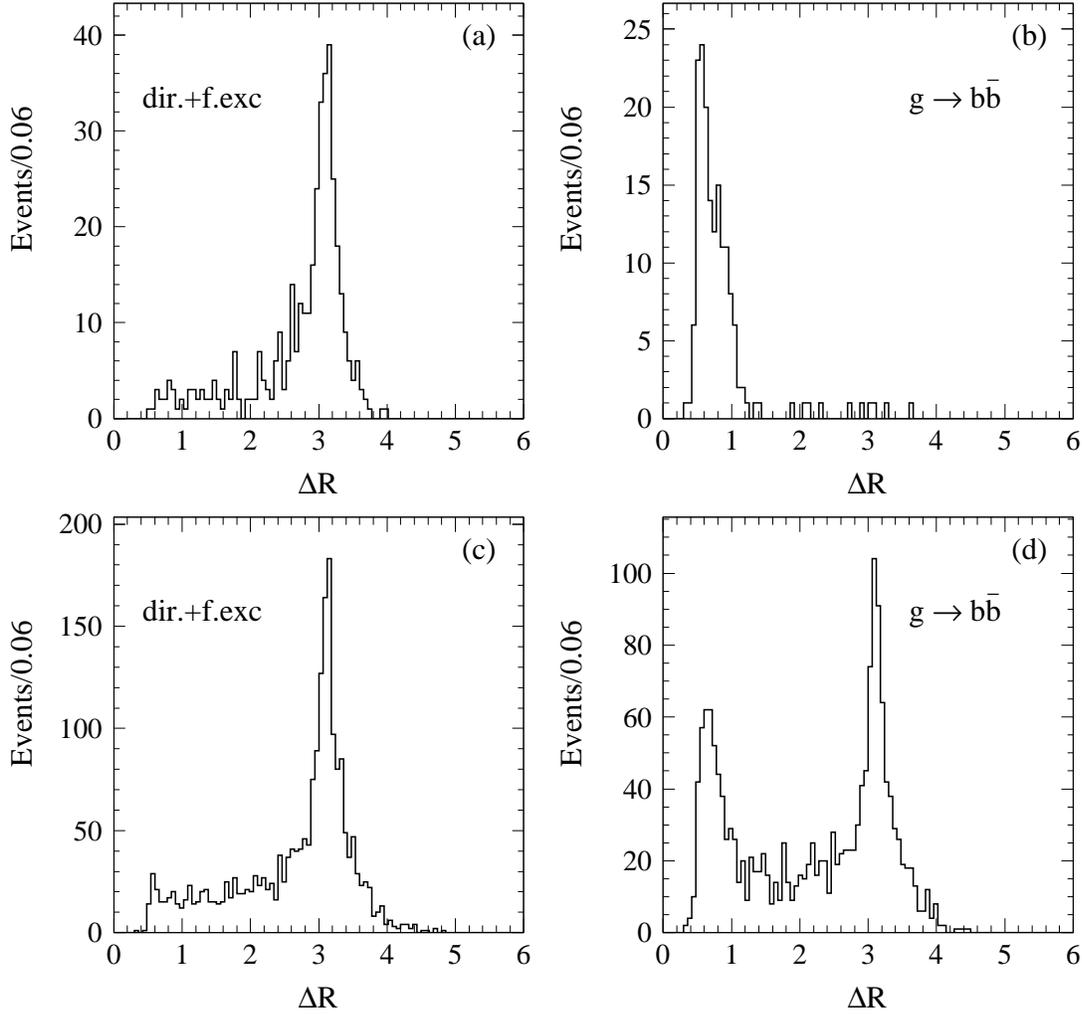}
 \caption[]{Distributions of the distance $\Delta R$ between two $b$-jets 
            tagged by SECVTX in JET 50 simulated events contributed by
            (a) direct production and flavor excitation or (b) gluon splitting. 
            (c) and (d) are the distributions of the distance between a $b$-jet
            tagged by SECVTX and the closest jet in the event with 
            $E_T\geq$ 10 GeV.}
 \label{fig:fig_11_0}
 \end{center}
 \end{figure}

        In the data, in order to distinguish  gluon splitting from the rest of 
        the heavy flavor production  we use intuitive kinematical
        differences. Jets from heavy flavor direct production are expected to 
        be produced back-to-back and are more likely to produce double tags.
        In events produced by heavy flavor excitation,  jets 
        produced by the backward-evolution of the structure functions tend 
        to be at large pseudo-rapidities and out of the SVX acceptance.
        On the other hand, gluon splitting produces pairs of jets with heavy 
        flavor at small separation
        $\Delta R= \sqrt{ (\Delta \phi)^2 +(\Delta \eta)^2}$. Most of the 
        time the two hadrons with heavy flavor produced by gluon splitting
        reside in the same jet.
        Figure~\ref{fig:fig_11_0} shows  distributions of the distance 
        between two $b$-jets for the different production mechanisms
        in the simulated sample.
        In addition, the relative gluon splitting
        contribution increases with the jet multiplicity.

        This motivates us to compare data and simulation in the following
        classes of SECVTX tags:
\begin{itemize}
  \item (1): number of tagged jets per event with at least one taggable jet.
  \item (2): number of tagged jets per event with at least one taggable jet
             and with three or more jets with $E_T \geq$ 15 GeV
             and $|\eta| \leq$ 2.
  \item (3): twice the number of events with two tagged jets per event with 
             two or more taggable jets.
\end{itemize}
        We also compare the data to the simulation for: 
\begin{itemize}
  \item (4): the fraction of (1) in which the tagged jet
              has a companion jet with
             $E_T\geq10$ GeV in a cone of radius 1.2 around its axis.
  \item (5): the fraction of events with double tags where
             the two tagged jets are at a distance $\Delta R\leq$1.2.
 \end{itemize}
        Table~\ref{tab:tab_11_2} lists the yields of tags in the
        data and in the simulation.
\newpage

\widetext
 \begin{table}
 \begin{center}
\def\arraystretch{0.8}
 \caption[]{Yields ($\times 10^{-3}$) of SECVTX tags in generic-jet
 data and in simulated samples
            generated with Herwig. Rows $1$, $2$ and $3$ represent
    the average   number of tags per event; rows $4$ and $5$ represent
        the    fraction of $1$ and $3$, respectively.
         Rates of simulated tags are not yet corrected for the 
             tagging efficiency scale factor measured
            in Section~\protect\ref{sec:simu_mod}.}
{\small
 \begin{tabular}{@{}l@{\hspace{1ex}}c@{\hspace{1em}}c@{\hspace{0.9em}}c
      @{\hspace{1em}}c@{\hspace{0.9em}}c@{\hspace{1em}}c@{\hspace{0.9em}}c@{}}
 \multicolumn{8}{c}{\small JET 50} \\ 
{~} & {~} & \multicolumn{2}{c}{\small direct production} 
          & \multicolumn{2}{c}{\small flavor excitation}
          & \multicolumn{2}{c}{\small gluon splitting}  \\
 {\small Class} & {\small Data} & {\small $b$-jets} & {\small $c$-jets} 
& {\small $b$-jets} & {\small $c$-jets} & {\small $g \rightarrow \bbbar$} & 
{\small $g \rightarrow \ccbar$}\\
\hline
 ~~~1   & 34.20$\pm$1.05 & 2.90$\pm$0.11  & 0.72$\pm$0.02  
                              & 6.53$\pm$0.17  & 2.63$\pm$0.11
                              & 7.37$\pm$0.18  & 4.17$\pm$0.14 \\
 ~~~2   & 43.00$\pm$1.37 & 2.31$\pm$0.16  & 0.53$\pm$0.08
                              & 6.36$\pm$0.26  & 2.26$\pm$0.16 
                              & 9.71$\pm$0.33  & 5.32$\pm$0.24 \\ 
 ~~~3   & ~7.50$\pm$0.65 & 2.00$\pm$0.18  & 0.16$\pm$0.04
                              & 0.94$\pm$0.13  & 0.07$\pm$0.03  
                              & 0.65$\pm$0.10  & 0.09$\pm$0.02 \\
 ~~~4   & ~5.60$\pm$0.38 & 0.23$\pm$0.03  & 0.04$\pm$0.01
                              & 0.71$\pm$0.06  & 0.25$\pm$0.03  
                              & 2.17$\pm$0.10  & 0.86$\pm$0.06 \\ 
 ~~~5   & ~0.58$\pm$0.08 & 0.00$\pm$0.08  & 0.00
                              & 0.08$\pm$0.03  & 0.00
                              & 0.26$\pm$0.05  & 0.00            \\
 \hline
\hline
 \multicolumn{8}{c}{\small JET 100}\\ 
{~} & {~} & \multicolumn{2}{c}{\small direct production} 
          & \multicolumn{2}{c}{\small flavor excitation}
          & \multicolumn{2}{c}{\small gluon splitting}  \\
 {\small Class} & {\small Data} & {\small $b$-jets} & {\small $c$-jets} 
& {\small $b$-jets} & {\small $c$-jets} & {\small $g \rightarrow \bbbar$} & 
{\small $g \rightarrow \ccbar$}\\
\hline

 ~~~1  & 42.05$\pm$1.84 & 4.31$\pm$0.22 & 1.23$\pm$0.12
                              & 5.57$\pm$0.25 & 2.24$\pm$0.16 
                              &11.85$\pm$0.37 & 6.88$\pm$0.28 \\ 
 ~~~2  & 51.50$\pm$2.04 & 3.51$\pm$0.27 & 0.86$\pm$0.13 
                              & 5.71$\pm$0.35 & 1.97$\pm$0.20 
                              &15.06$\pm$0.56 & 8.40$\pm$0.42 \\ 
 ~~~3  & 15.50$\pm$0.92 & 2.68$\pm$0.29 & 0.26$\pm$0.09
                              & 1.08$\pm$0.18 & 0.05$\pm$0.03 
                              & 1.42$\pm$0.21 & 0.10$\pm$0.05 \\
 ~~~4  & ~6.36$\pm$0.41 & 0.64$\pm$0.09 & 0.09$\pm$0.03
                              & 0.97$\pm$0.10 & 0.31$\pm$0.06 
                              & 5.03$\pm$0.24 & 2.10$\pm$0.15 \\ 
 ~~~5  & ~1.10$\pm$0.11 & 0.00$\pm$0.03 & 0.00
                              & 0.06$\pm$0.03 & 0.00             
                              & 0.54$\pm$0.09 & 0.00          \\ 
 \end{tabular}
 }
 \label{tab:tab_11_2}          
 \end{center}
 \end{table}
\narrowtext

\clearpage
        In the simulation,
        one notes that after tagging with SECVTX the contribution of 
        $c$-jets is reduced by more than a factor of four and becomes 
        negligible in events with double tags. However, the ratio
        of double to single SECVTX tags does not discriminate between 
        $b\bar{b}$ and $c\bar{c}$ production for this ratio is similarly 
        small for $b\bar{b}$ production through flavor excitation and 
        gluon splitting.

        We discriminate the flavor type with the additional comparison of
        rates of JPB tags (JPB has about the same tagging efficiency of 
        SECVTX for $b$-jets and is more than twice as efficient for 
        tagging $c$-jets). Since we use 
        JPB tags only to disentangle between $b$ and $c$-production, we 
        compare data and simulations in only two classes of JPB tags:
 \begin{itemize}
   \item (6): number of tagged jets per event with at least one taggable jet.
   \item (7): twice the number of events with two tagged jets per event with 
              two or more taggable jets.
 \end{itemize}
        Table~\ref{tab:tab_11_3} lists the yields of JPB tags in the data 
        and in the simulation.
\widetext
 \begin{table}
 \begin{center}
\def\arraystretch{0.8}
 \caption[]{Fractions ($\times 10^{-3}$) of JPB tags
per event  in generic-jet
            data and in simulated samples generated with Herwig. 
            Fractions of tags are not yet corrected for the tagging efficiency 
            scale factor measured in Section~\protect\ref{sec:simu_mod}.}
 {\small
 \begin{tabular}{@{}l@{\hspace{0.9ex}}c@{\hspace{0.75em}}c@{\hspace{0.65em}}c
      @{\hspace{0.75em}}c@{\hspace{0.65em}}c@{\hspace{0.85em}}c@{\hspace{0.7em}}c@{}}
 \multicolumn{8}{c}{\small JET 50}\\
 {~} & {~} & \multicolumn{2}{c}{\small direct production} 
           & \multicolumn{2}{c}{\small flavor excitation}
           & \multicolumn{2}{c}{\small gluon splitting}  \\
  {\small Class} & {\small Data} & {\small $b$-jets} & {\small $c$-jets} 
& {\small $b$-jets} & {\small $c$-jets} & {\small $g \rightarrow \bbbar$} & 
{\small $g \rightarrow \ccbar$}\\
\hline
 ~~~6   & 45.20$\pm$3.19   & 3.84$\pm$0.13  & 1.87$\pm$0.09 
                                 & 7.97$\pm$0.19  & 6.28$\pm$0.17  
                                 & 9.11$\pm$0.21  & 8.67$\pm$0.20 \\ 
 ~~~7   & ~4.75$\pm$0.28   & 1.62$\pm$0.12  & 0.26$\pm$0.05 
                                 & 0.81$\pm$0.09  & 0.23$\pm$0.05  
                                 & 0.89$\pm$0.09  & 0.52$\pm$0.06 \\ 
 \hline
\hline
 \multicolumn{8}{c}{\small JET 100} \\
{~} & {~} & \multicolumn{2}{c}{\small direct production} 
          & \multicolumn{2}{c}{\small flavor excitation}
          & \multicolumn{2}{c}{\small gluon splitting}  \\
 {\small Class} & {\small Data} & {\small $b$-jets} & {\small $c$-jets} 
& {\small $b$-jets} & {\small $c$-jets} & {\small $g \rightarrow \bbbar$} & 
{\small $g \rightarrow \ccbar$}\\
\hline
 ~~~6  & 53.07$\pm$5.09 & ~5.72$\pm$0.26 & ~2.66$\pm$0.18
                              & ~6.86$\pm$0.29 & ~5.69$\pm$0.26 
                              & 14.22$\pm$0.42 & 13.13$\pm$0.40 \\ 
 ~~~7  & ~5.50$\pm$0.34 & ~2.11$\pm$0.19 & ~0.39$\pm$0.08 
                              & ~0.78$\pm$0.11 & ~0.25$\pm$0.06 
                              & ~1.69$\pm$0.17 & ~1.06$\pm$0.13 \\ 
 \end{tabular}
 }
 \label{tab:tab_11_3}
 \end{center}
 \end{table}
\narrowtext

        We proceed to fit the data with the simulation in order to extract
        the rate of $g \rightarrow b\bar{b}$ and $g \rightarrow c\bar{c}$.      
        When fitting the simulation to the data, the yield of simulated
        SECVTX and JPB tags is corrected for the tagging efficiency scale 
        factors measured in Section~\ref{sec:simu_mod}. The 10\% uncertainty 
        in the scale factor determination is included in the error of the
        simulated rates of tags.
        In the fit, we also compare five distributions in each generic-jet sample
        and in the corresponding simulation:
 \begin{enumerate}
    \item the yield of the fraction of SECVTX tags per taggable 
          jet as a function of the jet-$E_T$.
    \item The distributions of the distance $\Delta R$ between a 
          jet tagged by SECVTX and a companion jet as defined above.
    \item The distributions of the distance $\Delta R$ between a 
          jet tagged by JPB and a companion jet as defined above.
    \item The distributions of the distance $\Delta R$ between two
          jets tagged by SECVTX.
    \item The distributions of the distance $\Delta R$ between two 
          jets tagged by JPB.
 \end{enumerate}
        In the comparison, the area of each distribution is normalized to unity.
        For each distribution we compute a reduced $\chi^{2}$
\[ \chi_D^2 = \frac{1}{N} \sum_{i=1}^{N} \frac { (d(i) - sd(i))^{2} }
              { ed(i)^{2} + esd(i)^{2}} \]
        where $N$ is the number of bins in each distribution,
        $d(i)$ and $sd(i)$ are the contents of the bin $i$
        of the distribution in the data and  the simulation respectively,
        with $ed(i)$ and $esd(i)$  their errors.
        The simulated jet-$E_T$ distributions have a systematic
        uncertainty due to the trigger simulation which is cumbersome to 
        account for in the fit. 
        Simulated distribution of distances between tagged
        jets have unaccounted systematic uncertainties due to how well
        the parton shower generator models gluon splitting at distances
        $\Delta R \geq$  1.2. 
        We use the reduced $\chi_D$  to diminish the importance of these
        comparisons with respect to the classes of absolute tagging rates.
        The data are fitted to the simulation  using a minimum $\chi^{2}$ method.
        We minimize the function
\[ \chi^{2}= \sum_{JET 50}^{JET 100} 
             ( \sum_{j=1}^{7} \frac {(F(j) - SF(j))^{2}}
             {EF(j)^{2}+ESF(j)^{2}} +\sum_{k=1}^{5} \chi_{D}^{k} ) \]
        where the index $k$ runs over the 5 kinematic distributions described
        in the previous paragraph, $F(j)$ are the yields of tags observed in 
        the data for the classes listed in Tables~\ref{tab:tab_11_2} and 
        ~\ref{tab:tab_11_3}, and
 \[ SF(j)= \sum_{n=1}^{6} P(n) \cdot C_{H}(j,n)*SF^{\alpha} \]
        is the corresponding yield of simulated tags. The contributions 
        $C_{H}(j,n)$  of different flavor types and 
        production mechanisms, as listed in Tables~\ref{tab:tab_11_2} and
        ~\ref{tab:tab_11_3}, are weighted with the  fit  parameters $P(n)$. 
        $SF$ is the tagging efficiency scale factor and $\alpha$=0 for $c$-jet, 
        1 for events with one tagged $b$-jet and 2 for events
        with two tagged $b$-jets.

        In the fit, the $b$-to-$c$ ratio for direct production is constrained
        to the default value with a 14\% Gaussian error. Option 1500 of Herwig
        evaluates the direct production cross section of massless quarks.
        The 14\% uncertainty accounts for having neglected the quark masses
        (estimated using option 1700 of Herwig) and for the 
        uncertainty in the fragmentation process
        (estimated using the Pythia generator).

        The $b$-to-$c$ ratio for flavor excitation is also constrained to 
        the default value with a 28\% Gaussian error. This uncertainty accounts
        for the largest variation of this ratio observed using a wide range
        of structure functions in the PDF library~\cite{pdflib}.

        The ratio of the $g \rightarrow b\bar{b}$ to $g \rightarrow c\bar{c}$
        is also constrained to the default value with a 28\% Gaussian error.
        The uncertainty accounts for a $\pm$0.5 GeV change of the $b$ and 
        $c$-quark masses around the default value.

        The fit has 21 degrees of freedom and yields a $\chi^2$ of 22. 
        The fit results are shown in Table~\ref{tab:tab_11_4}. 
        The weights of the gluon splitting cross sections  
        will be used to rescale the fraction of $W+$ jet events with heavy
        flavor predicted by Herwig. 

\begin{figure}
 \begin{center}
 \leavevmode
 \epsffile{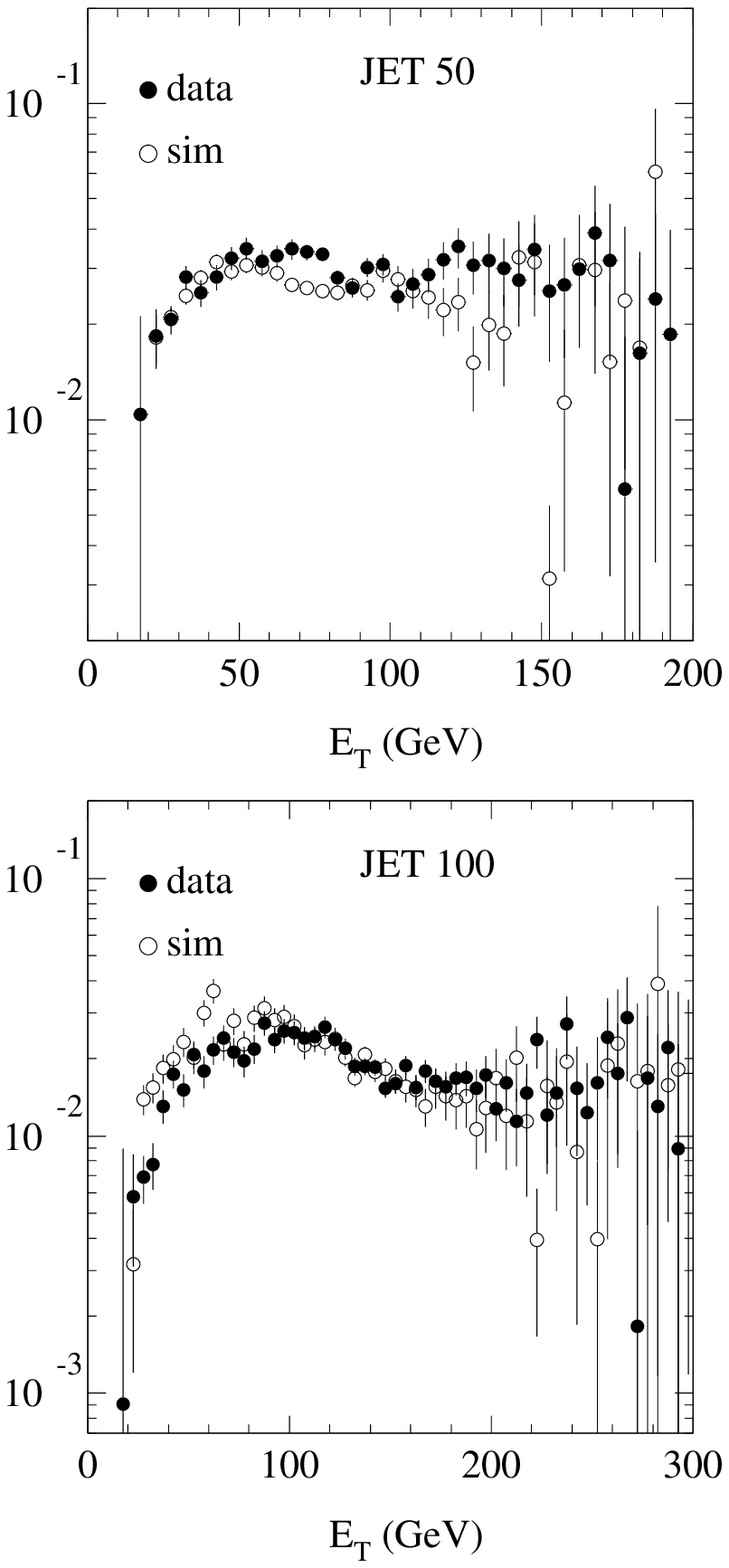}
 \caption[]{Fractions of taggable jets with a SECVTX tag as a function 
            of jet $E_T$ in the data and in the fitted simulation.
            The distributions of the data and the fitted simulation
            are normalized to unit area.}
 \label{fig:fig_11_1}
 \end{center}
 \end{figure}

\begin{figure}
 \begin{center}
 \leavevmode
 \epsffile{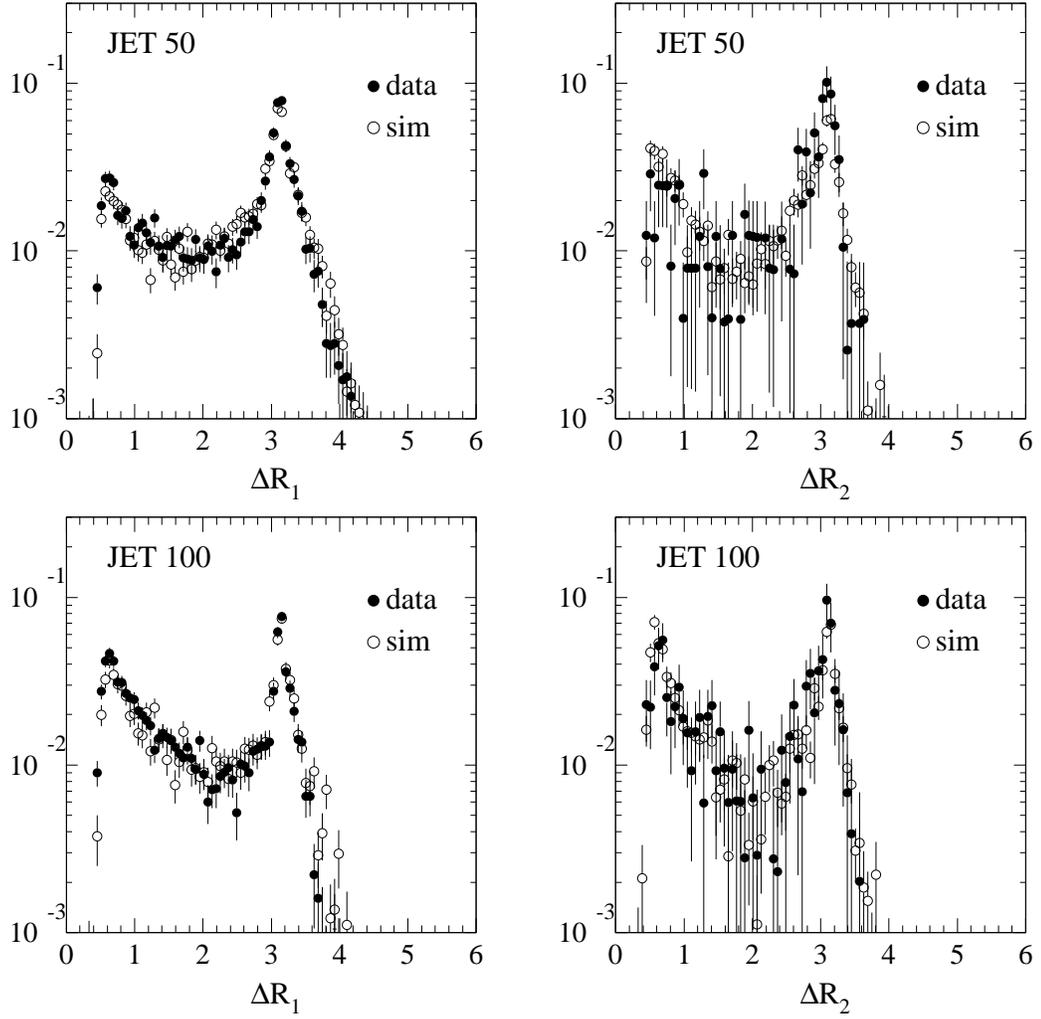}
 \caption[]{Distributions of the distance $\Delta R_1$ between a jet 
            tagged by SECVTX tag and the closest jet in the event and 
            of the distance $\Delta R_2$ between two jets tagged by 
            SECVTX. The distributions of the data and the fitted 
            simulation are normalized to unit area.}
 \label{fig:fig_11_2}
 \end{center}
 \end{figure}

\begin{figure}
 \begin{center}
 \leavevmode
 \epsffile{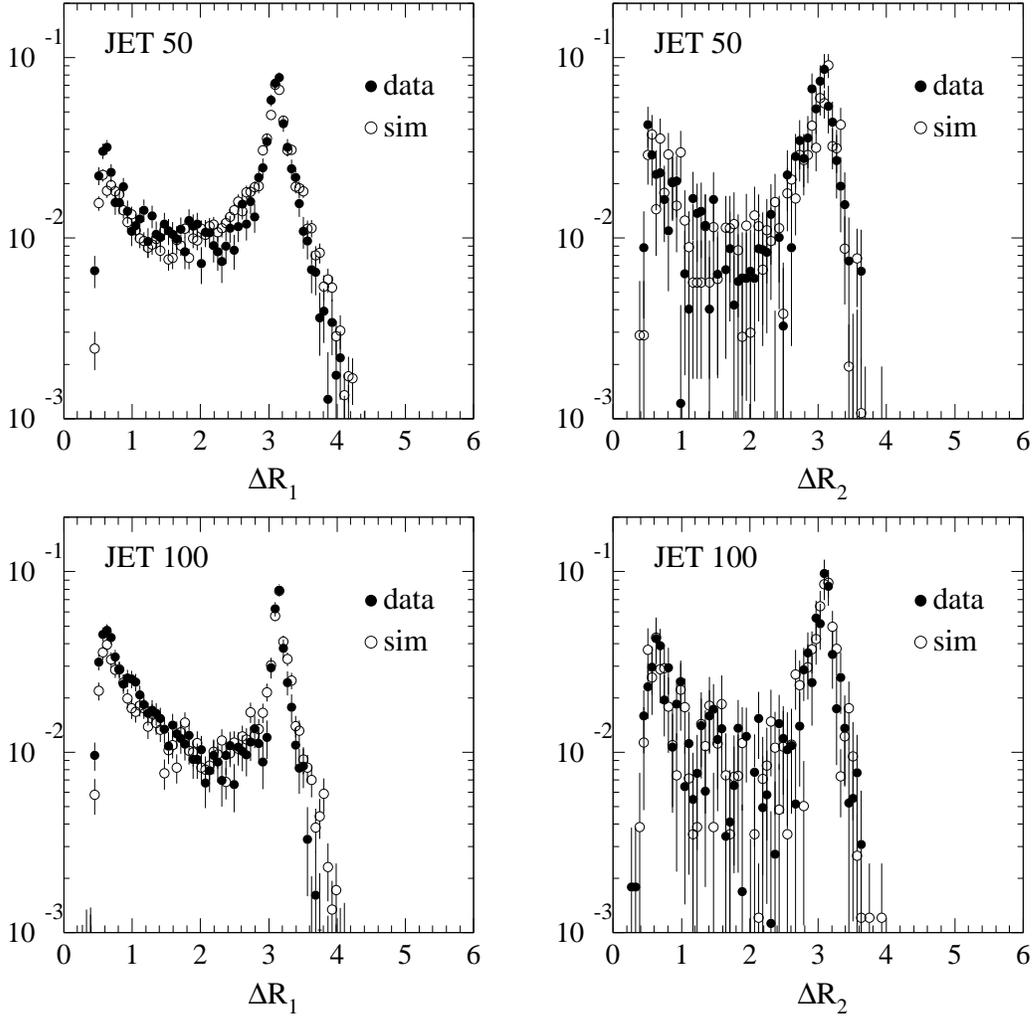}
 \caption[]{ Distributions of the distance $\Delta R_1$ between a jet with a 
             JPB tag and the closest jet in the event and of the distance
             $\Delta R_2$ between two jets with a JPB tag. The distributions 
             of the data and the fitted simulation are normalized to unit area.}
 \label{fig:fig_11_3}
 \end{center}
 \end{figure}

        Figure~\ref{fig:fig_11_1} compares the $E_T$ distributions
        of tagged jets in the data and in the fitted simulation.
        Similarly, Figures~\ref{fig:fig_11_2} and~\ref{fig:fig_11_3}
        compare distributions of distances between tagged jets.
\newpage

\widetext
 \begin{table}[p]
 \begin{center}
\def\arraystretch{0.8}
 \caption[]{Result of the fit of the Herwig simulation to 
 the JET 50 and 
            JET 100 data (see text).}
 \begin{tabular}{lc}
 {Process}                                          & {Cross section weight} \\
\hline
 $b$ {\small direct production+flavor excitation}   & 1.09  $\pm$  0.15  \\
 $g \rightarrow b\bar{b}$                           & 1.40  $\pm$  0.19  \\
 $b$ Total		                                      & 1.22  $\pm$  0.12  \\ 
\hline
 $c$ {\small direct production+flavor excitation}   & 1.12  $\pm$  0.28  \\
 $g \rightarrow c\bar{c}$                           & 1.35  $\pm$  0.36  \\
 $c$ Total	          	                            & 1.25  $\pm$  0.20  \\ 
\hline
 $b+c$ {\small direct production+flavor excitation} & 1.11  $\pm$  0.16  \\
 $g \rightarrow b\bar{b},c\bar{c}$                  & 1.36  $\pm$  0.22  \\
 $b+c$ Total		                                    & 1.24  $\pm$  0.12  \\
 \end{tabular}
 \label{tab:tab_11_4}
 \end{center}
 \end{table}
\narrowtext

\clearpage

       Table~\ref{tab:tab_11_5} compares rates of tags in generic-jet data 
       and in the Herwig simulation calibrated according to 
       Table~\ref{tab:tab_11_4}. The JET 20 sample was not used to calibrate 
       the simulation package. Similarly, the SLT algorithm was not used in 
       the Herwig calibration. The comparison of the number of SLT 
       tags in the data and simulation serves to check independently the
       calibration of the Herwig production cross section 
       and the SLT tagging efficiency in the simulation.
%

\widetext
 \begin{table}[htp]
 \begin{center}
\def\arraystretch{0.8}
 \caption[]{Number of tagged jets with heavy flavor 
  in generic-jet data and 
 in the calibrated Herwig simulation. Data and simulation are normalized to the
same number of events. The second and third columns list the number
 of tags and removed mistags in the data.}
 \begin{tabular}{lcccc}
  \multicolumn{5}{c}{ JET 20 (194009 events)}   \\ 
 Tag type   & Tags & Mistags  & Data  & Simulation \\
SECVTX      & 4674             &  616       & 4058$\pm$92  &  4052$\pm$143 \\
JPB         & 8343             & 2801       & 5542$\pm$295 &  5573$\pm$173 \\
SLT         & 4994             & 3962       & 1032$\pm$402 &  826$\pm$122  \\
\hline
  \multicolumn{5}{c}{ JET 50 (151270 events)} \\ 
 Tag type   & Tags & Mistags  & Data  & Simulation \\
SECVTX      &  6536            & 1360       & 5176$\pm$158 & 5314$\pm$142 \\
JPB         & 11533            & 4700       & 6833$\pm$482 & 6740$\pm$171 \\
SLT         &  6408            & 5241       & 1167$\pm$530 & 1116$\pm$111 \\
\hline
  \multicolumn{5}{c}{ JET 100 (129434 events)}  \\ 
 Tag type   & Tags & Mistags  & Data  & Simulation \\
SECVTX      & 7682          & 2227        & 5455$\pm$239     & 5889$\pm$176 \\
JPB         & 13365         & 6494        & 6871$\pm$659     & 7263$\pm$202 \\
SLT         & 7483          & 6367        & 1116$\pm$642     & 1160$\pm$168 \\
 \end{tabular}
 \label{tab:tab_11_5}
 \end{center}
 \end{table}
\narrowtext

\newpage
\section{Composition of the $W+\geq$1 jet sample}
       ~\label{sec:phys_bkgd}
       The background to the $t\bar{t}$ production 
       is determined using the data or the simulation calibrated as  
       described in Sections~\ref{sec:simu_mod} and~\ref{sec:herwig_norm}. 
       The $\ttbar$ production cross-section is determined by attributing
       the excess of tagged  $W+\geq$ 3 jet events to $\ttbar$ production.
       $\W+1$ and $\W+2$ jet events provide a check of the background
       calculation.
       The evaluation of the backgrounds to $t\bar{t}$ events
       is detailed in the subsections~A-H.
       The results of these background determinations are listed in the tables
       in Section~\ref{sec:xsec} where the cross section is calculated. 
       The following two Sections~\ref{sec:zjets_Tags} and~\ref{sec:w_neg_Tags}
       provide checks of the background estimates.

\subsection{non-$W$ background}
\begin{figure}
 \begin{center}
 \leavevmode
 \epsffile{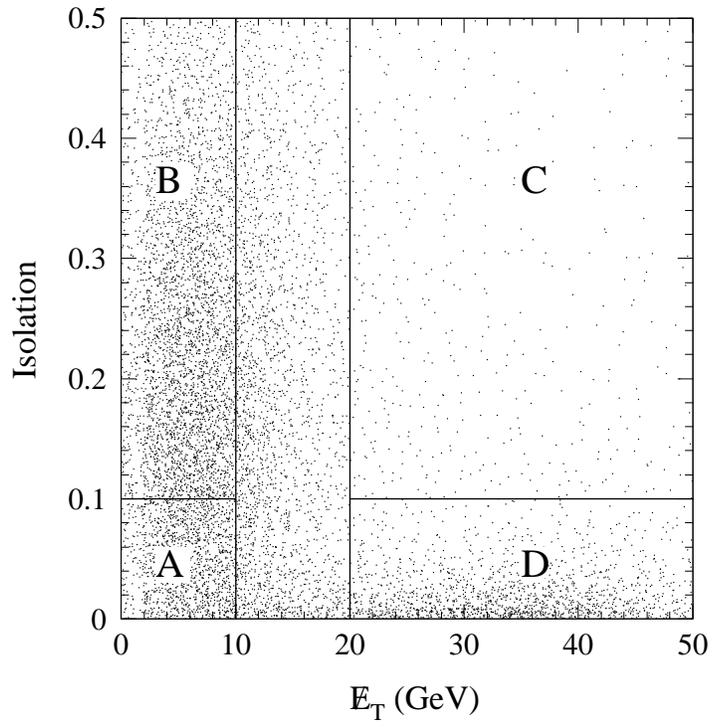}
 \caption[]{ Distributions of the primary lepton isolation vs. $\met$ in 
             $W+\geq$1 jet candidate events. The three regions A, B, C
             are used to evaluate the non-$\W$ contribution in the region D,
 which defines the $\W$ signal.}
 \label{fig:fig_12_0}
 \end{center}
 \end{figure}

\begin{figure}
 \begin{center}
 \leavevmode
 \epsffile{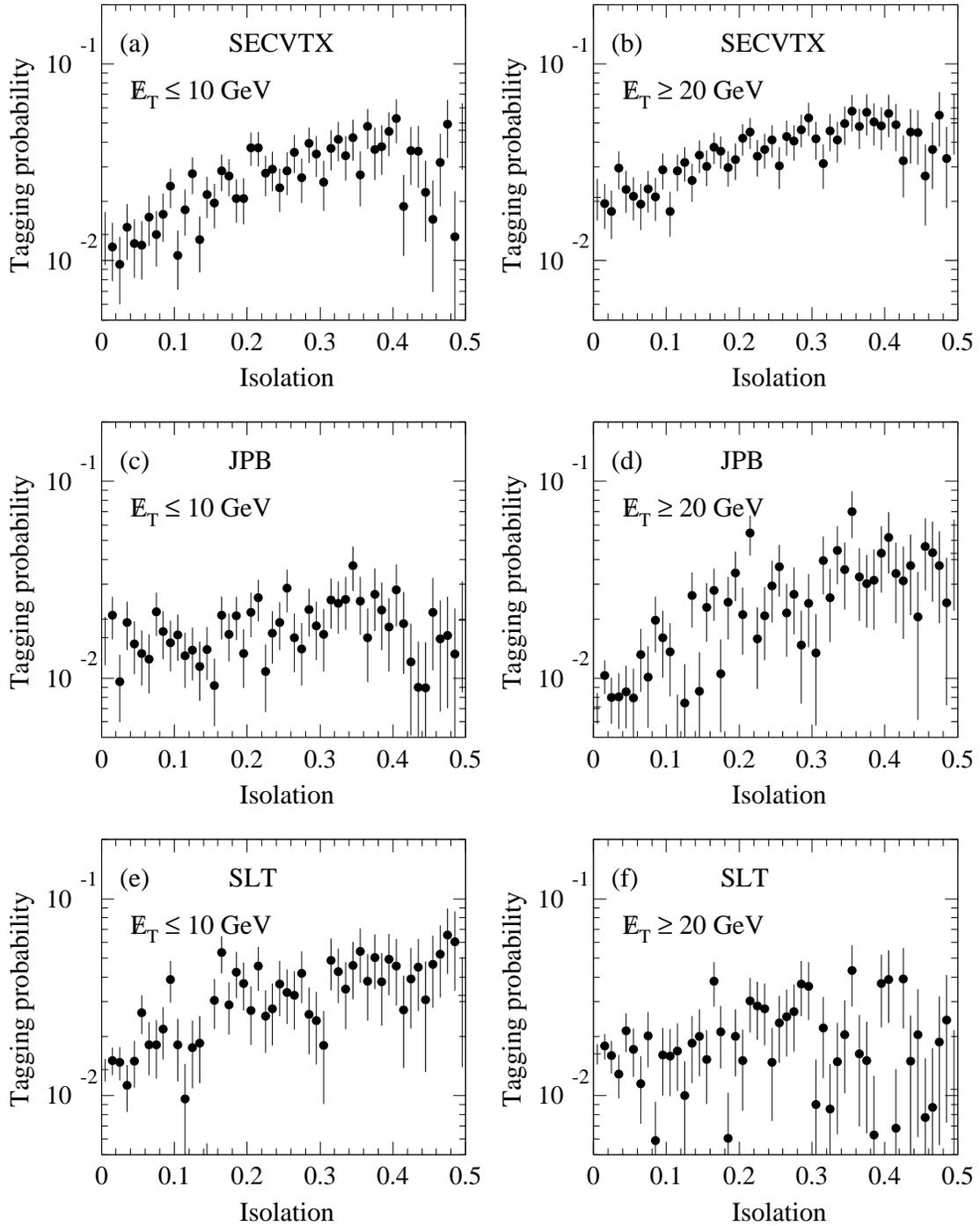}
 \caption[]{The tagging probability as a function of the isolation 
            of the primary lepton.}
 \label{fig:fig_12_1}
 \end{center}
 \end{figure}

       As in previous analyses~\cite{w_sigma}, the background from non-$W$
       sources, including $b\bar{b}$ production, is determined directly 
       from the data by studying the isolation of primary lepton candidates in
       the low ($\met \leq$ 10 GeV) and in the high ($\met \geq$ 20 GeV)
       $\met$ region. The number of non-$W$ events in each jet-bin is evaluated 
       as
      \[ N_{non-W} = N_C \times \frac{N_A}{N_B} \]
       where $N_A$, $N_B$ and $N_C$ are the number of events in regions A, B 
       and C of Figure~\ref{fig:fig_12_0}. The  corresponding number of tagged 
       events is
      \[ N_{\rm non-W}^{\rm tag} = N_{\rm non-W} \times P^{\rm tag} \]
       where $P^{\rm tag}$ is the tagging probability measured in region A.
       The yield of $P^{\rm tag}$ as a function of the lepton isolation is
       shown in Figure~\ref{fig:fig_12_1}.
%
\subsection{ $Z\rightarrow \tau^+\tau^-$ events}~\label{sec:ztau}
        The $Z\rightarrow \tau^+\tau^-$ contribution is estimated using the 
        Pythia generator (option MSEL=13). The simulation is normalized 
        to the same number of $Z \rightarrow \mu^+\mu^-,\; e^+e^-$ events 
        observed in the data.
\subsection{Single top production}~\label{sec:single_Top}
        The single top contribution via the $W-$gluon fusion channel is
        estimated with Herwig using the process 2000.
        The single top production for the annihilation process
        $q\bar{q} \rightarrow W^{\ast} \rightarrow t\bar{b}$
        is estimated using the Pythia generator (option MSEL=12).
        We use the cross sections 
        $\sigma_{W-g}$=1.5$\pm$0.4 pb for  $W-g$ fusion
        derived using the NLO calculation in Ref.~\cite{single_top} 
        and  $\sigma_{W^{\ast}\rightarrow t\bar{b}}$=0.74$\pm$0.05 pb 
        for the annihilation process~\cite{single_top}.
\subsection{Diboson production}~\label{sec:diboson}
        The contribution of  the $ZZ$, $WZ$ and $WW$ production is estimated 
        using the Pythia generator (options MSEL=15 and ISUB=22, 23 and 25, 
        respectively). We use the  cross sections
        $\sigma(WW)$=9.5$\pm$0.7 pb,
        $\sigma(WZ)$=2.6$\pm$0.34 pb and 
        $\sigma(ZZ)$=1.0$\pm$0.2 pb~\cite{diboson_xsec}.
\subsection{Mistags}~\label{sec:mstg}
        The SECVTX  and SLT mistags are calculated weighting each jet in the
        $W$ sample with the mistag probability matrices derived in 
        Section~\ref{sec:mistags} and Section~\ref{sec:slt_fake},
        respectively.
        The reevaluation of the SECVTX mistag matrix has resulted in a reduced
        estimate of this background in the signal region by (50 $\pm$ 5)\%
        compared with the previous estimates of Ref.~\cite{cdf_xsec,topxs}.

        For the jet-probability algorithm, each simulated background
        also includes the contribution of mistags. 
        The number of JPB mistags is evaluated only for the 
        fraction of $W+$ jet events which is not simulated, i.e. $W+$ jet
        direct production without heavy flavor.
\subsection{ The $W + b\bar{b}$ and $W + c\bar{c}$ contribution}
~\label{sec:w_hf}
        We use the Herwig generator (process 2100 with hard scattering 
        $p_T^{\rm min} \geq$ 10 GeV/c) to estimate the fraction of 
        $W+ \geq$ n jet events, $F_1^{\alpha}$, in which  only one jet 
        contains hadrons with heavy flavor resulting from gluon splitting
        ($\alpha$ refers to the flavor type). The fraction of $W+ \geq$ 2 
        jet events, $F_2^{\alpha}$, in which two different jets contain 
        hadrons with heavy flavor is calculated using the Vecbos generator 
        (see Section~\ref{sec:degrade}). 
        The fractions of $W+$ jet events with heavy flavor content are
        listed in Table~\ref{tab:tab_12_1}.
        We use the Herwig and Vecbos simulations also to determine
        the efficiency for finding events with one or two
        tagged jets, as listed in Table~\ref{tab:tab_12_2}.

        It follows that the number of tagged $Wb\bar{b}$
        and $Wc\bar{c}$  events is 
 \begin{eqnarray*}
   N_{\rm 1tag}^{\alpha}& = & N_W \cdot  
     (F_1^{\alpha} \cdot \varepsilon_{\rm 1tag}^{1\alpha} + 
     F_2^{\alpha} \cdot \varepsilon_{\rm 1tag}^{2\alpha})  \\ 
   N_{\rm 2tag}^{\alpha}& = & N_W \cdot
     F_2^{\alpha} \cdot \varepsilon_{\rm 2tag}^{2\alpha}  
\end{eqnarray*}
        where $N_W$ is the number of $W$ events in the data after
        removing the predicted number of non-$W$, di-bosons, single top, 
        unidentified $Z$  and $t\bar{t}$ events.
\newpage
\widetext
 \begin{table}[p]
 \begin{center}
\def\arraystretch{0.8}
 \caption[]{ Fraction of $W+ \geq$ 1 jet events with heavy
             flavor jets as a function of the jet multiplicity.}
 \begin{tabular}{lcccc} 
 & \multicolumn{2}{c}{$W\bbbar$} &      \multicolumn{2}{c}{$W\ccbar$}\\ 
{Sample} & {$F_1^{b}$ (\%)} & {$F_2^{b}$ (\%)} & {$F_1^{c}$ (\%)}
         & {$F_2^{c}$ (\%)}\\
\hline
 $W+1 \,{\rm jet}$      & 0.80$\pm$0.11 &     {~}
                                & 2.01$\pm$0.54 &     {~}       \\
 $W+2 \,{\rm jet}$      & 1.28$\pm$0.18 & 1.20$\pm$0.38
                                & 3.73$\pm$1.00 & 1.40$\pm$0.52 \\
 $W+3 \,{\rm jet}$      & 1.88$\pm$0.31 & 1.90$\pm$0.62
                                & 5.31$\pm$1.48 & 2.30$\pm$0.91 \\
 $W+\geq4 \,{\rm jet}$  & 3.54$\pm$1.06 & 2.40$\pm$0.77
                                & 6.08$\pm$2.45 & 3.00$\pm$1.13 \\
\end{tabular}
\label{tab:tab_12_1}
\end{center}
\end{table}
\narrowtext

\widetext
 \begin{table}[p]
 \begin{center}
\def\arraystretch{0.8}
 \caption[]{Tagging efficiencies ($\varepsilon$)
  in $W\bbbar$ and $W\ccbar$
            simulated events.}
 \begin{tabular}{lcccccc}
 \multicolumn{7}{c}{SECVTX} \\
 {Sample} & $\varepsilon_{\rm 1tag}^{1b}$ (\%) & $\varepsilon_{\rm 1 tag}^{2b}$ (\%) 
          & $\varepsilon_{\rm 2 tag}^{2b}$ (\%) & $\varepsilon_{\rm 1 tag}^{1c}$ (\%) 
          & $\varepsilon_{\rm 1 tag}^{2c}$ (\%) & $\varepsilon_{\rm 2tag}^{2c}$ (\%) \\
 $W+1 \,{\rm jet}$     & 24.6$\pm$0.8   &              &  
                       &  4.56$\pm$0.29 &              &              \\ 
 $W+2 \,{\rm jet}$     & 21.6$\pm$1.7   & 45.8$\pm$1.8 & 10.6$\pm$1.2
                       &  3.59$\pm$0.49 & 11.7$\pm$1.1 &  0.4$\pm$0.2 \\
 $W+\geq3 \,{\rm jet}$ & 20.6$\pm$4.4   & 46.8$\pm$4.0 & 10.7$\pm$2.8
                       &  3.59$\pm$1.26 & 14.3$\pm$2.3 &  0.0$\pm$0.0 \\
 \multicolumn{7}{c}{JPB} \\ 
 {Sample} & $\varepsilon_{\rm 1tag}^{1b}$ (\%) & $\varepsilon_{\rm 1 tag}^{2b}$
(\%)
          & $\varepsilon_{\rm 2 tag}^{2b}$ (\%) & $\varepsilon_{\rm 1
tag}^{1c}$ (\%)
          & $\varepsilon_{\rm 1 tag}^{2c}$ (\%) & $\varepsilon_{\rm 2tag}^{2c}$
(\%) \\
 $W+1 \,{\rm jet}$     & 23.8$\pm$0.7   &              &    
                       & ~9.8$\pm$0.4   &              &              \\
 $W+2 \,{\rm jet}$     & 20.3$\pm$1.4   & 40.7$\pm$1.5 & 10.0$\pm$0.9
                       & ~7.8$\pm$0.7   & 25.0$\pm$1.4 &  2.8$\pm$0.5 \\
 $W+\geq3 \,{\rm jet}$ & 21.7$\pm$3.8   & 43.2$\pm$3.4 & ~9.3$\pm$1.9
                       & 13.0$\pm$2.2   & 25.6$\pm$2.9 & 1.7$\pm$0.8  \\
 \multicolumn{7}{c}{ SLT} \\
 {Sample} & $\varepsilon_{\rm 1tag}^{1b}$ (\%) & $\varepsilon_{\rm 1 tag}^{2b}$
(\%)
          & $\varepsilon_{\rm 2 tag}^{2b}$ (\%) & $\varepsilon_{\rm 1
tag}^{1c}$ (\%)
          & $\varepsilon_{\rm 1 tag}^{2c}$ (\%) & $\varepsilon_{\rm 2tag}^{2c}$
(\%) \\
 $W+1 \,{\rm jet}$     &  7.7$\pm$0.9 &              & 
                       &  3.7$\pm$0.5 &              &                \\
 $W+2 \,{\rm jet}$     &  6.9$\pm$1.2 & 13.2$\pm$1.7 &  0.6$\pm$0.3
                       &  3.7$\pm$0.6 &  6.2$\pm$1.0 &  0.1$\pm$0.1   \\
 $W+\geq3 \,{\rm jet}$ &  7.1$\pm$2.6 & ~9.6$\pm$2.3 &  0.5$\pm$0.5
                       &  5.2$\pm$1.6 &  8.0$\pm$2.0 &  0.0$\pm$0.0   \\
 \end{tabular}
 \label{tab:tab_12_2}
 \end{center}
 \end{table}
\narrowtext

\clearpage
\subsection{The $Wc$ contribution}~\label{sec:wc}
        The fraction $F_{Wc}$ of $g s \rightarrow Wc$ and 
        $g d \rightarrow Wc$ events
        is  evaluated  using the Herwig $W+$ 1 jet simulation
        and is shown in Table~\ref{tab:tab_12_3}.
        The estimated uncertainty
        on $F_{Wc}$ is dominated by the uncertainty in the strange sea
        content in the proton, computed by examining a wide range of 
        different structure functions in Ref.~\cite{top_prd}.
        The tagging efficiencies for this process 
        are listed in Table~\ref{tab:tab_12_4}.
\newpage

\narrowtext
 \begin{table}[p]
 \begin{center}
\def\arraystretch{0.8}
 \caption[]{Fractions  of $Wc$ events.}
\begin{tabular}{lc} 
   {Sample}               & {$F_{Wc}$ (\%)} \\
\hline
 $W+1 \, {\rm jet}$       & 4.8$\pm$1.4 \\

 $W+2 \, {\rm jet}$       & 7.2$\pm$2.2 \\

 $W+3 \, {\rm jet}$       & 7.5$\pm$2.3 \\

 $W+\geq 4 \, {\rm jet}$  & 7.5$\pm$2.3 \\
\end{tabular}
\label{tab:tab_12_3}
\end{center}
\end{table}
\widetext

\widetext
 \begin{table}[p]
 \begin{center}
\def\arraystretch{0.8}
 \caption[]{Tagging efficiencies in $Wc$ events.}
 \begin{tabular}{lccc} 
 {Sample} & {$\varepsilon_{Wc}^{SEC}$ (\%)} & {$\varepsilon_{Wc}^{JPB}$ (\%)}
          & {$\varepsilon_{Wc}^{SLT}$ (\%)} \\
 \hline
  $W+1 \, {\rm jet}$      & 4.1$\pm$0.4 & ~8.7$\pm$0.4 & 3.3$\pm$0.4 \\

  $W+2 \, {\rm jet}$      & 4.2$\pm$0.6 & 10.8$\pm$1.0 & 5.2$\pm$0.7 \\

  $W+\geq3 \, {\rm jet}$  & 4.5$\pm$0.6 & 16.7$\pm$2.9 & 6.9$\pm$2.0 \\
 \end{tabular}
 \label{tab:tab_12_4}
 \end{center}
 \end{table}
\widetext

\clearpage
\subsection{Direct production of $Z+$ jet with heavy flavor }~\label{sec:z_hf}
        We use the Pythia generator (option MSEL=13) to 
        estimate the fraction of unidentified  $Z+$ jet events
        passing our selection. 
        The simulation is normalized to the number of $Z \rightarrow \ell\ell$
        observed in the data.
        The fraction  of $Z+$ jet events containing heavy flavor is determined
        by rescaling the Herwig prediction for the fraction of $W+$ jet events 
        with heavy flavor. The rescaling factor is derived by
        comparing   $W+$ 1 jet and $Z+$ 1 
        jet events generated with Pythia (option MSEL=14 and 13, respectively). 
        For the same number of generated events, we find 
        $0.3\pm0.15$ $Zc$ events for every $Wc$ event, 
        $1 \pm 0.3$ $Zc\bar{c}$ events for every $Wc\bar{c}$ event, and 
        $2.0 \pm 0.5$ $Zb\bar{b}$ events for every $Wb\bar{b}$ event. 
\section{Check of the background calculation using the
         $Z+\geq 1$ jet sample}~\label{sec:zjets_Tags}
      The production mechanisms of $W$ and $Z$ bosons in association with jets
      are very similar. The contribution of $\ttbar$ events to the $Z+$ jet 
      events is negligible. This sample  provides a good benchmark for our 
      background calculation.
      The selection of the $Z+\geq$1 jet event sample is described in 
      Section~\ref{sec:zremoval}. Table~\ref{tab:tab_13_1} shows the predicted 
      composition of the $Z+\geq$1 jet sample before tagging. 
      In this Table the number of $Z+$ jet events is derived from the data 
      by subtracting the $WW$, $WZ$, $ZZ$, $t\bar{t}$, and single top
      contribution. 
      The measured and predicted rates of events with SECVTX, JPB and SLT tags
      are shown in Tables~\ref{tab:tab_13_2} to~\ref{tab:tab_13_4}.
      The  product of the four Poisson probabilities, one per jet bin,
      for the observed numbers given the prediction 
      is $P_0=1.2\times 10^{-3}$
      for Table~\ref{tab:tab_13_2},
      $P_0=2.1\times 10^{-4}$ for Table~\ref{tab:tab_13_3}, and
      $P_0=1.0 \times 10^{-3}$  for 
      Table~\ref{tab:tab_13_4}.
      With a Monte Carlo simulation, in which we fluctuate the predicted rates 
      by their uncertainty according to a gaussian distribution, we 
      estimate that the likelihood of observing a probability 
      no larger than $P_0$  is 33.8\% for
      events with SECVTX tags, 17.9\% for events with  JPB tags and
      41.1\% for events with SLT tags. In $Z+$ jet events the background 
      prediction agrees with the data.
\newpage

\widetext
 \begin{table}[p]
 \begin{center}
\def\arraystretch{0.8}
 \caption[]{Composition of the $Z+\geq$1 jet sample before tagging. 
            We use $\sigma_{t\overline{t}}$= 5 pb  
            from Ref~\cite{catani}.}
 \begin{tabular}{lcccc}
  Source & $Z+1 \,{\rm jet}$&$Z+2 \,{\rm jet}$ & $Z+3 \,{\rm jet}$
                 & $Z+\geq4 \,{\rm jet}$     \\
 \hline
 Data           &   1148       &    159      &     16      &   4 \\
 $WW$           & 0.8$\pm$0.2 & 0.2$\pm$0.1 & 0.0$\pm$0.0  & 0.0$\pm$0.0 \\
 $WZ$           & 2.2$\pm$0.5 & 1.7$\pm$0.4 & 0.3$\pm$0.1  & 0.1$\pm$0.0 \\
 $ZZ$           & 1.2$\pm$0.3 & 1.6$\pm$0.4 & 0.3$\pm$0.1  & 0.0$\pm$0.0 \\
 $Zc$           & 16.5$\pm$4.9 & 3.3$\pm$1.0 & 0.3$\pm$0.1 & 0.1$\pm$0.0 \\
 $Z\bbbar$        & 18.3$\pm$2.5 & 7.6$\pm$1.3 & 1.1$\pm$0.2 & 0.4$\pm$0.1 \\
 $Z\ccbar$        & 23.0$\pm$6.1 & 7.9$\pm$1.7 & 1.1$\pm$0.3 & 0.3$\pm$0.1 \\
 $Z+$ jet without h.f.&
 1085.3$\pm$8.3 & 135.3$\pm$2.5 &12.2$\pm$0.4 & 2.9$\pm$0.1 \\
 Single top     & 0.1$\pm$0.0 & 0.0$\pm$0.0 & 0.0$\pm$0.0 & 0.0$\pm$0.0 \\
 $\ttbar$         & 0.6$\pm$0.1 & 1.4$\pm$0.3 & 0.5$\pm$0.1 & 0.2$\pm$0.0 \\
 \end{tabular}
 \label{tab:tab_13_1}
 \end{center}
 \end{table}
\narrowtext

\newpage

\narrowtext
 \begin{table}[p]
 \begin{center}
\def\arraystretch{0.8}
 \caption[]{Summary of observed and predicted number of $Z$ events 
            with one (ST) and two (DT) SECVTX tags.}
 \begin{tabular}{lcccc}
  Source & $Z+1 \,{\rm jet}$&$Z+2 \,{\rm jet}$ & $Z+3 \,{\rm jet}$
                 & $Z+\geq4 \,{\rm jet}$     \\
 \hline
 Mistags           & 1.27$\pm$0.13 & 0.34$\pm$0.03 & 0.08$\pm$0.01 & 0.01$\pm$0.01 \\
 $WW,WZ,ZZ$        & 0.09$\pm$0.03 & 0.18$\pm$0.05 & 0.03$\pm$0.01 & 0.00$\pm$0.00 \\
 $Zc$              & 0.67$\pm$0.21 & 0.15$\pm$0.05 & 0.02$\pm$0.00 & 0.00$\pm$0.00 \\
 $Z\ccbar$, $Z\bbbar$ (ST) & 5.56$\pm$0.70 & 2.59$\pm$0.46 & 0.40$\pm$0.08 & 0.14$\pm$0.03 \\
 $Z\ccbar$, $Z\bbbar$ (DT) & & 0.39$\pm$0.13 & 0.06$\pm$0.03 & 0.02$\pm$0.01 \\
 Single top        & 0.01$\pm$0.01 & 0.00$\pm$0.00 & 0.00$\pm$0.00 & 0.00$\pm$0.00 \\
 $\ttbar$ (ST)       & 0.22$\pm$0.05 & 0.44$\pm$0.09 & 0.20$\pm$0.05 & 0.03$\pm$0.01 \\
 $\ttbar$ (DT)       &  & 0.23$\pm$0.06 & 0.07$\pm$0.02 & 0.03$\pm$0.01 \\
 \hline
 Prediction (ST)  & 7.83$\pm$0.74 & 3.70$\pm$0.47 & 0.73$\pm$0.10 & 0.20$\pm$0.03 \\
 Prediction (DT)  & & 0.62$\pm$0.14 & 0.13$\pm$0.03 & 0.04$\pm$0.01 \\
 \hline
 Data  (ST)  &     10 &      3 &      0 &      1 \\
 Data with (DT) &        &      2 &      0 &      0 \\
 \end{tabular}
 \label{tab:tab_13_2}
 \end{center}
 \end{table}
\widetext

\newpage
\widetext
 \begin{table}[p]
 \begin{center}
\def\arraystretch{0.8}
 \caption[]{Summary of observed and predicted number of $Z$ events 
            with one (ST) and two (DT) JPB tags.}
 \begin{tabular}{lcccc}
  Source & $Z+1 \,{\rm jet}$&$Z+2 \,{\rm jet}$ & $Z+3 \,{\rm jet}$
                 & $Z+\geq4 \,{\rm jet}$     \\
 \hline
 Mistags       & 5.65$\pm$0.57 & 1.51$\pm$0.15 & 0.34$\pm$0.04 & 0.05$\pm$0.01 \\
 $WW,WZ,ZZ$    & 0.13$\pm$0.03 & 0.24$\pm$0.06 & 0.02$\pm$0.01 & 0.02$\pm$0.01 \\
 $Zc$          & 1.39$\pm$0.44 & 0.35$\pm$0.11 & 0.05$\pm$0.02 & 0.01$\pm$0.00 \\
 $Z\ccbar$, $Z\bbbar$ (ST) & 6.63$\pm$ 0.87 & 2.85$\pm$ 0.45 & 0.50$\pm$0.09 & 0.17$\pm$0.03 \\
 $Z\ccbar$, $Z\bbbar$ (DT) & & 0.42$\pm$0.13 & 0.06$\pm$0.02 & 0.02$\pm$0.01 \\
 Single top       & 0.01$\pm$0.00 & 0.00$\pm$0.00 & 0.00$\pm$0.00 & 0.00$\pm$0.00 \\
 $\ttbar$ (ST)      & 0.17$\pm$0.02 & 0.43$\pm$0.06 & 0.19$\pm$0.03 & 0.04$\pm$0.01 \\
 $\ttbar$ (DT)      & & 0.17$\pm$0.02 &0.06$\pm$0.01 & 0.03$\pm$0.00 \\
 \hline
 Prediction (ST) & 13.98$\pm$1.13 & 5.37$\pm$0.49 & 1.11$\pm$0.10 & 0.30$\pm$0.04 \\
 Prediction (DT) & & 0.59$\pm$0.13 & 0.12$\pm$0.02 &0.05$\pm$0.01 \\
 \hline 
  Data  (ST)  &     11 &      5 &      1 &      2 \\
  Data  (DT) &        &      0 &      0 &      0 \\
 \end{tabular}
 \label{tab:tab_13_3}
 \end{center}
 \end{table}
\narrowtext

\newpage

\widetext
 \begin{table}[p]
 \begin{center}
\def\arraystretch{0.8}
 \caption[]{Summary of observed and predicted number of $Z$ events 
            with one (ST)  and two (DT) SLT tags.}
 \begin{tabular}{lcccc}
  Source & $Z+1 \,{\rm jet}$&$Z+2 \,{\rm jet}$ & $Z+3 \,{\rm jet}$
                 & $Z+\geq4 \,{\rm jet}$     \\
 \hline
 Mistags              & 12.65$\pm$1.27 & 3.66$\pm$0.37 & 0.57$\pm$0.06 & 0.15$\pm$0.02 \\
 $WW,WZ,ZZ$           & 0.04$\pm$0.02  & 0.09$\pm$0.03 & 0.01$\pm$0.01 & 0.01$\pm$0.01 \\
 $Zc$                 & 0.55$\pm$0.17  & 0.17$\pm$0.05 & 0.02$\pm$0.01 & 0.01$\pm$0.00 \\
 $Z\ccbar$, $Z\bbbar$ (ST) & 2.26$\pm$0.36 & 1.10$\pm$0.19 & 0.16$\pm$0.03 & 0.06$\pm$0.01 \\
 $Z\ccbar$, $Z\bbbar$ (DT) & &0.02$\pm$0.01 & 0.00$\pm$0.00 & 0.00$\pm$0.00 \\
 Single top           & 0.00$\pm$0.00 & 0.00$\pm$0.00 & 0.00$\pm$0.00 & 0.00$\pm$0.00 \\
 $\ttbar$ (ST)          & 0.04$\pm$0.00 & 0.19$\pm$0.02 & 0.08$\pm$0.01 & 0.01$\pm$0.00 \\
 $\ttbar$ (DT)          & & 0.00$\pm$0.00 & 0.00$\pm$0.00 & 0.00$\pm$0.00 \\
 \hline
 Prediction (ST) & 15.54$\pm$1.33 & 5.21$\pm$0.42 & 0.85$\pm$0.07 & 0.24$\pm$0.02 \\
 Prediction (DT) & & 0.03$\pm$0.01 & 0.01$\pm$0.00 & 0.00$\pm$0.00 \\
 \hline
 Data  (ST) &     16 &      3 &      0 &      1 \\
 Data  (DT)&        &      0 &      0 &      0 \\
 \end{tabular}
 \label{tab:tab_13_4}
 \end{center}
 \end{table}
\narrowtext

 \section{Rates of negative tags in the $W+\geq$ 1 jet sample}
         ~\label{sec:w_neg_Tags}
      As shown in Section~\ref{sec:mistags}, the mistag rates plus 
      the estimated heavy flavor contribution to the negative tags 
      account for the observed rates of negative tags in all 
      generic-jet data. A similar test in the  $W+\geq$1 jet sample 
      offers an additional check of the mistag rate predictions 
      and a complementary test of the method used to estimate the 
      background contribution to the $\ttbar$ signal. 
   
      The rate of negative tags for  each  process is calculated
      from the corresponding  simulation or using the 
      data as we do for positive tags. We use the sample composition
      before tagging listed in Tables~\ref{tab:tab_15_1} 
      and~\ref{tab:tab_15_3}  for SECVTX and JPB, respectively.
      Table~\ref{tab:tab_14_1} compares numbers of observed and
      predicted negative SECVTX tags as a function of the jet multiplicity.
      The analogous comparison for negative JPB tags is shown in 
      Table~\ref{tab:tab_14_2}. Data and predictions agree within the 
      estimated uncertainties.
\newpage

\widetext
 \begin{table}[p]
 \begin{center}
\def\arraystretch{0.8}
 \caption[]{ Summary of the predicted and observed number of $\W+$ jet events
             with 
             negative SECVTX tags.
             The contribution of each  process before tagging  is taken from
             Table~\protect\ref{tab:tab_15_2}.}
 \begin{tabular}{lcccc}
 Source                 &  $W+1 \,{\rm jet}$      &  $W+2 \,{\rm jet}$      &  $W+3 \,{\rm jet}$      &  $W+\geq4 \,{\rm jet}$   \\
 \hline
 Mistags                        &   10.82$\pm$ 1.08 &    3.80$\pm$ 0.38 &    0.99$\pm$ 0.10 &    0.35$\pm$ 0.04 \\
 Non-$W$                        &    0.30$\pm$ 0.15 &    0.30$\pm$ 0.21 &    0.00$\pm$ 0.35 &    0.00$\pm$ 0.14 \\
 $WW,~WZ,~ZZ$                   &    0.00$\pm$ 0.00 &    0.04$\pm$ 0.04 &    0.00$\pm$ 0.00 &    0.00$\pm$ 0.00 \\
 Single top                     &    0.07$\pm$ 0.02 &    0.05$\pm$ 0.02 &    0.01$\pm$ 0.00 &    0.00$\pm$ 0.00 \\
 $Wc$                           &    0.69$\pm$ 0.32 &    0.34$\pm$ 0.15 &    0.12$\pm$ 0.09 &    0.02$\pm$ 0.02 \\
 $Wc\bar{c}$ (ST)               &    0.34$\pm$ 0.15 &    0.18$\pm$ 0.07 &    0.07$\pm$ 0.05 &    0.01$\pm$ 0.01 \\
 $Wc\bar{c}$ (DT)               &                   &    0.00$\pm$ 0.00 &    0.00$\pm$ 0.00 &    0.00$\pm$ 0.00 \\
 $Wb\bar{b}$ (ST)               &    1.42$\pm$ 0.26 &    0.32$\pm$ 0.09 &    0.08$\pm$ 0.05 &    0.02$\pm$ 0.02 \\
 $Wb\bar{b}$ (DT)               &                   &    0.00$\pm$ 0.00 &    0.00$\pm$ 0.00 &    0.00$\pm$ 0.00 \\
 $Z\rightarrow \tau\tau$        &    0.00$\pm$ 0.00 &    0.00$\pm$ 0.00 &    0.00$\pm$ 0.00 &    0.00$\pm$ 0.00 \\
 $Zc$                           &    0.01$\pm$ 0.00 &    0.00$\pm$ 0.00 &    0.00$\pm$ 0.00 &    0.00$\pm$ 0.00 \\
 $Zc\bar{c}$ (ST)               &    0.01$\pm$ 0.00 &    0.00$\pm$ 0.00 &    0.00$\pm$ 0.00 &    0.00$\pm$ 0.00 \\
 $Zc\bar{c}$ (DT)               &                   &    0.00$\pm$ 0.00 &    0.00$\pm$ 0.00 &    0.00$\pm$ 0.00 \\
 $Zb\bar{b}$ (ST)               &    0.08$\pm$ 0.01 &    0.02$\pm$ 0.01 &    0.01$\pm$ 0.01 &    0.00$\pm$ 0.00 \\
 $Zb\bar{b}$ (DT)               &                   &    0.00$\pm$ 0.00 &    0.00$\pm$ 0.00 &    0.00$\pm$ 0.00 \\
 $t\bar{t}$ (ST)                &    0.01$\pm$ 0.00 &    0.12$\pm$ 0.03 &    0.31$\pm$ 0.08 &    0.27$\pm$ 0.07 \\
 $t\bar{t}$ (DT)                &                   &    0.00$\pm$ 0.00 &    0.00$\pm$ 0.00 &    0.00$\pm$ 0.00 \\
 \hline
 Prediction (ST)   &   13.74$\pm$ 1.18 &    5.18$\pm$ 0.48 &    1.60$\pm$ 0.39 &    0.69$\pm$ 0.17 \\
 Prediction (DT)   &                   &    0.00$\pm$ 0.00 &    0.00$\pm$ 0.00 &    0.00$\pm$ 0.00 \\
 \hline
 Data  (ST)           &   19  &    7  &    2  &    0 \\
 Data  (DT)           &       &    0  &    0  &    0 \\
 \end{tabular}

 \label{tab:tab_14_1}
 \end{center}
 \end{table}
\narrowtext

\newpage
\widetext
 \begin{table}[p]
 \begin{center}
 \def\arraystretch{0.8}
 \caption[]{Summary of the predicted and observed number of $W+$ jet events 
            with negative JPB tags.
            The contribution of each  process before tagging  is taken from
            Table~\protect\ref{tab:tab_15_4}.}
 \begin{tabular}{lcccc}
 Source                 &  $W+1 \,{\rm jet}$      &  $W+2 \,{\rm jet}$      &  $W+3 \,{\rm jet}$      &  $W+\geq4 \,{\rm jet}$   \\
 \hline
 Mistags                        &   41.81$\pm$ 4.24 &   12.99$\pm$ 1.35 &    2.25$\pm$ 0.28 &    0.25$\pm$ 0.19 \\
 Non-$W$                        &    2.74$\pm$ 0.45 &    1.42$\pm$ 0.43 &    0.39$\pm$ 0.19 &    0.16$\pm$ 0.08 \\
 $WW,~WZ,~ZZ$                   &    0.50$\pm$ 0.15 &    0.74$\pm$ 0.19 &    0.36$\pm$ 0.13 &    0.02$\pm$ 0.01 \\
 Single top                     &    0.23$\pm$ 0.05 &    0.34$\pm$ 0.08 &    0.09$\pm$ 0.03 &    0.02$\pm$ 0.01 \\
 $Wc$                           &    9.31$\pm$ 2.91 &    1.82$\pm$ 0.67 &    0.46$\pm$ 0.21 &    0.04$\pm$ 0.03 \\
 $Wc\bar{c}$ (ST)               &    4.55$\pm$ 1.27 &    0.71$\pm$ 0.24 &    0.26$\pm$ 0.12 &    0.03$\pm$ 0.02 \\
 $Wc\bar{c}$ (DT)               &                   &    0.00$\pm$ 0.00 &    0.00$\pm$ 0.00 &    0.00$\pm$ 0.00 \\
 $Wb\bar{b}$ (ST)               &    3.14$\pm$ 0.50 &    1.77$\pm$ 0.36 &    0.39$\pm$ 0.11 &    0.06$\pm$ 0.04 \\
 $Wb\bar{b}$ (DT)               &                   &    0.01$\pm$ 0.01 &    0.01$\pm$ 0.01 &    0.00$\pm$ 0.00 \\
 $Z\rightarrow \tau\tau$        &    0.44$\pm$ 0.20 &    0.52$\pm$ 0.21 &    0.09$\pm$ 0.09 &    0.00$\pm$ 0.00 \\
 $Zc$                           &    0.08$\pm$ 0.02 &    0.02$\pm$ 0.01 &    0.01$\pm$ 0.00 &    0.00$\pm$ 0.00 \\
 $Zc\bar{c}$ (ST)               &    0.12$\pm$ 0.04 &    0.02$\pm$ 0.01 &    0.01$\pm$ 0.01 &    0.00$\pm$ 0.00 \\
 $Zc\bar{c}$ (DT)               &                   &    0.00$\pm$ 0.00 &    0.00$\pm$ 0.00 &    0.00$\pm$ 0.00 \\
 $Zb\bar{b}$ (ST)               &    0.17$\pm$ 0.03 &    0.10$\pm$ 0.02 &    0.04$\pm$ 0.02 &    0.01$\pm$ 0.01 \\
 $Zb\bar{b}$ (DT)               &                   &    0.00$\pm$ 0.00 &    0.00$\pm$ 0.00 &    0.00$\pm$ 0.00 \\
 $t\bar{t}$ (ST)                &    0.12$\pm$ 0.03 &    1.21$\pm$ 0.26 &    2.98$\pm$ 0.65 &    3.35$\pm$ 0.73 \\
 $t\bar{t}$ (DT)                &                   &    0.06$\pm$ 0.01 &    0.09$\pm$ 0.02 &    0.26$\pm$ 0.06 \\
 \hline
 Prediction (ST)   &   63.21$\pm$ 5.34 &   21.65$\pm$ 1.67 &    7.35$\pm$ 0.80 &    3.93$\pm$ 0.76 \\
 Prediction (DT)   &                   &    0.08$\pm$ 0.02 &    0.10$\pm$ 0.02 &    0.26$\pm$ 0.06 \\
 \hline
 Data  (ST)           &   66  &   23  &    8 &    5 \\
 Data (DT)            &       &    1  &    0 &    1 \\
 \end{tabular}
 \label{tab:tab_14_2}
 \end{center}
 \end{table}
\narrowtext

\section{Measurement of the $\ttbar$ production cross section}~\label{sec:xsec}
      The $\ttbar$ production cross section is derived using
 \[ \sigma_{t\bar{t}} = \frac{N^{\rm obs}_{\rm tag} - N^{\rm bkg}_{\rm tag}}
            { {\cal A}_{\ttbar}  \epsilon_{\rm tag} \int {\cal L}dt} \]
      where $N^{\rm obs}_{\rm tag}$ is the number of tagged $W+\geq$ 3 jet 
      events, $N^{\rm bkg}_{\rm tag}$ is the background prediction,
      ${\cal A}_{\ttbar}$ is the detector acceptance for $\ttbar$ events,
      $\epsilon_{\rm tag}$ is the efficiency for tagging top events and
      $\int {\cal L}dt$ = 105.1 $\pm$ 4.0 pb$^{-1}$ is the total integrated luminosity.

      The acceptance for $t\bar{t}$ events is evaluated with a simulation
      which uses  the Pythia generator
      and is (7.8$\pm$1.3)\% for a top mass of 175 $\gevcc$. 
      The 17\% systematic error accounts for all  uncertainties in the simulation 
      which come from the following:
      lepton identification and trigger simulation ($\pm$ 14\%), 
      jet energy scale ($\pm$ 5\%), modeling of initial state
      gluon  radiation ($\pm$ 2\%), final state gluon  radiation ($\pm$ 5\%),
      Monte Carlo modeling of the $t\bar{t}$ production ($\pm$ 5\%),
      detector resolution effects ($\pm$ 2\%), and instantaneous luminosity
      dependence ($\pm$ 2\%).

      The tagging efficiencies are evaluated using the same simulation
      and are $0.505\pm 0.051$, $0.455\pm 0.046$ and 
      $0.157\pm 0.016$ for SECVTX, JPB and SLT respectively.

      In the background calculation the rate of $W+$ jet events with heavy 
      flavor is estimated from the number of events due to $W+$ jet direct
      production using the fraction of heavy flavor determined in 
      Section~\ref{sec:w_hf}. Therefore the contribution of $\ttbar$ events
      must be removed from the data.
      This is done by iterating.
      The $\ttbar$ cross section is estimated from the excess of 
      tagged  $W+\geq$ 3 jet events over the background
      calculated  assuming $\sigma_{\ttbar}=0$. The 
      resulting $\sigma_{\ttbar}$  is used to evaluate the number  
      of $\ttbar$ events before tagging; this contribution is subtracted 
      from the data to obtain the contribution of the  $W+$ jet direct production. 
      The amount of $W+$ jet with heavy flavor is recalculated and
      $\sigma_{\ttbar}$ is updated.
      The procedure is repeated until $\sigma_{\ttbar}$
      is stable to within 1\%.

      In the sample of 252 $W+\geq$ 3 jet events, there are 29 events with 
      at least one jet tagged by the SECVTX algorithm. 
      Using the  procedure described above, the background estimate 
      is 8.0 $\pm$ 1.0 events. Assuming that all the excess is due to $t\bar{t}$
      production, the resulting $\ttbar$ cross section is
      5.08 $\pm$ 1.54 pb (the statistical error is $\pm$ 1.30 pb  and 
      the systematic $\pm$ 0.82 pb). The estimated break-down of the $W+\geq$ 1
      jet sample before and after tagging 
      is shown in Tables~\ref{tab:tab_15_1} and~\ref{tab:tab_15_2}. 
      
      As a cross-check, we calculate $\sigma_{\ttbar}$ using rates of JPB tags. 
      There are 41 events with at least one jet tagged by the JPB algorithm
      with a background of 11.1 $\pm$ 1.3 events. The observed excess of
      events yields $\sigma_{\ttbar}$ = 8.02 $\pm$ 2.16 pb. 
      The estimated  break-down of the $W+\geq$ 1 jet sample before and after tagging
      is shown in Tables~\ref{tab:tab_15_3} and~\ref{tab:tab_15_4}.

      There are 25 events with at least one jet tagged by the SLT algorithm
      with a background of 13.2 $\pm$ 1.2 events. The observed excess
      of events yields 
      $\sigma_{\ttbar}$ = 9.18 $\pm$ 4.26 pb (the statistical error is 
      $\pm$  3.89 pb and the systematic $\pm$ 1.72 pb).
      The estimated break-down of the $W+\geq$ 1 jet sample before and 
      after tagging
      is shown in Tables~\ref{tab:tab_15_5} and~\ref{tab:tab_15_6}.

      Because of a small dependence of the acceptance and the tagging efficiencies
      on the top mass, for a $\pm$5 $\gevcc$ change in the top mass the cross 
      sections calculated using SECVTX and JPB tags change by $\pm$1.8\%
      and the cross section calculated using SLT tags changes by $\pm$2.3\%.
\newpage

\widetext
 \begin{table}[p]
 \begin{center}
\def\arraystretch{0.8}
 \caption{Composition of the $W+\geq$1 jet sample before tagging
using $\sigma_{\ttbar}$ = 5.08 $\pm$ 1.54 pb.}
 \begin{tabular}{lcccc}
 Source                 &  $W+1 \,{\rm jet}$      &  $W+2 \,{\rm jet}$      &  $W+3 \,{\rm jet}$      &  $W+\geq4 \,{\rm jet}$   \\
 \hline
 Data                           &   9454 &   1370 &    198 &     54 \\
 Non-$W$                        &   560.1$\pm$ 14.9 &    71.2$\pm$  2.7 &    12.4$\pm$  2.0 &     5.1$\pm$  1.7 \\
 $WW$                           &    31.2$\pm$  5.4 &    31.1$\pm$  5.4 &     5.2$\pm$  1.0 &     0.8$\pm$  0.2 \\
 $WZ$                           &     4.4$\pm$  0.9 &     4.8$\pm$  1.0 &     0.9$\pm$  0.2 &     0.1$\pm$  0.0 \\
 $ZZ$                           &     0.3$\pm$  0.1 &     0.4$\pm$  0.1 &     0.1$\pm$  0.0 &     0.0$\pm$  0.0 \\
 Unidentified-$Z$         &   234.8$\pm$ 14.5 &    38.5$\pm$  5.9 &     7.9$\pm$  2.4 &     0.7$\pm$  0.7 \\
 Single top                     &    14.1$\pm$  2.1 &     7.9$\pm$  1.7 &     1.7$\pm$  0.4 &     0.3$\pm$  0.1 \\
 $Wc$                           &   413.1$\pm$123.9 &    86.8$\pm$ 26.1 &    11.2$\pm$  3.4 &     1.9$\pm$  0.7 \\
 $Wb\bar{b}$                    &    69.0$\pm$  9.5 &    29.7$\pm$  5.1 &     5.7$\pm$  1.1 &     1.5$\pm$  0.5 \\
 $Wc\bar{c}$                    &   173.1$\pm$ 46.2 &    61.9$\pm$ 13.6 &    11.4$\pm$  2.6 &     2.3$\pm$  0.9 \\
 $W+$ jet without h.f.               &  7952.0$\pm$133.6 &  1027.7$\pm$ 31.1 &   121.1$\pm$  7.7 &    19.9$\pm$  6.1 \\
 $t\bar{t}$                     &     1.8$\pm$  0.5 &    10.1$\pm$  2.8 &    20.3$\pm$  5.7 &    21.3$\pm$  5.9 \\
 \end{tabular}
 \label{tab:tab_15_1}
 \end{center}
\end{table}
\narrowtext

\clearpage
\newpage

\widetext
 \begin{table}[p]
 \begin{center}
\def\arraystretch{0.8}
 \caption{Summary of the predicted and observed number of $W$ events with 
          one (ST) or two (DT) SECVTX tags.}
 \begin{tabular}{lcccc}
 Source                 &  $W+1 \,{\rm jet}$      &  $W+2 \,{\rm jet}$      &  $W+3 \,{\rm jet}$      &  $W+\geq4 \,{\rm jet}$   \\
 \hline
 Mistags                        &   10.82 $\pm$ 1.08 &  ~3.80 $\pm$ 0.38 & ~0.99 $\pm$ 0.10 &    0.35 $\pm$ 0.04 \\
 Non-$W$                        &   ~8.18 $\pm$ 0.78 &  ~1.49 $\pm$ 0.47 & ~0.76 $\pm$ 0.38 &    0.31 $\pm$ 0.16 \\
 $WW,~WZ,~ZZ$                   &   ~0.52 $\pm$ 0.14 &  ~1.38 $\pm$ 0.28 & ~0.40 $\pm$ 0.13 &    0.00 $\pm$ 0.00 \\
 Single top                     &   ~1.36 $\pm$ 0.35 &  ~2.38 $\pm$ 0.54 & ~0.63 $\pm$ 0.14 &    0.14 $\pm$ 0.03 \\
 $Wc$                           &   16.89 $\pm$ 5.38 &  ~3.94 $\pm$ 1.30 & ~0.51 $\pm$ 0.17 &    0.09 $\pm$ 0.04 \\
 $Wc\bar{c}$ (ST)               &   ~7.89 $\pm$ 2.17 &  ~3.54 $\pm$ 0.88 & ~0.77 $\pm$ 0.25 &    0.16 $\pm$ 0.07 \\
 $Wc\bar{c}$ (DT)               &                    &  ~0.06 $\pm$ 0.04 & ~0.00 $\pm$ 0.00 &    0.00 $\pm$ 0.00 \\
 $Wb\bar{b}$ (ST)               &   17.00 $\pm$ 2.41 &  ~8.35 $\pm$ 1.74 & ~1.62 $\pm$ 0.40 &    0.41 $\pm$ 0.14 \\
 $Wb\bar{b}$ (DT)               &                    &  ~1.51 $\pm$ 0.52 & ~0.31 $\pm$ 0.13 &    0.07 $\pm$ 0.03 \\
 $Z\rightarrow \tau\tau$        &   ~0.96 $\pm$ 0.30 &  ~0.70 $\pm$ 0.25 & ~0.17 $\pm$ 0.12 &    0.00 $\pm$ 0.00 \\
 $Zc$                           &   ~0.14 $\pm$ 0.04 &  ~0.03 $\pm$ 0.01 & ~0.01 $\pm$ 0.00 &    0.00 $\pm$ 0.00 \\
 $Zc\bar{c}$ (ST)               &   ~0.22 $\pm$ 0.06 &  ~0.10 $\pm$ 0.03 & ~0.04 $\pm$ 0.02 &    0.00 $\pm$ 0.00 \\
 $Zc\bar{c}$ (DT)               &                    &  ~0.00 $\pm$ 0.00 & ~0.00 $\pm$ 0.00 &    0.00 $\pm$ 0.00 \\
 $Zb\bar{b}$ (ST)               &   ~0.93 $\pm$ 0.14 &  ~0.46 $\pm$ 0.12 & ~0.17 $\pm$ 0.06 &    0.02 $\pm$ 0.02 \\
 $Zb\bar{b}$ (DT)               &                    &  ~0.08 $\pm$ 0.03 & ~0.03 $\pm$ 0.02 &    0.00 $\pm$ 0.00 \\
 Total background (ST)          &   64.90 $\pm$ 6.45 &  26.26 $\pm$ 2.51 & ~6.11 $\pm$ 0.68 &    1.50 $\pm$ 0.23 \\
 Total background (DT)          &                    &  ~1.65 $\pm$ 0.52 & ~0.34 $\pm$ 0.13 &    0.07 $\pm$ 0.03 \\
 $t\bar{t}$ (ST)                &   ~0.54 $\pm$ 0.14 &  ~3.34 $\pm$ 0.87 & ~6.76 $\pm$ 1.76 &    7.42 $\pm$ 1.93 \\
 $t\bar{t}$ (DT)                &                    &  ~0.76 $\pm$ 0.20 & ~2.88 $\pm$ 0.75 &    3.96 $\pm$ 1.03 \\
 \hline
 $t\bar{t}$ + background (ST)   &   65.44 $\pm$ 6.45 &   29.61 $\pm$ 2.66 &   12.87 $\pm$ 1.89 &    8.92 $\pm$ 1.95 \\
 $t\bar{t}$ + background (DT)   &                    &   ~2.41 $\pm$ 0.56 &   ~3.23 $\pm$ 0.76 &    4.03 $\pm$ 1.03 \\
 \hline
 Data  (ST)           &   66  &   35 &   10 &   11 \\
 Data  (DT)           &       &    5 &    6 &    2 \\
 \end{tabular}
 \label{tab:tab_15_2}
 \end{center}
 \end{table}
\narrowtext

\clearpage
\newpage

\widetext
 \begin{table}[p]
 \begin{center}
\def\arraystretch{0.8}
 \caption{ Composition of the $W+\geq$ 1 jet sample before
           tagging using $\sigma_{\ttbar}$ = 8.02 $\pm$ 2.16 pb.}
 \begin{tabular}{lcccc}
 Source                 &  $W+1 \,{\rm jet}$      &  $W+2 \,{\rm jet}$    
  &  $W+3 \,{\rm jet}$      &  $W+\geq4 \,{\rm jet}$   \\
 \hline
 Data                           &   9454 &   1370 &    198 &     54 \\
 Non-$W$                        &   560.1$\pm$ 14.9 &    71.2$\pm$  2.7 &    12.4$\pm$  2.0 &     5.1$\pm$  1.7 \\
 $WW$                           &    31.2$\pm$  5.4 &    31.1$\pm$  5.4 &     5.2$\pm$  1.0 &     0.8$\pm$  0.2 \\
 $WZ$                           &     4.4$\pm$  0.9 &     4.8$\pm$  1.0 &     0.9$\pm$  0.2 &     0.1$\pm$  0.0 \\
 $ZZ$                           &     0.3$\pm$  0.1 &     0.4$\pm$  0.1 &     0.1$\pm$  0.0 &     0.0$\pm$  0.0 \\
 Unidentified-$Z$               &   234.8$\pm$ 14.5 &    38.5$\pm$  5.9 &     7.9$\pm$  2.4 &     0.7$\pm$  0.7 \\
 Single top                     &    14.1$\pm$  2.1 &     7.9$\pm$  1.7 &     1.7$\pm$  0.4 &     0.3$\pm$  0.1 \\
 $Wc$                           &   413.1$\pm$123.9 &    86.4$\pm$ 25.9 &    10.3$\pm$  3.2 &     1.0$\pm$  0.7 \\
 $Wb\bar{b}$                    &    69.0$\pm$  9.5 &    29.5$\pm$  5.1 &     5.3$\pm$  1.0 &     0.8$\pm$  0.5 \\
 $Wc\bar{c}$                    &   173.1$\pm$ 46.2 &    61.6$\pm$ 13.5 &    10.5$\pm$  2.5 &     1.2$\pm$  0.8 \\
 $W+$ jets without h.f.          &  7951.0$\pm$133.5 &  1022.7$\pm$ 31.1 &   111.6$\pm$  9.2 &    10.3$\pm$  8.3 \\
 $t\bar{t}$                     &     2.9$\pm$  0.7 &    15.9$\pm$  3.8 &    32.1$\pm$  7.7 &    33.6$\pm$  8.1 \\
 \end{tabular}
 \label{tab:tab_15_3}
 \end{center}
 \end{table}
\narrowtext

\clearpage
\newpage
\widetext
 \begin{table}[p]
 \begin{center}
\def\arraystretch{0.8}
 \caption{Summary of the predicted and observed number of $W$ events with one (ST)
 or two (DT) 
          jet-probability tags.}
 \begin{tabular}{lcccc}
 Source                 &  $W+1 \,{\rm jet}$      &  $W+2 \,{\rm jet}$      &  $W+3 \,{\rm jet}$      &  $W+\geq4 \,{\rm jet}$   \\
 \hline
 Mistags                        &  ~41.80 $\pm$ 4.24 & 12.78 $\pm$ 1.33 & ~2.19 $\pm$ 0.27 & ~0.25 $\pm$ 0.19 \\
 Non-$W$                        &  ~12.55 $\pm$ 0.95 & ~2.53 $\pm$ 0.61 & ~0.57 $\pm$ 0.33 & ~0.24 $\pm$ 0.14 \\
 $WW,~WZ,~ZZ$                   &  ~~1.15 $\pm$ 0.26 & ~2.39 $\pm$ 0.43 & ~0.74 $\pm$ 0.19 & ~0.05 $\pm$ 0.04 \\
 Single top                     &  ~~1.32 $\pm$ 0.32 & ~2.19 $\pm$ 0.51 & ~0.59 $\pm$ 0.14 & ~0.11 $\pm$ 0.03 \\
 $Wc$                           &  ~34.80 $\pm$10.58 & ~9.02 $\pm$ 2.84 & ~1.67 $\pm$ 0.59 & ~0.16 $\pm$ 0.11 \\
 $Wc\bar{c}$ (ST)               &  ~17.02 $\pm$ 4.60 & ~7.24 $\pm$ 1.73 & ~1.70 $\pm$ 0.45 & ~0.20 $\pm$ 0.14 \\
 $Wc\bar{c}$ (DT)               &                    & ~0.47 $\pm$ 0.20 & ~0.05 $\pm$ 0.03 & ~0.01 $\pm$ 0.01 \\
 $Wb\bar{b}$ (ST)               &  ~16.43 $\pm$ 2.32 & ~7.47 $\pm$ 1.52 & ~1.47 $\pm$ 0.35 & ~0.21 $\pm$ 0.14 \\
 $Wb\bar{b}$ (DT)               &                    & ~1.42 $\pm$ 0.48 & ~0.25 $\pm$ 0.10 & ~0.03 $\pm$ 0.02 \\
 $Z\rightarrow \tau\tau$        &  ~~2.35 $\pm$ 0.47 & ~1.13 $\pm$ 0.32 & ~0.17 $\pm$ 0.12 & ~0.09 $\pm$ 0.09 \\
 $Zc$                           &  ~~0.28 $\pm$ 0.09 & ~0.08 $\pm$ 0.03 & ~0.03 $\pm$ 0.01 & ~0.00 $\pm$ 0.00 \\
 $Zc\bar{c}$ (ST)               &  ~~0.46 $\pm$ 0.13 & ~0.20 $\pm$ 0.06 & ~0.09 $\pm$ 0.04 & ~0.01 $\pm$ 0.01 \\
 $Zc\bar{c}$ (DT)               &                    & ~0.01 $\pm$ 0.01 & ~0.00 $\pm$ 0.00 & ~0.00 $\pm$ 0.00 \\
 $Zb\bar{b}$ (ST)               &  ~~0.90 $\pm$ 0.14 & ~0.42 $\pm$ 0.10 & ~0.16 $\pm$ 0.06 & ~0.02 $\pm$ 0.02 \\
 $Zb\bar{b}$ (DT)               &                    & ~0.07 $\pm$ 0.03 & ~0.03 $\pm$ 0.01 & ~0.00 $\pm$ 0.00 \\
 Total background (ST)          &  129.08 $\pm$12.56 & 45.53 $\pm$ 4.00 & ~9.43 $\pm$ 0.97 & ~1.34 $\pm$ 0.34 \\
 Total background (DT)          &                    & ~1.97 $\pm$ 0.52 & ~0.33 $\pm$ 0.10 & ~0.04 $\pm$ 0.02 \\
 $t\bar{t}$ (ST)                &  ~~0.80 $\pm$ 0.17 & ~4.77 $\pm$ 1.04 & ~9.93 $\pm$ 2.17 & 10.61 $\pm$ 2.32 \\
 $t\bar{t}$ (DT)                &                    & ~1.10 $\pm$ 0.24 & ~3.90 $\pm$ 0.85 & ~5.46 $\pm$ 1.19 \\
 \hline
 $t\bar{t}$ + background (ST)   &  129.87 $\pm$12.56 & 50.30 $\pm$ 4.14 & 19.37 $\pm$ 2.38 & 11.95 $\pm$ 2.35 \\
 $t\bar{t}$ + background (DT)   &                    & ~3.07 $\pm$ 0.57 & ~4.23 $\pm$ 0.86 & ~5.50 $\pm$ 1.20 \\
 \hline
 Data  (ST)           &  124 &   62 &   21 &   12 \\
 Data  (DT)           &      &    6 &    5 &    3 \\
 \end{tabular}
 \label{tab:tab_15_4}
 \end{center}
 \end{table}
\narrowtext

\clearpage
\newpage
\widetext
 \begin{table}[p]
 \begin{center}
 \caption{The composition of the $W+\geq$1 jet sample before
          tagging using $\sigma_{\ttbar}$ = 9.18 $\pm$ 4.26 pb.}

 \begin{tabular}{lcccc}
 Source                 &  $W+1 \,{\rm jet}$      &  $W+2 \,{\rm jet}$      &  $W+3 \,{\rm jet}$      &  $W+\geq4 \,{\rm jet}$   \\
 \hline
 Data                           &   9454 &   1370 &    198 &     54 \\
 Non-$W$                        &   560.1$\pm$ 14.9 &    71.2$\pm$  2.7 &    12.4$\pm$  2.0 &     5.1$\pm$  1.7 \\
 $WW$                           &    31.2$\pm$  5.4 &    31.1$\pm$  5.4 &     5.2$\pm$  1.0 &     0.8$\pm$  0.2 \\
 $WZ$                           &     4.4$\pm$  0.9 &     4.8$\pm$  1.0 &     0.9$\pm$  0.2 &     0.1$\pm$  0.0 \\
 $ZZ$                           &     0.3$\pm$  0.1 &     0.4$\pm$  0.1 &     0.1$\pm$  0.0 &     0.0$\pm$  0.0 \\
 Unidentified-$Z$         &   234.8$\pm$ 14.5 &    38.5$\pm$  5.9 &     7.9$\pm$  2.4 &     0.7$\pm$  0.7 \\
 Single top                     &    14.1$\pm$  2.1 &     7.9$\pm$  1.7 &     1.7$\pm$  0.4 &     0.3$\pm$  0.1 \\
 $Wc$                           &   413.1$\pm$123.9 &    86.3$\pm$ 25.9 &    10.0$\pm$  3.2 &     0.6$\pm$  1.3 \\
 $Wb\bar{b}$                    &    69.0$\pm$  9.5 &    29.5$\pm$  5.1 &     5.1$\pm$  1.1 &     0.5$\pm$  1.0 \\
 $Wc\bar{c}$                    &   173.1$\pm$ 46.2 &    61.5$\pm$ 13.5 &    10.1$\pm$  2.6 &     0.8$\pm$  1.6 \\
 $W+$ jet without h.f.                &  7950.6$\pm$133.5 &  1020.8$\pm$ 31.8 &   107.8$\pm$ 17.3 &     6.6$\pm$ 17.5 \\
 $t\bar{t}$                     &     3.3$\pm$  1.5 &    18.2$\pm$  8.2 &    36.7$\pm$ 16.5 &    38.5$\pm$ 17.3 \\
 \end{tabular}
 \label{tab:tab_15_5}
 \end{center}
 \end{table}
 \widetext

\clearpage
\newpage
\widetext
 \begin{table}[p]
 \begin{center}
\def\arraystretch{0.8}
 \caption{ Summary of the predicted and observed number of $W$ events with 
        one (ST) or two (DT)   SLT tags.}
 \begin{tabular}{lcccc}
 Source                 &  $W+1 \,{\rm jet}$      &  $W+2 \,{\rm jet}$      &  $W+3 \,{\rm jet}$      &  $W+\geq4 \,{\rm jet}$   \\
 \hline
 Mistags                        &  101.92 $\pm$10.19 & 30.90 $\pm$ 3.09 & ~7.34 $\pm$ 0.73 &    3.01 $\pm$ 0.30 \\
 Non-$W$                        &  ~~8.96 $\pm$ 0.84 & ~2.09 $\pm$ 0.56 & ~0.38 $\pm$ 0.27 &    0.16 $\pm$ 0.11 \\
 $WW,~WZ,~ZZ$                   &  ~~0.50 $\pm$ 0.16 & ~0.88 $\pm$ 0.22 & ~0.10 $\pm$ 0.05 &    0.00 $\pm$ 0.00 \\
 Single top                     &  ~~0.38 $\pm$ 0.10 & ~0.67 $\pm$ 0.15 & ~0.18 $\pm$ 0.05 &    0.05 $\pm$ 0.01 \\
 $Wc$                           &  ~13.12 $\pm$ 4.27 & ~4.26 $\pm$ 1.45 & ~0.65 $\pm$ 0.29 &    0.04 $\pm$ 0.09 \\
 $Wc\bar{c}$ (ST)               &  ~~6.41 $\pm$ 1.89 & ~2.68 $\pm$ 0.66 & ~0.61 $\pm$ 0.21 &    0.05 $\pm$ 0.10 \\
 $Wc\bar{c}$ (DT)               &                    & ~0.02 $\pm$ 0.02 & ~0.00 $\pm$ 0.00 &    0.00 $\pm$ 0.00 \\
 $Wb\bar{b}$ (ST)               &  ~~5.31 $\pm$ 0.96 & ~2.84 $\pm$ 0.67 & ~0.41 $\pm$ 0.13 &    0.04 $\pm$ 0.08 \\
 $Wb\bar{b}$ (DT)               &                    & ~0.09 $\pm$ 0.05 & ~0.01 $\pm$ 0.01 &    0.00 $\pm$ 0.00 \\
 $Z\rightarrow \tau\tau$        &  ~~0.43 $\pm$ 0.20 & ~0.09 $\pm$ 0.09 & ~0.09 $\pm$ 0.09 &    0.00 $\pm$ 0.00 \\
 $Zc$                           &  ~~0.11 $\pm$ 0.04 & ~0.04 $\pm$ 0.01 & ~0.01 $\pm$ 0.01 &    0.00 $\pm$ 0.00 \\
 $Zc\bar{c}$ (ST)               &  ~~0.17 $\pm$ 0.05 & ~0.08 $\pm$ 0.02 & ~0.03 $\pm$ 0.01 &    0.00 $\pm$ 0.00 \\
 $Zc\bar{c}$ (DT)               &                    & ~0.00 $\pm$ 0.00 & ~0.00 $\pm$ 0.00 &    0.00 $\pm$ 0.00 \\
 $Zb\bar{b}$ (ST)               &  ~~0.29 $\pm$ 0.06 & ~0.16 $\pm$ 0.04 & ~0.05 $\pm$ 0.02 &    0.01 $\pm$ 0.01 \\
 $Zb\bar{b}$ (DT)               &                    & ~0.00 $\pm$ 0.00 & ~0.00 $\pm$ 0.00 &    0.00 $\pm$ 0.00 \\
 Total background (ST)          &  137.60 $\pm$11.29 & 44.66 $\pm$ 3.60 & ~9.86 $\pm$ 0.88 &    3.35 $\pm$ 0.36 \\
 Total background (DT)          &                    & ~0.10 $\pm$ 0.05 & ~0.01 $\pm$ 0.01 &    0.00 $\pm$ 0.00 \\
 $t\bar{t}$ (ST)                &  ~~0.25 $\pm$ 0.11 & ~2.44 $\pm$ 1.07 & ~5.14 $\pm$ 2.25 &    6.08 $\pm$ 2.66 \\
 $t\bar{t}$ (DT)                &                    & ~0.07 $\pm$ 0.03 & ~0.24 $\pm$ 0.10 &    0.32 $\pm$ 0.14 \\
 \hline
 $t\bar{t}$ + background (ST)   &  137.85 $\pm$11.29 & 47.10 $\pm$ 3.75 & 15.00 $\pm$ 2.41 &    9.43 $\pm$ 2.68 \\
 $t\bar{t}$ + background (DT)   &                    & ~0.17 $\pm$ 0.06 & ~0.25 $\pm$ 0.10 &    0.32 $\pm$ 0.14 \\
 \hline
 Data  (ST)           &  146 &   56 &   17 &    8 \\
 Data  (DT)           &      &    0 &    0 &    0 \\
 \end{tabular}

 \label{tab:tab_15_6}
 \end{center}
 \end{table}
\narrowtext

\clearpage
\section{Combined $t\bar{t}$ production cross section}
\label{sec:comb}
      The best measurement of the $t\bar{t}$ cross section comes from combining
      the results of this analysis with the dilepton and all-hadronic
      analyses~\cite{topdil,allhad}.
      The revised $b$-tagging efficiency reported in this
      paper effects the cross section measured in the all-hadronic channel.
      The details of this analysis have not changed from those reported in
      Ref.~\cite{allhad}.
      The cross section measurement from the dilepton channel~\cite{topdil} 
      does not require $b$-tagging information and so is unchanged by the 
      revisions reported here. It is affected slightly by the revised determination
      of the total integrated luminosity as were all measurements.
      A comparison between the previously published results and the revised
      cross sections used for the new combined result is shown in
      Table~\ref{tab:tab_16.0}.

      We combine the measurements from the SVX and SLT tagged lepton + jets,
      all-hadronic, and dilepton channels, using a maximum likelihood
      technique similar to that described in Refs.~\cite{topxs,top_prd}.
      This procedure properly accounts for correlated systematic uncertainties, 
      such as the uncertainties in the $b$-tagging efficiency, the luminosity, 
      the kinematic acceptance, and some of the calculated backgrounds.  
      In all channels the acceptances are calculated with $M_{top} = 175$ $\gevcc$. 
      The resulting combined $t\bar{t}$ production cross section is
 \[ \sigma_{t\bar{t}} = 6.5^{+1.7}_{-1.4}\,{\rm pb} \]
      where the quoted uncertainties include both statistical and systematic
      effects, which are approximately equal in magnitude.
\begin{table}[h]
 \begin{center}
\caption[]{Summary of old and new CDF $t\bar{t}$ production cross section
results.}
\begin{tabular}{lcc}
Channel  & Previous result  &  New result  \\
\hline
Lepton + jets (SVX)   & $6.2^{+2.1}_{-1.7}$ pb  & $5.1 \pm 1.5$ pb \\
Lepton + jets (SLT)   & $9.2^{+4.3}_{-3.6}$ pb  & $9.2 \pm 4.3$ pb \\
Dilepton              & $8.2^{+4.4}_{-3.4}$ pb  & $8.4^{+4.5}_{-3.5}$ pb \\
All-hadronic          & $10.1^{+4.5}_{-3.6}$ pb & $7.6^{+3.5}_{-2.7}$ pb \\
 \end{tabular}
 \label{tab:tab_16.0}
 \end{center}
 \end{table}

\section{Conclusions} 
\label{sec:concl}
       Having improved the method for determining the $b$-tagging efficiency, and
       the method for calculating the backgrounds to $t\bar{t}$ production, we
       revise our previous measurements of $\sigma_{t\bar{t}}$ in the lepton + jets
       channel~\cite{cdf_xsec}. We find $\sigma_{t\bar{t}} = 5.08 \pm 1.54\,{\rm pb}$
       and  $\sigma_{t\bar{t}} = 9.18 \pm 4.26\,{\rm pb}$ using events with
       SECVTX and SLT tags, respectively. We have used the jet-probability 
       algorithm as a cross-check and find that it gives a result consistent with 
       these measurements.
       The  measurement of the $t\bar{t}$ cross section, obtained by combining
       the results of this analysis with the dilepton and all-hadronic
       analyses, is  $\sigma_{t\bar{t}} = 6.5^{+1.7}_{-1.4}\,{\rm pb} $, in agreement
       with the SM predictions~\cite{catani} and the measurement performed by the
       D\O~collaboration~\cite{d0}.
\acknowledgments
       We thank the Fermilab staff and the technical staff of the participating
       Institutions for their contributions. This work was supported by the 
       U.S.~Department of Energy and National Science Foundation; the Istituto 
       Nazionale di Fisica Nucleare; the Ministry of Education and Culture of Japan; 
       the Natural Science and Engineering
       Research Council of Canada; the National Science Council of the republic 
       of China;
       the Swiss National Science Foundation; the A.P.~Sloan Foundation; the
       Bundesministerium fuer Bildung und Forschung; and the  Korea Science and 
       Engineering Foundation.


\begin{thebibliography}{99}
\bibitem{isoltr} The track isolation requirement is that the sum of the 
                 transverse momentum of all tracks in a cone of radius around 
                 the candidate track direction  be less than 4 $\gevc$.
\bibitem{herw_setup} We use Herwig {\sc v5.6} with $m_b$ = 4.75 GeV/c$^2$ and 
                    $m_c$ = 1.50 GeV/c$^2$ for the $b$ and $c$-quark masses, 
                    respectively. The value of the parameter VGCUT in the 
                    parton shower is set to VGCUT=0.06 and the value of the 
                    parameter CLPOW in the cluster hadronization is set to 
                    CLPOW=1.3. 
                    The generic-jet and low-$p_T$ Monte Carlo samples use
   	 	            the MRS(G) structure functions, while all the other
                    Monte Carlo use the MRS~D$_0^{\prime}$ structure functions.
\end{thebibliography}
\end{document}